\PassOptionsToPackage{unicode}{hyperref}
\PassOptionsToPackage{bookmarks=false}{hyperref}
\documentclass[preprint,12pt,num,dvipsnames,usenames,sort&compress]{elsarticle}
\usepackage{lineno}
\usepackage{verbatim}
\usepackage[utf8]{inputenc}
\usepackage{subfig}
\usepackage{wrapfig}
\usepackage{graphicx}
\usepackage{float}
\usepackage{upgreek}
\usepackage{multicol}
\usepackage{longtable}
\usepackage[table]{xcolor}
\usepackage{tabularx,colortbl}
\definecolor{lightgray}{gray}{0.9}
\usepackage{geometry,fullpage}
\usepackage{units}
\usepackage{amsmath,amssymb}
\usepackage{fancybox}
\usepackage{paralist,relsize}
\usepackage{xspace}

\newcommand{\nnbe}{\stackrel{{\scriptscriptstyle (} - {\scriptscriptstyle)}}{\nu}\!\!\!\! _{\rm e}}
\newcommand{\dm}{{\rm DM}}

\newcommand{\atm}{{\rm atm}}

\newcommand{\tr}{{\rm tr}}
\newcommand{\CS}{\mathcal{S}}
\newcommand{\cnb}{C$\nu$B}
\newcommand{\rnb}{C$\nu_{keV}$B}
\newcommand{\snb}{S$\nu$B}

\newcommand{\slashed}[1]{\displaystyle{\not}{#1}}
\newcommand{\p}{p}
\newcommand{\pp}{p}
\newcommand{\q}{q}
\newcommand{\x}{x}
\newcommand{\M}{M}
\newcommand{\N}{N}
\newcommand{\qq}{q}
\newcommand{\GeV}{{\rm GeV}}
\newcommand{\TeV}{\rm TeV}
\newcommand{\keV}{\rm keV}
\newcommand{\nustar}{\textit{NuStar}\xspace}
\newcommand{\xmm}{\textit{XMM-Newton}\xspace}
\newcommand{\chandra}{\textit{Chandra}\xspace}
\newcommand{\suzaku}{\textit{Suzaku}\xspace}
\newcommand{\Lnu}{l}

\usepackage{hyperref}
\newcommand\myshade{85}
\colorlet{mylinkcolor}{violet}
\colorlet{mycitecolor}{Aquamarine}
\colorlet{myurlcolor}{YellowOrange}

\hypersetup{
  linkcolor  = mylinkcolor,
  citecolor  = mycitecolor!\myshade!black,
  urlcolor   = myurlcolor!\myshade!black,
  colorlinks = true,
}
\journal{Progress in Particle and  Nuclear Physics}

\begin{document}

\begin{frontmatter}

\title{Sterile Neutrino Dark Matter}

\author[1]{A.~Boyarsky}
\author[2,4]{M.~Drewes}
\author[5,6,7,8]{T.~Lasserre}
\author[7,9]{S.~Mertens}
\author[12]{O.~Ruchayskiy}
\address[1]{Universiteit Leiden - Instituut Lorentz for Theoretical Physics, P.O. Box 9506, NL-2300 RA Leiden, Netherlands}
\address[2]{Centre for Cosmology, Particle Physics and Phenomenology, Universit\'{e} catholique de Louvain, Louvain-la-Neuve B-1348, Belgium}
\address[4]{Excellence Cluster Universe, Boltzmannstr. 2, D-85748, Garching, Germany}
\address[5]{Commissariat \`{a} l'\'energie atomique et aux \'energies alternatives, Centre de Saclay,DRF/IRFU, 91191 Gif-sur-Yvette, France}
\address[6]{Institute for Advance Study, Technische Universit\"at M\"unchen, James-Franck-Str. 1, 85748 Garching}
\address[7]{Physik-Department, Technische Universit\"at M\"unchen, James-Franck-Str. 1, 85748 Garching}
\address[8]{AstroParticule et Cosmologie, Universit\'e Paris Diderot, CNRS/IN2P3, CEA/IRFU, Observatoire de Paris, Sorbonne Paris Cit\'e, 75205 Paris Cedex 13, France}
\address[9]{Max-Planck-Institut f\"ur Physik (Werner-Heisenberg-Institut), Foehringer Ring 6, 80805 M\"unchen, Germany}
\address[12]{Discovery Center, Niels Bohr Institute, Copenhagen University, Blegdamsvej 17, DK-2100 Copenhagen, Denmark}

\address{}

\begin{abstract}
We review sterile neutrinos as possible Dark Matter candidates. After a short summary on the role of neutrinos in cosmology and particle physics, we give a comprehensive overview of the current status of the research on sterile neutrino Dark Matter. First we discuss the motivation and limits obtained through astrophysical observations. Second, we review different mechanisms of how sterile neutrino Dark Matter could have been produced in the early universe. Finally, we outline a selection of future laboratory searches for keV-scale sterile neutrinos, highlighting their experimental challenges and discovery potential.
\end{abstract}


\end{frontmatter}


\newpage
\begin{footnotesize}
\vspace{-0.5cm}
\tableofcontents
\end{footnotesize}

\section{Dark Matter in the Universe}
\label{sec:intro}

There is a body of strong and convincing evidence that most of the mass in the observable universe is not composed of known particles.
Indeed, numerous independent tracers of the gravitational potential (observations of the
motion of stars in galaxies and galaxies in clusters; emissions from hot
ionised gas in galaxy groups and clusters; 21 cm line in galaxies; both weak
and strong gravitational lensing measurements) demonstrate that the dynamics
of galaxies and galaxy clusters cannot be explained by the Newtonian potential
created by visible matter only. Moreover, cosmological data (analysis of the
cosmic microwave background anisotropies and of the statistics of galaxy
number counts) show that the cosmic large scale structure started to develop
long before decoupling of photons due to the recombination of hydrogen in the early
Universe and, therefore, long before ordinary matter could start
clustering. This evidence points at the existence of a new substance,
universally distributed in objects of all scales and providing a contribution
to the total energy density of the Universe at the level of about 27\%.
This hypothetical new substance is commonly known as "Dark Matter" (DM).
The DM abundance is often expressed in terms of the \emph{density parameter} $\Omega_{\rm DM}=\rho_{\rm DM}/\rho_0$, where $\rho_{\rm DM}$ is the comoving DM density and $\rho_0=3H^2 m_{Pl}^2/(8\pi)$ is the critical density of the universe, with $H$ the Hubble parameter and $m_{Pl}$ the Planck mass.
Current measurements suggest $\Omega_{\rm DM} h^2=0.1186
\pm0.0020$, where $h$ is $H$ in units $100 \ {\rm km}/({\rm s \ Mpc})$ \cite{Patrignani:2016xqp}.
Different aspects of the DM problem can be found in
reviews~\cite{Einasto:2011jw,Frenk:2012ph,Peebles:2013hla}, for historical
exposition of the problem see~\cite{Frenk:2012ph,Bertone:2016nfn,deSwart:2017heh}. 
Various attempts to explain this phenomenon by the presence of macroscopic compact objects (such
as, for example, old stars~\cite{Yoo:2003fr,Moniez:09,Alves:00,MACHO:00}) or
by modifications of the laws of gravity 
(for a review see~\cite{Starkman:2012mk,Milgrom:2012xw,McGaugh:2014nsa}) failed to provide a consistent description of all the above phenomena (see the overviews in~\cite{McGaugh:2014nsa,Bull:2015stt}).
Therefore, a microscopic  origin of DM phenomenon, i.e., a new particle or particles, remains the most plausible hypothesis.\footnote{Primordial black holes that formed prior to the baryonic acoustic oscillations visible in the sky are one of the few alternative explanations that are still viable, see e.g.  \cite{Carr:2017jsz}.}

\subsection{Standard Model Neutrino as Dark Matter Candidate?}
\label{sec:neutrino-dm}

The only electrically neutral and long-lived particles in the Standard Model (SM) of particle physics are the neutrinos, the properties of which are briefly reviewed in Sec.~\ref{sec:seesaw} (cf. e.g. \cite{Lesgourgues:2006nd} for a review of neutrinos in cosmology).
As experiments show that neutrinos have mass, they could in principle play the role of DM particles.  Neutrinos are involved in weak
interactions (\ref{WeakWW}) that keep them  in thermal equilibrium in the early Universe down to the
temperatures of few MeV. At smaller temperatures, the interaction rate of weak
reactions drops below the expansion rate of the Universe and neutrinos
``freeze out'' from the equilibrium.  Therefore, a background of relic
neutrinos was created just before primordial nucleosynthesis.
As interaction
strength and, therefore, decoupling temperature of these
particles are known, one can easily find that their number density, equal per each flavour to
\begin{equation}
  \label{eq:1}
  n_\alpha = \frac{6}{4}\frac{\zeta(3)}{\pi^2}T_\nu^3 
\end{equation}
where $T_\nu\simeq (4/11)^{1/3}T_\gamma\simeq 1.96 {\rm K} \simeq 10^{-4}$ eV.\footnote{We use natural units throughout the theoretical discussion in sections \ref{sec:intro}-\ref{exp}.}
The associated matter density of neutrinos at late stage (when neutrinos are non-relativistic) is determined by the sum of neutrino masses
\begin{equation}
  \label{eq:2}
  \rho_\nu \simeq n_\alpha \sum m_i
\end{equation}
To constitute the whole DM this sum
should be about $11.5$~eV (see e.g.~\cite{Lesgourgues:2006nd}). Clearly, this 
is in conflict with the existing experimental bounds: measurements of the
electron spectrum of $\beta$-decay put the combination of neutrino masses
below $\unit[2]{eV}$~\cite{Agashe:2014kda} while from the cosmological data one can
infer an upper bound of the sum of neutrino masses varies between $\unit[0.58]{eV}$ at
95\% CL ~\cite{Ade:2015xua} and $\unit[0.12]{eV}$ \cite{Vagnozzi:2017ovm,Lattanzi:2017ubx}, depending on the dataset included and assumptions made in the fitting.  
The fact that neutrinos could not constitute 100\% of
DM follows also from the study of phase space density of DM dominated objects
that should not exceed the density of degenerate Fermi gas: fermionic
particles could play the role of DM in dwarf galaxies only if their mass is
above few hundreds of eV (the so-called 'Tremaine-Gunn
bound'~\cite{Tremaine:1979we}, for review see \cite{Boyarsky:2008ju} and references
therein) and in galaxies if their mass is tens of eV.  Moreover, as the mass
of neutrinos is much smaller than their decoupling temperature, they decouple
relativistic and become non-relativistic only deeply in matter-dominated epoch
(``\emph{Hot Dark Matter}''). For such a DM candidate the history of structure
formation would be very different and the Universe would look rather
differently nowadays~\cite{White:1984yj}.  All these strong arguments prove
convincingly that the \emph{dominant fraction of DM} can not be made of
the known neutrinos and therefore \emph{the Standard Model of
  elementary particles does not contain a viable DM candidate.}

\subsection{Solution to Dark Matter Puzzle in Different Approaches to 
  BSM Physics}
\label{sec:bsm-dm}
The hypothesis that DM is made of particles necessarily implies an extension of the SM with new particles. 
This makes the DM problem part of a small number 
of observed phenomena in particle physics, astrophysics and cosmology that clearly point towards the existence of ``New Physics''.  These
major unsolved challenges are commonly known as ``beyond the Standard Model'' (BSM)
problems and include 
\begin{itemize}
\item[I)] \textbf{Dark Matter}: What is it composed of, and how was it produced? 
\item[II)] \textbf{Neutrino oscillations}: Which mechanism gives masses to the known neutrinos? 
\item[III)] \textbf{Baryon asymmetry of the Universe}: What mechanism created the tiny matter-antimatter asymmetry in the early Universe?  This \emph{baryon asymmetry of the universe} (BAU) is believed to be the origin of all baryonic matter in the present day universe after mutual annihilation of all other particles and antiparticles, cf. e.g. \cite{Canetti:2012zc}.
\item[IV)] \textbf{The hot big bang}: Which mechanism set the homogeneous and isotropic initial conditions of the radiation dominated epoch in cosmic history?
In particular, if the initial state was created during a stage of
accelerated expansion (\emph{cosmic inflation}), what was driving it?
\end{itemize}
In addition to these observational puzzles, there are also deep theoretical questions about the structure of the SM: the \emph{gauge hierarchy problem}, \emph{strong   CP-problem}, the \emph{cosmological constant problem}, the \emph{flavour puzzle} and the question why the SM gauge group is $SU(3)\times SU(2) \times U(1)$.  Some yet unknown particles or interactions would be needed to answer these questions.
 
\label{sec:wimp}
Perhaps, most of the research in BSM
  physics during the last decades was devoted to a solution of the \emph{gauge hierarchy problem},
  i.e.\ the problem of quantum stability of the mass of the Higgs boson
  against radiative corrections. The requirement of the absence of
  quadratically divergent corrections to the Higgs boson is an example of so-called ``naturalness'', cf. eg. \cite{Giudice:2008bi}.  
  Quite a number of different suggestions were proposed
  on how the ``naturalness'' of the electroweak symmetry breaking can be
  achieved. They are based on supersymmetry, technicolour, large extra
  dimensions or other ideas. Most of these approaches postulate the existence of new
  particles that participate in electroweak interactions. Therefore in these
  models \emph{weakly interacting massive particles} (WIMPs) appear as natural
  DM candidates. WIMPs generalise the idea of neutrino DM~\cite{Lee:1977ua}: they also interact with the SM sector with roughly electroweak strength, however their mass is large (from few GeV to hundreds of TeV), so that these particles are
already non-relativistic when they decouple from primordial plasma. This
suppresses their number density and they easily satisfy the lower mass  bound that ruled out the known neutrinos as DM.  In this case the present day
density of such particles depends very weakly (logarithmically) on the mass
of the particle as long as it is heavy enough.  This ``universal'' density
happens to be within the order of magnitude consistent with DM density (the
so-called ``\emph{WIMP miracle}''). WIMPs are usually stable or very long lived, but can annihilate with each other in the regions of large DM
  densities, producing a flux of $\gamma$-rays, antimatter and neutrinos.
However, to date, neither particle colliders nor any of the large number of direct and indirect DM searches could provide convincing evidence for the existence of WIMPs. This provides clear motivation to investigate alternatives to the WIMP paradigm.
Indeed, there exist many DM candidates that differ by their mass,
interaction types and interaction strengths by many order of magnitude (for
reviews see e.g.~\cite{Bertone:2004pz,Anchordoqui:2015uww,Taoso:07}).

\subsection{Heavy "Sterile" Neutrinos}
As we saw above, neutrinos in principle are a very natural DM candidate. The reasons why the known neutrinos cannot compose all of the observed DM are the smallness of their mass and the magnitude of their coupling to other particles. Hence, one obvious solution is to postulate the existence of heavier ``sterile'' neutrinos with weaker interactions that fulfil the constraints from cosmic structure formation and phase space densities.
Indeed, the existence of heavy neutrinos is predicted by many theories of particle physics, and they would provide a very simple explanation for the observed neutrino oscillations via the seesaw mechanism, which we briefly review in Sec,~ \ref{sec:SeeSawwMechanism}.
The implications of the existence of heavy neutrinos strongly depend on the magnitude of their mass, see e.g. \cite{Drewes:2013gca} for a review. 
For masses of a few keV, they are a viable DM candidate~\cite{Dodelson:1993je,Shi:1998km,Dolgov:2000ew,Abazajian:2001vt,Abazajian:2001nj,Asaka:2005an,Asaka:2005pn,Asaka:2006nq}. 
Sterile neutrino DM
interacts much weaker than ordinary neutrinos.
These particles can  leave imprints in X-ray spectra of galaxies and
galaxy clusters
\cite{Dolgov:2000ew,Abazajian:2001vt,Boyarsky:2005us,Boyarsky:2006fg,Herder:2009im}.
Moreover, they decay \cite{Shrock:1974nd} and X-ray observations provide bounds on their parameters.
There exist a larger number of models that accommodate this possibility, see e.g. \cite{Adhikari:2016bei} for a review.

The present article provides an update of the phenomenological constraints on sterile neutrino DM. 
In Sec.~\ref{sec:seesaw} we briefly review the role of neutrinos in particle physics and define our notation. We in particular address the idea of heavy sterile neutrinos in Sec.~\ref{sec:SeeSawwMechanism}. In Sec.~\ref{sec:dm} we provide an overview of the observational constraints on sterile neutrino DM. 
These partly depend on the way how the heavy neutrinos were produced in the early universe. Different production mechanisms are reviewed in Sec.~\ref{sec:production}. Sec.~\ref{exp} is devoted to laboratory searches for DM sterile neutrinos. We finally conclude in Sec.~\ref{Conclusion}.

\section{Neutrinos in the Standard Model and Beyond}
\label{sec:seesaw}

Neutrinos are the most elusive known particles. 
Their weak interactions make it very difficult to study their properties. At the same time, there are good reasons to believe that neutrinos may hold a key to resolve several mysteries in particle physics and cosmology. 
Neutrinos are unique in several different ways.
\begin{itemize} 
\item Neutrinos are the only fermions that appear only with left handed (LH) chirality in the SM. 
\item In the minimal SM, neutrinos are massless. The observed neutrino flavour oscillations clearly indicate that at least two neutrinos have non-vanishing mass. In the framework of renormalisable quantum field theory, the existence of neutrino masses definitely implies that some new states exist, see Sec.~\ref{NewPhysicsSec}.
This is why neutrino masses are often referred to as the only sign of New Physics that has been found in the laboratory.\footnote{This "New Physics" could in principle be fairly boring if it only consists of new neutrino spin orientations, cf.\ the discussion following Eq.~(\ref{diracmassterm}), or could provide a key to understand how the SM is embedded in a more fundamental theory of nature, cf. sec.~\ref{NewPhysicsSec}.}
\item The neutrino masses are much smaller than all other fermion masses in the SM. The reason for this separation of scales is unclear. This is often referred to as the \emph{mass puzzle}.
\item The reason why neutrinos oscillate is that the quantum states in which they are produced by the weak interaction (\emph{interaction eigenstates}) 
are not quantum states with a well defined energy (\emph{mass eigenstates}).
The misalignment between both sets of states can be described by a flavour mixing matrix $V_\nu$  in analogy to the mixing of quarks by the Cabibbo-Kobayashi-Maskawa (CKM) matrix.\footnote{In the basis where charged Yukawa couplings are diagonal, the mixing matrix $V_\nu$ is identical to the Pontecorvo-Maki-Nakagawa-Sakata matrix \cite{Maki:1962mu,Pontecorvo:1967fh}.} However, while the CKM matrix is very close to unity, the neutrino mixing matrix $V_\nu$ looks very different and shows no clear pattern. This is known as the \emph{flavour puzzle}.
\end{itemize}
We refer to the three known neutrinos that appear in the SM as \emph{active neutrinos} because they feel the weak interaction with full strength. This is in contrast to the hypothetical \emph{sterile neutrinos} that e.g.\ may compose the DM, which are gauge singlets.
In the following we very briefly review some basic properties of neutrinos before moving on to the specific topic of sterile neutrinos as DM candidates; for a more detailed treatment we e.g. refer the reader to ref.~\cite{Giunti:2007ry}.

\subsection{Neutrinos in the Standard Model}
Neutrinos can be produced in the laboratory in two different ways. On one hand, they appear as a by-product in nuclear reactions, and commercial nuclear power plants generate huge amounts of neutrinos "for free". The downside is that the neutrinos are not directed, and their energy spectrum is not known with great accuracy.
On the other hand, a neutrino can be produced by sending a proton beam on a fixed target. The pions that are produced in these collisions decay into muons and neutrinos. Both mechanisms are, in a similar way, also realised in nature. 
Neutrinos from nuclear reactions in nature include the
solar neutrinos are produced in fusion reactions in the core of the sun and a small neutrino flux due to the natural radioactivity in the soil. 
Atmospheric neutrinos, on the other hand, are produced in the cascade of decays following high energy collisions of cosmic rays with nuclei in the atmosphere.
A large number of experiments has been performed to study neutrinos from all these sources.  In the following we briefly summarise the combined knowledge that has been obtained from these efforts, without going into too experimental details.

In the SM, there exist three neutrinos.
They are usually classified in terms of their interaction eigenstates $\nu_{\alpha}$, where $\alpha=e, \mu, \tau$ refers to the ``family'' or ``generation'' each of them belongs to,\footnote{The generation that a neutrino belongs to is defined in the flavour basis where the weak currents have the form (\ref{WeakWW}).}
and referred to as ``electron neutrino'', ``muon neutrino'' and ``tau neutrino''. This convention is in contrast to the quark sector, where the particle names $u$, $d$, $s$, $c$, $b$ and $t$ refer to mass eigenstates. 
Neutrinos interact with other particles only via the weak interaction,\footnote{They may have additional interactions that are related to the mechanism of neutrino mass generation, such as the Yukawa interactions in the seesaw Lagrangian (\ref{Lseesaw}). These are usually not relevant at low energies.}
\begin{equation}\label{WeakWW}
-\frac{g}{\sqrt{2}}\overline{\nu_L}\gamma^\mu e_L W^+_\mu
-\frac{g}{\sqrt{2}}\overline{e_L}\gamma^\mu \nu_L W^-_\mu  
- \frac{g}{2\cos\theta_W}\overline{\nu_L}\gamma^\mu\nu_L Z_\mu ,
\end{equation}
where $g$ is the gauge coupling constant and $\theta_W$ the Weinberg angle  and $\nu_L=(\nu_{L e}, \nu_{L \mu}, \nu_{L \tau})^T$ is a flavour vector of left handed (LH) neutrinos.
Neutrino oscillation data implies that the three interacting states $\nu_{\alpha}$ are composed of at least three different mass eigenstates $\nu_i$,  and the corresponding spinors are related by the transformation
\begin{equation}\label{NeutrinoMixing}
\nu_{L \alpha} = (V_\nu)_{\alpha i}\nu_i. 
\end{equation}
While the number of active neutrinos is known to be three,\footnote{Adding a fourth neutrino that is charged under the weak interaction would imply that one has to add a whole fourth generation of quarks and lepton to the SM to keep the theory free of anomalies. This possibility is (at least in simple realisations) strongly disfavoured by data \cite{Lenz:2013iha}. }
it is not known how many mass states are contained in these because the $\nu_{L \alpha}$ could mix with an unknown number of sterile neutrinos. In Eq.~(\ref{NeutrinoMixing}) we assume that there are three $\nu_i$ and take the possibility of additional states into account by allowing a non-unitarity $\eta$ in the mixing matrix,
\begin{equation}\label{nonunitarity}
V_\nu = \left(1 + \eta\right)U_\nu.
\end{equation} 
The matrix $U_\nu$ can be parametrised as
\begin{equation}
U_\nu=V^{(23)}U_{\delta}V^{(13)}U_{-\delta}V^{(12)}{\rm diag}(e^{i\alpha_1 /2},e^{i\alpha_2 /2},1)
\end{equation}
with $U_{\pm\delta}={\rm diag}(e^{\mp i\delta/2},1,e^{\pm i\delta/2})$.  
The matrices $V^{(ij)}$ can be expressed as
\begin{eqnarray}
V^{(23)}=\left(\begin{tabular}{ccc}
$1$ & $0$ & $0$ \\ $0$ & $c_{23}$ & $s_{23}$ \\ $0$ & $-s_{23}$ & $c_{23}$
\end{tabular}\
\right) \ &,& \
V^{(13)}=\left(
\begin{tabular}{ccc}
$c_{13}$ & 0 & $s_{13}$ \\ 0 & 1 & 0 \\ $-s_{13}$ & 0 & $c_{13}$
\end{tabular}
\right),\nonumber\\
V^{(12)}&=&\left(
\begin{tabular}{ccc}
$c_{12}$ & $s_{12}$ &  0\\ $-s_{12}$ & $c_{12}$ & 0 \\ 0 & 0 & 1
\end{tabular}
\right), 
\end{eqnarray}
where $c_{ij}=\cos(\uptheta_{ij})$ and $s_{ij}\sin(\uptheta_{ij})$. $\uptheta_{ij}$ are the neutrino mixing angles, $\alpha_1$, $\alpha_2$ and $\delta$ are $CP$-violating phases.

Under the assumption $\eta=0$, neutrino oscillation data constrains most parameters in $U_\nu$.\footnote{This assumption is in principle not necessary \cite{Antusch:2006vwa}, but only leads to small errors if $\eta$ is small. It is mainly justified by the fact that it gives a good fit to the data from almost all neutrino oscillation experiments.}
The values of the mixing angles are $\uptheta_{12}\simeq 34^{\circ}$, $45^{\circ}<\uptheta_{23}<50^{\circ}$ and $\uptheta_{13}\simeq 9^{\circ}$.
The masses  $m_i$ of the $\nu_i$ are unknown, as neutrino oscillations are only sensitive to difference $m_i^2-m_j^2$. In particular, two mass square differences have been determined as $\Delta m_{\rm sol}^2\equiv m_2^2-m_1^2\simeq 7.4\times10^{-5}{\rm eV}^2$ and $\Delta m_{\rm atm}^2\equiv|m_3^2-m_1^2|\simeq 2.5\times10^{-3}{\rm eV}^2$, meaning that at least two $\nu_i$ have non-zero masses. 
The precise best fit values differ for normal and inverted hierarchy (in particular for $\uptheta_{23}$). 
They can e.g. be found at \href{http://www.nu-fit.org/}{http://www.nu-fit.org/}, see also \cite{Bergstrom:2015rba,Esteban:2016qun,deSalas:2018bym}.
What remains unknown are 
\begin{itemize}
\item{\it The hierarchy of neutrino masses} - One can distinguish between 
two different orderings amongst the $m_i$. The case $m_1<m_2<m_3$, with $\Delta m_{\rm sol}^2=m_2^2-m_1^2 $ and $\Delta m_{\rm atm}^2=m_3^2-m_1^2\simeq m_3^2-m_2^2\gg  \Delta m_{\rm sol}^2 $, is called {\it normal hierarchy}. 
The case $m_3^2<m_1^2<m_2^2$, with $\Delta m_{\rm sol}^2=m_2^2-m_1^2 $ and $\Delta m_{\rm atm}^2=m_1^2-m_3^2\simeq m_2^2-m_3^2\gg  \Delta m_{\rm sol}^2 $, is referred to as {\it inverted hierarchy}. The next generation of neutrino oscillation experiments is expected to determined the hierarchy.
\item{\it The CP-violating phases} - The Dirac phase $\delta$ is the analogue to the CKM phase. Global fits to neutrino data tend to prefer $\delta\neq 0$, but are not conclusive yet. $\delta$ may be measured by the DUNE or NOvA experiments in the near future.
The Majorana phases $\alpha_1$ and $\alpha_2$ have no equivalent in the quark sector. They are only physical if neutrinos are Majorana particles (see below).
For Dirac neutrinos they can be absorbed into redefinitions of the fields.
\item{\it The absolute mass scale} - The mass $m_{\rm lightest}$ of the lightest neutrino is unknown, but the sum of masses is bound from above as $\sum_i m_i< 0.23$ eV \cite{Ade:2015xua} by cosmological data from Cosmic Microwave Background (CMB) observations with the Planck satellite.\footnote{
During the final stage of writing this document new Planck results were make public that limit the sum of light neutrino masses to $0.12$ eV \cite{Aghanim:2018eyx}.
} 
A stronger bound $\sum_i m_i< 0.12$ eV \cite{Palanque-Delabrouille:2015pga}  can be derived if other cosmological datasets are included \cite{Lattanzi:2017ubx}. It is also bound from below by the measured mass squares, $\sum_i m_i>0.06$ eV for normal and $\sum_i m_i>0.1$ eV for inverted hierarchy. 
\item{\it The type of mass term} - It is not clear whether neutrinos are Dirac or Majorana particles, i.e., if their mass term is of the type (\ref{diracmassterm}) or (\ref{majoranamassterm}). 
\end{itemize}

\subsection{The Origin of Neutrino Mass}
All fermions in the SM with the exception of neutrinos have two properties in common: They are Dirac fermions, and their masses are generated by the Higgs mechanism. It is not clear whether this also applies to neutrinos. Since they are neutral, they could in principle be their own antiparticles (Majorana fermions), and it may be that their mass is not solely generated 
by the Higgs mechanism.

\paragraph{Dirac neutrinos.} 
We first consider the possibility that neutrinos are Dirac particles.
This necessarily requires the existence of right handed (RH) neutrinos $\nu_R$  to construct
mass term
\begin{equation}\label{diracmassterm}
\overline{\nu_L} m_D\nu_R + h.c.
\end{equation}
Though this means adding new degrees of freedom to the SM, 
there are no new particles (i.e., no new mass eigenstates); adding the $\nu_R$ just leads to additional spin states for the light neutrinos and antineutrinos. 
A bi-unitary transformation $m_D=U_\nu {\rm diag}(m_1,m_2,m_3)\tilde{U}_\nu^\dagger$ can be used to diagonalise the mass term (\ref{diracmassterm}), with real and positive $m_i$, and  one can define a Dirac spinor $\Psi_\nu\equiv\tilde{U}_\nu^\dagger\nu_R+U_\nu^\dagger\nu_L$ with a diagonal mass term $m_\nu^{\rm diag}={\rm diag}(m_1,m_2,m_3)$, such that the free neutrino Lagrangian can be written as $\overline{\Psi_\nu}(i\slashed{\partial}-m_\nu^{\rm diag})\Psi_\nu$.  
The matrix $\tilde{U}_\nu$ is not physical and can be absorbed into a redefinition of the flavour vector $\nu_R$.
The neutrino mixing matrix  $U_\nu$ then appears in the coupling of $\Psi_\nu$ to $W_\mu$ if we substitute $\nu_L=P_L \Psi_\nu$ in (\ref{WeakWW}), where $P_L$ is the LH chiral projector. 
If the mass term (\ref{diracmassterm}) is generated from a Yukawa interaction $\bar{\ell_L}F\tilde{\Phi}\nu_R + h.c.$ in the same way as all SM fermion masses, them Yukawa couplings $F$ has to be very small ($F\sim 10^{-12}$) in order to be consistent with the observed $m_i^2$. This is one reason why many theorists consider this possibility ``unnatural''.
Here $\Phi$ is the Higgs doublet and $\tilde{\Phi}=\epsilon\Phi^*$, where $\epsilon$ is the antisymmetric $SU(2)$-invariant tensor, and $F$ is a matrix of Yukawa interactions.  $\ell_{L}=(\nu_{L},e_{L})^{T}$ are the LH lepton doublets. 
Another reason that has been used to argue against this scenario  is that the symmetries of the SM allow a term of the form  $\overline{\nu_{R}}M_{M}^{\dagger}\nu^c_{R}$, 
where $\nu_R^c=C\bar{\nu_R}^T$ and $C=i\gamma_2\gamma_0$ is the charge conjugation matrix.\footnote{This is in contrast to all SM particles, for which such a term is forbidden by gauge symmetry unless it is generated by a spontaneous symmetry breaking.}
As described in detail in Sec.~\ref{sec:SeeSawwMechanism}, adding such a term to (\ref{diracmassterm}) implies that neutrinos are Majorana fermions, and there exist new particles with masses $\sim M_M$, one of which could be a DM candidate.

\begin{figure}
\begin{center}
\includegraphics[width=0.8\textwidth]{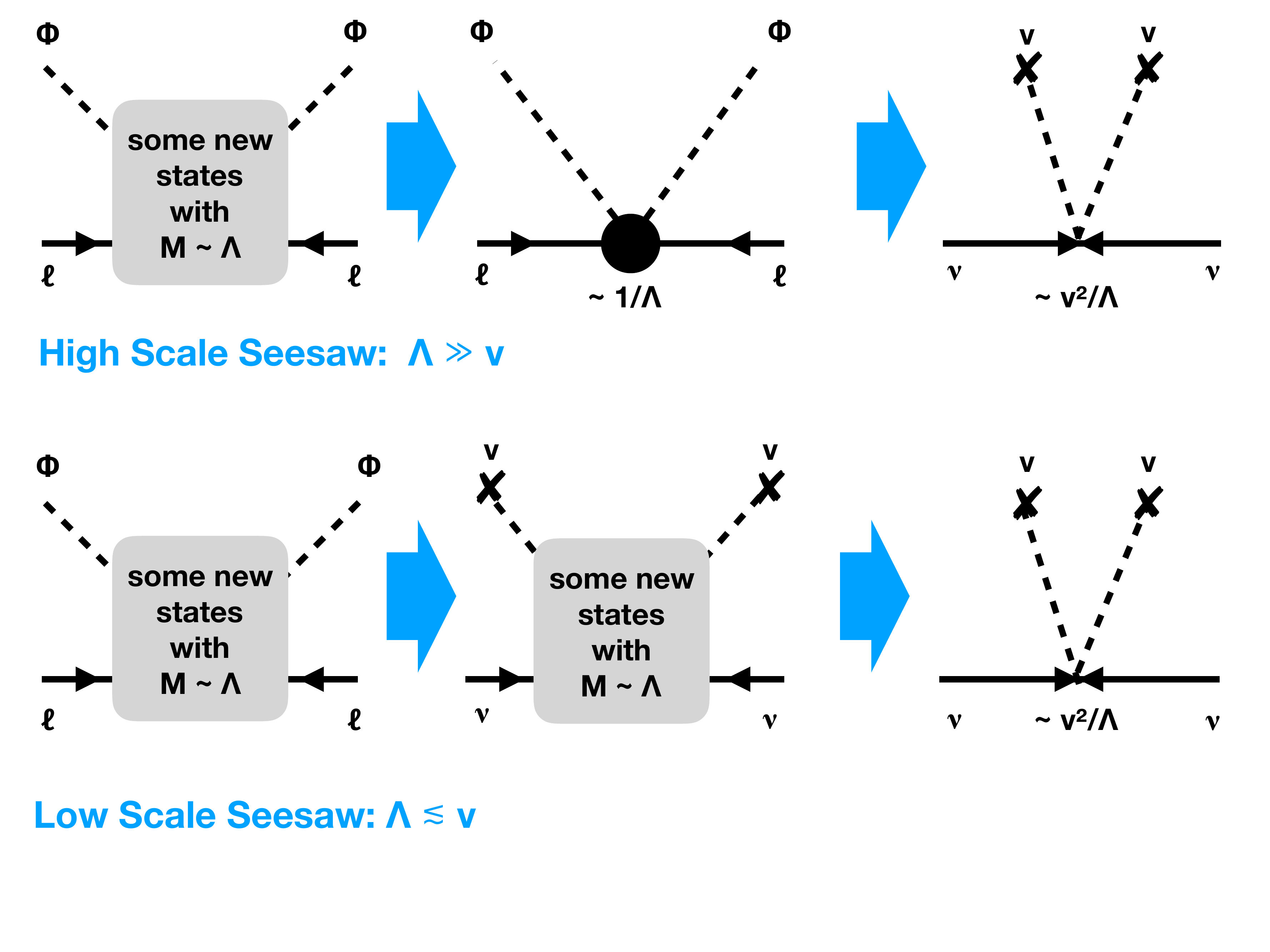}
\end{center}
\caption{If the new states associated with neutrino mass generation are much heavier than the energies $E$ in an experiment ($\Lambda\gg E$), then they do not propagate as real particles in processes, and the Feynman diagram on the left can in good approximation be replaced by a local ``contact interaction'' vertex that we e.g. represent by the black dot in the diagram in the middle in the first  row. 
This is analogous to the way the four fermion interaction $\propto G_F$ in Fermi's theory is obtained from integrating out the weak gauge bosons.
In \emph{high scale seesaw models} ($\Lambda\gg v \gg m_i$, upper row in the figure) this description holds for all laboratory experiments. 
At energies $E\ll v$, there are no real Higgs particles, and one can replace the Higgs field by its expectation value, $\Phi\rightarrow(0,v)^T$. Then the operator (\ref{weinbergoperator}) reduces to the mass term (\ref{majoranamassterm}) with $m_\nu=-v^2 c^{\rm [5]}\Lambda^{-1}$, which can be represented by the diagram on the right. Here the dashed Higgs lines that end in a cross represent the insertion of a Higgs vev $v$. They and the effective vertex are often omitted, so that the Majorana mass term is simply represented by ``clashing arrows''.
In \emph{low scale seesaw} models ($v > \Lambda \gg m_i$, cf. sec.~\ref{LSSS}) the order of the two steps (in terms of energy scales)  is reversed, as illustrated in the lower row of the figure,  but the result at energies much smaller than $v$ and $\Lambda$ is the same.\label{WeinbergFig} 
}
\end{figure}

\paragraph{Majorana neutrinos.} 
A Majorana mass term of the form 
\begin{equation}\label{majoranamassterm}
\frac{1}{2}\overline{\nu_{L}}m_\nu \nu_{L}^{c} + h.c.
\end{equation} 
can be constructed without adding any new degrees of freedom to the SM.
Such term, however, breaks the gauge invariance.
It can be generated from a gauge invariant term  \cite{Weinberg:1979sa}
\begin{equation}\label{weinbergoperator}
\frac{1}{2}\overline{l_{L}}\tilde{\Phi} c^{\rm [5]}\Lambda^{-1}\tilde{\Phi}^{T}l_{L}^{c} + h.c.
,\end{equation}
via the Higgs mechanism. In the unitary gauge this simply corresponds to the replacements $\Phi\rightarrow (0,v)^T$, which yields $m_\nu=-v^2 c^{\rm [5]}\Lambda^{-1}$. Here $v=174$ GeV is the Higgs field expectation value and $c^{\rm [5]}\Lambda^{-1}$ is some flavour matrix of dimension $1/{\rm mass}$. 
The dimension-5 operator (\ref{weinbergoperator}) is not renormalisable; 
in an effective field theory approach it can be understood as the low energy limit of renormalisable operators that is obtained after "integrating out" heavier degrees of freedom, see Fig.~\ref{WeinbergFig}.
The Majorana mass term (\ref{majoranamassterm}) can be diagonalised by a transformation 
\begin{equation}\label{MajoranaDiagonalization}
  m_{\nu}=U_\nu{\rm diag}(m_1,m_2,m_3)U_\nu^{T}.
\end{equation}

\subsection{Neutrino mass and New  Physics}\label{NewPhysicsSec}
The previous considerations show that the explanation of neutrino masses within the framework of renormalisable relativistic quantum field theory certainly requires adding new degrees of freedom to the SM. 
Leaving aside the somewhat boring possibility of Dirac neutrinos discussed after Eq.~(\ref{diracmassterm}),\footnote{It should be emphasised that there exist more interesting ways to generate Dirac neutrinos than simply adding a small Yukawa coupling. While (\ref{weinbergoperator}) is the only dimension five operator  that exists in the SM violates lepton number and necessarily turns neutrinos into Majorana particles, an effective Dirac mass term can be generated from higher dimensional operators (cf. e.g. \cite{Broncano:2002rw} and \cite{CentellesChulia:2018gwr,CentellesChulia:2018bkz}) that preserve lepton number. Such operators would also be signs of new physics.
}
this implies the existence of new particles.
Hence, neutrino masses may act as a "portal" to a (possibly more complicated) unknown/hidden sector that may yield the answer to deep questions in cosmology, such as the origin of matter and DM.	
There are different possible explanations why these new particles have not been found. One possibility is that their masses are larger than the energy of collisions at the LHC. Another possibility is that the new particles couple only feebly to ordinary matter, leading to tiny branching ratios. 

Any model of neutrino masses should explain the ``mass puzzle'', i.e.\ the fact that the $m_i$  are many orders of magnitude smaller than any other fermion masses in the SM. 
A key unknown is the energy scale $\Lambda$ of the New Physics that is primarily responsible for the neutrino mass generation. 
Typically (though not necessarily) one expects the new particles to have masses of this order,\footnote{
Light (pseudo) Goldstone degrees of freedom can e.g. appear if the smallness of neutrino masses is caused be an approximate symmetry.
Moreover, some dimensionful quantities (e.g. the expectation value of a symmetry breaking field) can take very small values while the masses of the associated particles remain well above the energy scale of neutrino oscillation experiments.
} 
and we assume this in the following.
The fact that current neutrino oscillation data is consistent with the minimal hypothesis of three light neutrino mass eigenstates suggests that $\Lambda$ 
is much larger than the typical energy of neutrino oscillation experiments,\footnote{Several comments are in place here. First, while the good agreement of the three light neutrino picture with data clearly \emph{suggests} that $\Lambda\gg$ eV, this is not a necessity, cf. e.g. \cite{deGouvea:2005er}. Moreover, while the lack of evidence for deviations $\eta$  from unitarity in $V_\nu$ or exotic signatures in the near detectors of neutrino experiments hints at $\Lambda>$ MeV \cite{Blennow:2016jkn}, no definite conclusion can be drawn from this alone, and lower bounds on $\Lambda$ in specific models usually involve data from precision test of the SM, beam bump or fixed target experiments or cosmological considerations.} 
so that neutrino oscillations can be described in the framework of Effective Field Theory (EFT) in terms of operators $\mathcal{O}^{ [n]}_i = c^{[n]}_i \Lambda^{n-4}$ of mass dimension $n>4$ that are suppressed by powers of $\Lambda^{n-4}$.
Here the $c^{[n]}_i$ are flavour matrices of so-called Wilson coefficients. 
If the underlying theory is required to preserve unitarity at the perturbative level, $\Lambda$ should be below the Planck scale \cite{Maltoni:2000iq}.
The only operator with $n=5$ is given by eq.~(\ref{weinbergoperator}).
The smallness of the $m_i$ can then be explained by one or several of the following reasons:
a) $\Lambda$ is large, b) the entries of the matrices $c^{[n]}_i$ are small, c) there are cancellations between different terms in $m_\nu$.

\begin{itemize}
\item[a)] 
\textbf{High scale seesaw mechanism}:  
Values of $\Lambda\gg v$ far above the electroweak scale automatically lead to small $m_i$, which has earned this idea the name \emph{seesaw mechanism} ("as $\Lambda$ goes up, the $m_i$ go down"). 
The three tree level implementations of the seesaw idea \cite{Ma:1998dn} are known as type-I~\cite{Minkowski:1977sc,Glashow:1979nm,GellMann:1980vs,Mohapatra:1979ia,Yanagida:1980xy,Schechter:1980gr} 
  type-II~\cite{Schechter:1980gr,Magg:1980ut,Cheng:1980qt,Lazarides:1980nt,Mohapatra:1980yp} and type-III~\cite{Foot:1988aq} seesaw.
The type I seesaw is the extension of the SM by $n$ right handed neutrinos $\nu_{R i}$ with Yukawa couplings $\overline{\ell_L}F\nu_R\tilde{\Phi}$ and a Majorana mass $\overline{\nu_R}M_M\nu_R^{\rm c}$ (hence $\Lambda\sim M_M$). 
It is discussed below in~\ref{sec:SeeSawwMechanism} and gives rise to the operator (\ref{weinbergoperator}) with $ c^{[5]}\Lambda^{-1}=F M_M^{-1} F^T$, cf. (\ref{Leff}).
In the type-II seesaw, $m_\nu$ is directly generated by an additional Higgs field $\Delta$ that transforms as a SU(2) triplet, while the type-III mechanism involves a fermionic SU(2) triplet $\Sigma_L$.
\item[b)]
\textbf{Small numbers}: 
The $m_i$ can be made small for any value of $\Lambda$ if the Wilson coefficients $c^{[n]}_i$ are small, which can be explained in different ways.
One possibility is that it is the consequence of a small coupling constant. 
The Dirac neutrino scenario discussed below Eq.~(\ref{diracmassterm}) is of this kind. A popular way of introducing very small coupling constants without tuning them "by hand" is to assume that they  are created due to the spontaneous symmetry breaking of a flavour symmetry by one or several flavons \cite{Froggatt:1978nt}. This may also help to address the flavour puzzle. 
Small $c^{[n]}_i$ can further be justified if the $\mathcal{O}^{ [n]}_i$ are generated radiatively, cf. e.g. \cite{Zee:1980ai,Witten:1979nr,Zee:1985id,Babu:1988ki,Ma:2006km} and many subsequent works.
The suppression due to the ``loop factor'' $(4\pi)^2$ alone is not sufficient to explain the smallness of the $m_i$, 
and the decisive factor is usually the suppression due to a small coupling of the new particles in the loop, possibly accompanied by a seesaw-like suppression due to the heavy masses of the particles in the loop.
More exotic explanations e.g. involve extra dimensions \cite{Dienes:1998sb,ArkaniHamed:1998vp}, string effects \cite{Blumenhagen:2006xt,Antusch:2007jd} or the gravitational anomaly \cite{Dvali:2016uhn}.
\item[c)]
\textbf{Protecting symmetry}: 
The physical neutrino mass squares $m_i^2$ are given by the eigenvalues of $m_\nu^\dagger m_\nu$. Individual entries of the light neutrino mass matrix $m_\nu$ are not directly constrained by neutrino oscillation experiments and  may be much larger than the $m_i$ if there is a symmetry that leads to cancellations in $m_\nu^\dagger m_\nu$. 
These cancellations can either be "accidental" (which often requires come fine-tuning) or may be explained by an approximate global symmetry.
A comparably low New Physics scale can e.g. be made consistent with small $m_i$ in a natural way if a generalised B-L symmetry is approximately conserved by the New Physics.
Popular models that can implement this idea include the inverse \cite{Mohapatra:1986aw,Mohapatra:1986bd,Bernabeu:1987gr} and linear \cite{Akhmedov:1995ip,Akhmedov:1995vm} seesaw, scale invariant models \cite{Khoze:2013oga} and the $\nu$MSM \cite{Shaposhnikov:2006nn}.\footnote{Strictly speaking lepton number $L$ is already broken by anomalies in the SM \cite{Adler:1969gk,Bell:1969ts}, and it is more correct to refer to these scenarios as $B-L$ conserving. Due to the strong suppression of $B$ violating processes in experiments the common jargon is to refer to these models as ``approximately lepton number conserving''.   In the early universe this makes a big difference, as $B+L$ violating processes occur frequently at temperatures above the electroweak scale \cite{Kuzmin:1985mm}.}
\end{itemize}
 Of course, any combination of these concepts may be realised in nature, including the simultaneous presence of different seesaw mechanisms. For instance, a combination of moderately large $\Lambda\sim $ TeV and moderately small Yukawa couplings $F\sim 10^{-5}$ (similar to that of the electron) is sufficient to generate a viable low scale type-I seesaw mechanism without the need of any protecting symmetry. If a protecting symmetry as added,  a TeV scale seesaw is feasible with $\mathcal{O}[1]$ Yukawa couplings
 or, alternatively, even values of $\Lambda$ below the electroweak scale can explain the observed neutrino oscillations in a technically natural way \cite{Shaposhnikov:2006nn}.
Models with $\Lambda$ at or below the TeV scale are commonly referred to as \emph{low scale seesaw} cf. sec.~\ref{LSSS}.
The typical seesaw behaviour that $m_i$ decreases if $\Lambda$ is increased holds in such scenarios as long as $\Lambda\gg m_i$, in spite of the fact that $v/\Lambda$ is not a small quantity.
The EFT treatment in terms of $\mathcal{O}^{ [n]}_i$ may still hold for neutrino oscillation experiments, but the collider phenomenology of low scale seesaw models \cite{Cai:2017mow} has to be studied in the full theory if $\Lambda$ is near or below the collision energy, cf. fig.~\ref{WeinbergFig}.

In addition to the smallness of the neutrino masses, it is desirable to find an explanation for the ``flavour puzzle'', i.e., the observed mixing pattern of neutrinos. 
Numerous attempts have been made to identify discrete or continuous symmetries in $m_\nu$. An overview of scenarios that are relevant in the context of sterile neutrino DM is  e.g. given in Ref.~\cite{King:2014nza}. 
The basic problem is that the reservoir of possible symmetries to choose from is practically unlimited. 
For any possible observed pattern of neutrino masses and mixings one can find a symmetry that ``predicts'' it. 
Models can only be convincing if they either predict  observables that have not been measured at the time when they were proposed, such as sum rules for the mass~\cite{Barry:2010yk,Dorame:2011eb,King:2013psa,King:2014nza,Agostini:2015dna} or mixing~\cite{Petcov:2004rk,Marzocca:2013cr,Ballett:2013wya,King:2014nza}, or are ``simple'' and aesthetically appealing from some (subjective) viewpoint. 
Prior to the measurement of $\uptheta_{13}$, models predicting $\uptheta_{13}=0$ appeared very well-motivated, such as those with tri/bi-maximal mixing~\cite{Harrison:1999cf,Harrison:2002er,Ma:2004zv,Altarelli:2005yx,Xing:2002sw}. The observed $\uptheta_{13}\neq0$, however, makes it difficult to explain $m_\nu$ in terms of a simple symmetry and a small number of parameters, as the number of parameters that have to be introduced to break the underlying symmetries is usually comparable to or even larger than the number of free parameters in $m_\nu$ that one wants to explain.
An interesting alternative way to look at this is to compare the predictivity of  different models to the possibility that the values in $m_\nu$ are simply random~\cite{Hall:1999sn}.

\subsection{Sterile Neutrinos}
\label{sec:SeeSawwMechanism}
\subsubsection{Right Handed Neutrinos and Type-I Seesaw} 
The terms "sterile neutrino", "right handed neutrino", "heavy neutral lepton" and "singlet fermion" are often used interchangeably in the literature.
In what follows we apply the term ``sterile neutrino'' to singlet fermions that mix with the neutrinos $\nu_L$, i.e., we consider this mixing as a defining feature that distinguishes a sterile neutrino from a generic singlet fermion. 
Such particles are predicted by many extensions of the SM, and in particular in the type-I seesaw model, which is defined by adding $n$ RH neutrinos $\nu_R$ to the SM.
The Lagrangian reads\footnote{Here we represent the fields with left and right chirality by four component spinors $\nu_L$ and $\nu_R$ with $P_{L,R}\nu_{R,L}=0$, so that no explicit chiral projectors $P_{L,R}$ are required in the interaction terms. }  
\begin{equation}
    \mathcal{L}
  = \mathcal{L}_\text{SM} + i \, \overline{\nu_{R i}}\slashed\partial\nu_{R i}
  - \frac{1}{2} \left( \overline{\nu_{R i}^c}(M_M)_{ij}\nu_{R j} + \overline{\nu_{R i}}(M_M^\dagger)_{ij}\nu_{R j}^c \right)
  - F_{a i}\overline{\ell_{L a}}\varepsilon\phi^* \nu_{R i}
  - F_{a i}^*\overline{\nu_{R i}}\phi^T \varepsilon^\dagger \ell_{L a}
\ . \label{Lseesaw}
\end{equation}

$\mathcal{L}_{\rm SM}$ is the Lagrangian of the SM. 
The $F_{a i}$ are Yukawa couplings between the $\nu_{R i}$, the Higgs field $\phi$ and the SM leptons $\ell_a$. Here we have suppressed SU(2) indices; $\varepsilon$ is the totally antisymmetric SU(2) tensor.
$M_M$ is a Majorana mass matrix for the singlet fields $\nu_{R i}$.
The Majorana mass term $M_{M}$ is allowed for $\nu_R$ because the $\nu_R$ are gauge singlets. The lowest New Physics scale $\Lambda$ should here be identified with the smallest of the eigenvalues $M_I$ (with $I=1,\ldots n$) of $M_M$. 
At energies $E\ll \Lambda$, the $\nu_R$ can be ``integrated out'' and (\ref{Lseesaw}) effectively reduces to 
\begin{eqnarray}
\label{Leff}
	\mathcal{L}_{\rm eff} &=&\mathcal{L}_{\rm SM}+ \frac{1}{2}\bar{\ell_{L}}\tilde{\Phi} F M_M^{-1}F^T \tilde{\Phi}^{T}\ell_{L}^{c}
	\end{eqnarray}
and thus generates the term (\ref{weinbergoperator}) with $c^{\rm [5]}\Lambda^{-1}=F M_M^{-1}F^T$.
The Higgs mechanism generates the Majorana mass term (\ref{majoranamassterm}) from (\ref{Leff}), and $m_\nu$ is given by 
\begin{eqnarray}
m_\nu= -v^2 F M_M^{-1}F^T,\label{preactivemass} 
\end{eqnarray}
where $v=174$ GeV is the Higgs field expectation value.
The full neutrino mass term after electroweak symmetry breaking reads
\begin{eqnarray}\label{neutrinomassfull}
\frac{1}{2}
(\overline{\nu_L} \  \overline{\nu_R^c})
\mathfrak{M}
\left(
\begin{tabular}{c}
$\nu_L^c$\\
$\nu_R$
\end{tabular}
\right) + h.c.
\equiv
\frac{1}{2}
(\overline{\nu_L} \ \overline{\nu_R^c})
\left(
\begin{tabular}{c c}
$0$ & $m_D$\\
$m_D^T$ & $M_M$
\end{tabular}
\right)
\left(
\begin{tabular}{c}
$\nu_L^c$\\
$\nu_R$
\end{tabular}
\right) + h.c. ,
\end{eqnarray}
where $m_D\equiv Fv$. 
The magnitude of the $M_I$ is experimentally almost unconstrained, and different choices have very different implications for particle physics, cosmology and astrophysics, see e.g.~\cite{Drewes:2013gca} for a review.
For $M_I\gg 1$ eV there is a hierarchy $m_D\ll M_M$ (in terms of eigenvalues), 
and one finds two distinct sets of mass eigenstates: Three light neutrinos $\nu_i$ that can be identified with the known neutrinos, and $n$ states that have masses $\sim M_I$.
Mixing between the active and sterile neutrinos is suppressed by elements of the mixing matrix
\begin{equation}\label{thetaDef}
\theta\equiv m_D M_M^{-1}.
\end{equation}  
This allows to rewrite (\ref{preactivemass}) as
\begin{eqnarray}
m_\nu=  -v^2 F M_M^{-1}F^T = -m_D M_M^{-1} m_D^T=-\theta M_M \theta^T. \label{activemass} 
\end{eqnarray}
All $3+n$ mass eigenstates are Majorana fermions and can be represented by the elements of the flavour vectors\footnote{
In principle there is no qualitative difference between the mass states $\upnu_i$ and $N_I$, the difference is a quantitative one in terms of the values of the masses and mixing angles. One could simply use a single index $i=1\ldots 3+n$  and refer to $N_I$ as $\upnu_{i=3+I}$,  with mass $m_{i=3+I}=M_I$.  This is indeed often done in experimental papers.  Here we adopt the convention to use different symbols for the light and heavy states  which emphasises the fact that, due to the many orders of magnitude difference in the masses and mixing angles, these particles play very different roles in cosmology.
}
\begin{equation}  \upnu=V_{\nu}^{\dagger}\nu_L-U_{\nu}^{\dagger}\theta \nu_R^c+V_{\nu}^T\nu_L^c-U_{\nu}^T\theta^{\ast} \nu_R \label{LightMassEigenstates}
\end{equation}
and
\begin{equation}
  N=V_N^{\dagger}\nu_R+\Theta^T \nu_L^c+V_N^T\nu_R^c+\Theta^{\dagger} \nu_L,
\end{equation}

The unitary matrices $U_\nu$ and $U_N$ diagonalise the mass matrices $m_\nu$ and $M_N\equiv M_M + \frac{1}{2}\big(\theta^{\dagger} \theta M_M + M_M^T \theta^T \theta^{*}\big)$ of the light and heavy neutrinos, respectively, as
$M_N^{\rm diag}=U_N^T M_N U_N=\text{diag}(M_1,M_2,M_3)$
and
$m_{\nu}^{\rm diag}= U_{\nu}^{\dagger}m_{\nu}U_{\nu}^{\ast}=\text{diag}(m_1,m_2,m_3)$.
The eigenvalues of $M_M$ and $M_N$ coincide in good approximation, we do not distinguish them in what follows and  refer to both as $M_I$. 
The light neutrino mixing matrix in Eq.~(\ref{NeutrinoMixing}) and its heavy equivalent $V_N$ are given by

\begin{align}\label{VnuDef}
  V_{\nu}\equiv \left( \mathbb{I}-\frac{1}{2}\theta\theta^{\dagger} \right) U_{\nu} \quad {\rm and}
  V_N\equiv \left( \mathbb{I}-\frac{1}{2}\theta^T\theta^{\ast} \right) U_N,
\end{align}

and comparison with (\ref{nonunitarity}) reveals $\eta=-\frac{1}{2}\theta\theta^{\dagger}$.\footnote{Unfortunately, unitarity violation due to heavy neutrinos does not appear to be able to resolve the long-standing issues of short-baseline anomalies~\cite{Kopp:2013vaa}.}
The active-sterile mixing is determined by the matrix
\begin{equation}
\Theta\equiv\theta U_N^*.
\end{equation}

An important implication of the relation (\ref{activemass}) is that one $N_I$ with non-vanishing mixing $\theta_{\alpha I}$ is needed for each non-zero light neutrino mass $m_i$. 
Hence, if the minimal seesaw mechanism is the only source of light neutrino masses, there must be at least $n=2$ RH neutrinos because two mass splittings $\Delta m_{\rm sol}$ and $\Delta m_{\rm atm}$ have been observed. 
If the lightest neutrino is massive, i.e.\ $m_{\rm lightest}\equiv{\rm min}(m_1,m_2,m_3)\neq 0$, 
then this implies $n\geq3$, irrespectively of the magnitude of the $M_I$.
A heavy neutrino that is a DM candidate (let us call it $N_1$) would not count in this context \cite{Boyarsky:2006jm}: to ensure its longevity, the three mixing angles $\theta_{\alpha 1}$ must be so tiny that their effect on the light neutrino masses in Eq.~(\ref{activemass}) is negligible. 
This has an interesting consequence for the scenario in which the number $n$ of sterile flavours equals the number of generations in the SM ($n=3$): If one of the heavy neutrinos composes the DM, then the lightest neutrino is effectively massless ($m_{\rm lightest}\simeq 0$). 
If, on the other hand, $m_{\rm lightest}>10^{-3}$ eV, then all $N_I$ must have sizeable mixings $\theta_{\alpha I}$, which implies that they are too short lived to be the DM , cf. Eq.~(\ref{gamma}).
These conclusions can of course be avoided for $n>3$, or if there is another source of neutrino mass. 

\subsubsection{Low Scale Seesaw}\label{LSSS}
In conventional seesaw models, it is assumed that the scale $\Lambda$ is not only larger than the $m_i$, but even much larger than the electroweak scale. 
This hierarchy is 
suggested in Eq.~(\ref{Leff}).
In this case the smallness of the $m_i$ is basically a result of the smallness of $v/\Lambda$ or, more precisely, $v/M_I\ll 1$.  
 If they are produced via the weak interactions of their $\theta_{\alpha I}$-suppressed components $\nu_{L \alpha}$, then heavy neutrinos $N_I$ that compose the DM usually must have masses below the electroweak scale ($M_I<v$). 
 Otherwise the upper bound on the magnitude of the $\theta_{\alpha I}$ from the requirement that their lifetime exceeds the age of the universe prohibits an efficient thermal production in the early universe.
$M_M$ therefore has at least one eigenvalue below the electroweak scale in such scenarios. 
A particular motivation for so-called \emph{low scale seesaws} comes from the fact that they avoid the \emph{hierarchy problem} due to contributions from superheavy $N_I$ to the Higgs mass \cite{Vissani:1997ys}. 
Moreover, the fact that the properties of the Higgs boson and top quark appear to be precisely in the narrow regime where the electroweak vacuum is metastable \cite{Degrassi:2012ry} and a vacuum decay catastrophe \cite{Krive:1976sg} can be avoided may suggest the absence of any new scale between the electroweak and Planck scale \cite{Shaposhnikov:2007nj,Shaposhnikov:2009pv}.

Popular low scale seesaw scenarios include the inverse \cite{Mohapatra:1986aw,Mohapatra:1986bd,Bernabeu:1987gr} and linear \cite{Akhmedov:1995ip,Akhmedov:1995vm} seesaw, the $\nu$MSM \cite{Shaposhnikov:2006nn} and scenarios based on on classical scale invariance \cite{Khoze:2013oga}. They often involve some implementation of a generalised $B-L$ symmetry, and the smallness of the light neutrino masses $m_i$ is explained as a result of the smallness of the symmetry breaking parameters (rather than $v/M_I$). Sterile neutrino DM candidates can be motivated in the minimal model (\ref{Lseesaw}) \cite{Asaka:2005an,Shaposhnikov:2006nn} as well as various different extensions, cf. ref.~\cite{Merle:2013gea,Adhikari:2016bei} for reviews and e.g. refs.  \cite{Dev:2012bd,Mavromatos:2012cc,Heeck:2015qra,Dev:2016xcp} for specific examples.

For $n=3$, and in the basis where $M_M$ and the charged lepton Yukawa couplings are diagonal in flavour space, one can express $M_M$ and $F$ as
\begin{equation}
M_M=\Lambda\begin{pmatrix} \mu' & 0 & 0 \\ 0 & 1 - \mu  &0  \\ 0 &0 & 1 + \mu    \end{pmatrix} \quad , \quad
F=\begin{pmatrix} \epsilon'_e & F_e + \epsilon_e & i(F_e - \epsilon_e)  \\ \epsilon'_\mu & F_\mu + \epsilon_\mu & i(F_\mu - \epsilon_\mu)  \\ \epsilon'_\tau & F_\tau + \epsilon_\tau & i(F_\tau - \epsilon_\tau)  \end{pmatrix}\label{FullNeutrinoMassSymm}
\end{equation}
without any loss of generality.
Here $\Lambda$ is the characteristic seesaw scale, the $F_\alpha$ are generic Yukawa couplings and the dimensionless quantities $\epsilon_\alpha, \epsilon'_\alpha, \mu, \mu'$ are symmetry breaking parameters. 
In the limit $\epsilon_\alpha, \epsilon'_\alpha, \mu, \mu' \to 0$ the quantity $B-\bar{L}$ is conserved, where the generalised lepton number
\begin{equation}\label{LbarDef}
\bar{L}=L + L_{\nu_R}
\end{equation}
is composed of the SM lepton number $L$ and a charge $L_{\nu_R}$ that can be associated with the right handed neutrinos in the symmetric limit; the combination $\frac{1}{\sqrt{2}}(\nu_{R 2} + i \nu_{R 3})$ carries $L_{\nu_R}=1$, 
the combination $\frac{1}{\sqrt{2}}(\nu_{R 2} - i \nu_{R 3})$ carries $L_{\nu_R}=-1$
and $\nu_{R 1}$ carries $L_{\nu_R}=0$.
Such scenarios can provide a sterile neutrino DM candidate (here $N_1$) in a technically natural way: The smallness of $\mu'$explains its low mass, while the small of the $\epsilon'_\alpha$ can ensure its longevity.
Recent reviews of models including keV sterile neutrinos can e.g. be found in refs.~\cite{Merle:2013gea,Adhikari:2016bei}.
The parametrisation (\ref{FullNeutrinoMassSymm}) is completely general; specific models can make predictions how exactly the limit  $\epsilon_\alpha, \epsilon'_\alpha, \mu, \mu' \to 0$ should be taken.

An appealing feature of low scale scenarios is that the heavy neutrinos can be searched for experimentally in fixed target experiments \cite{Gorbunov:2007ak} like SHiP \cite{Anelli:2015pba,Alekhin:2015byh} or NA62 \cite{CortinaGil:2017mqf,Drewes:2018gkc}, 
at the existing \cite{Helo:2013esa,Izaguirre:2015pga,Gago:2015vma,Dib:2015oka,Antusch:2017hhu,Cottin:2018nms} or future \cite{Chou:2016lxi,Kling:2018wct,Helo:2018qej,Curtin:2018mvb} LHC experiments or at future colliders \cite{Blondel:2014bra,Antusch:2016ejd,Antusch:2017pkq}.
Observable event rates 

\cite{Shaposhnikov:2006nn,Kersten:2007vk,Moffat:2017feq}. 
The interactions of the light and heavy neutrinos can be determined by inserting the relation $\nu_L=P_L(V_\nu \upnu + \Theta N)$ from eq.~\eqref{LightMassEigenstates} into eq.~\eqref{WeakWW}. 
The unitarity violation in $V_\nu$ implies a flavour-dependent suppression of the light neutrinos' weak interactions Eq.~\eqref{LightMassEigenstates}. 
The $N_I$ have $\theta$-suppressed weak interactions \cite{Shro80,Shrock:1980ct,Shrock:1981wq} due to the doublet component $\Theta^{T}_{I\alpha}\nu_{L\alpha}^{c}+\Theta^\dagger_{I\alpha}\nu_{L\alpha}$ in (\ref{LightMassEigenstates}).
In addition, the $N_I$ directly couple to the Higgs boson via their Yukawa coupling to the physical Higgs field ${\rm h}$ in (\ref{Lseesaw}) in unitary gauge.
The full $N_I$ interaction term at leading order in $\theta$ can be expressed as 
\begin{align}\label{NWW}
\mathcal L \supset& -\frac{g}{\sqrt{2}}\overline{N} \Theta^\dagger\gamma^\mu e_{L} W^+_\mu
-\frac{g}{\sqrt{2}}\overline{e_{L}}\gamma^\mu \Theta N W^-_\mu\notag\\
&- \frac{g}{2\cos\theta_W}\overline{N} \Theta^\dagger\gamma^\mu \nu_{L} Z_\mu
- \frac{g}{2\cos\theta_W}\overline{\nu_{L}}\gamma^\mu \Theta N Z_\mu\notag\\
&-\frac{g}{\sqrt{2}}\frac{M_N}{m_W}\Theta {\rm h} \overline{\nu_{L}}N
-\frac{g}{\sqrt{2}}\frac{M_N}{m_W}\Theta^\dagger {\rm h} \overline{N}\nu_{L}
\ .
\end{align}
If kinematically allowed, they appear in any process that involves light neutrinos, but with an amplitude suppressed by $\Theta_{\alpha I}$, cf. eq.~(\ref{NWW}).
An overview of the present constraints is e.g.\ given in refs. ~\cite{Atre:2009rg,Kusenko:2009up,Boyarsky:2009ix,Abazajian:2012ys,Drewes:2013gca,Alekhin:2015byh,Deppisch:2015qwa,Drewes:2015iva,deGouvea:2015euy,Adhikari:2016bei,Cai:2017mow}.
Most of the proposed searches cannot be applied to DM sterile neutrinos because the tiny mixing angle that is required to ensure their stability on cosmological time scales implies that the branching ratios are too small. We discuss dedicated experimental searches for DM sterile neutrinos in Sec.~\ref{exp}.

\subsubsection{The Neutrino Minimal Standard Model}
\label{sec:numsm}
Since all of the problems I)-IV) summarised in sec.~\ref{sec:bsm-dm} should be explained in a consistent extension of the SM, it is instructive to address them together.
One possible guideline in the search for a common solution is provided by the ``Ockham's razor'' approach, which has been highly successful in science in the past: one tries to minimise the number of new entities introduced but maximise the number of problems which can be addressed simultaneously.
A specific low scale seesaw model that realises this idea was suggested in 2005~\cite{Asaka:2005pn,Asaka:2005an}, see~\cite{Boyarsky:2009ix} for a review.
This \emph{Neutrino Minimal Standard Model} ($\nu$MSM) provides a new systematic approach that addresses the beyond-the-Standard-Model problems in a bottom-up fashion, not introducing particle heavier than the Higgs boson and attempting to be \emph{complete} \cite{Shaposhnikov:2007nj,Shaposhnikov:2008pf} and \emph{testable}~\cite{Gorbunov:2007ak,Gorbunov:2014ypa,Blondel:2014bra,Hernandez:2016kel,Drewes:2016jae,Antusch:2017pkq}.  
In this approach the three right-handed neutrinos are responsible for neutrino flavour oscillations, generate the BAU via low scale leptogenesis \cite{Akhmedov:1998qx,Asaka:2005pn} and provide a DM candidate. The DM candidate has a keV mass and very feeble interactions, while the other two have quasi-degenerate masses between $\sim 100$ MeV and the electroweak scale; these particles are responsible for the generation of light neutrino masses and baryogenesis. This os precisely the pattern predicted by eq.~(\ref{FullNeutrinoMassSymm}) if all symmetry breaking parameters are small.
The feasibility of this scenario was proven with a parameter space study in Refs.~\cite{Canetti:2012vf,Canetti:2012kh}.
In the $\nu$MSM the interactions of the DM particles with
the Standard model sector are so feeble that these particles never enter the
equilibrium with primordial plasma and therefore their number density is
greatly reduced (as compared to those of neutrinos). In this way they evade
the combination of Tremain-Gunn~\cite{Tremaine:1979we} and cosmological
bounds. 

\section{Properties of Sterile Neutrino Dark Matter}
\label{sec:dm}

As originally suggested in~\cite{Dodelson:1993je}, sterile neutrinos with the mass in the keV range can play the role of DM.\footnote{In this section we almost exclusively focus on the minimal scenario where sterile neutrino DM is produced thermally via the $\theta$-suppressed weak interactions, which favours the keV scale due to constraints on $\theta$ coming from X-ray observations discussed further below.
If the DM is produced via another mechanism which does not rely on this mixing, cf. sec.~\ref{sec:production}, then there is no lower bound on the $|\theta_{\alpha I}|$ and X-ray constraints can be avoided by making them arbitrarily small. This means that the DM particles can be much heavier than keV.
However, while these are perfectly viable models of singlet fermion DM, they do not fall under our definition of \emph{sterile neutrinos} because the new particles practically do not mix with the SM neutrinos, cf. sec.~\ref{sec:SeeSawwMechanism}.}
Indeed, these particles are neutral, massive and, while unstable, can have their lifetime longer than the age of the Universe (controlled by the active-sterile mixing parameter $\theta$).\footnote{
In this section we consider the case of one sterile neutrino $N$ with mass $\M$ and mixings $\theta_\alpha$ with the SM flavours $\alpha$. When this scenario is embedded into a seesaw model, we can identify $N$ with one of the heavy mass eigenstates, i.e., $N\equiv N_I$ and $\theta_\alpha\equiv\theta_{\alpha I}$.
} 
Such sterile neutrinos are produced in the early Universe at high temperatures. 
Unlike other cosmic relic particles (e.g.\ photons, neutrinos or hypothetical WIMPs) the feeble interaction strength of sterile neutrinos means that they \emph{were never in thermal equilibrium} in the early Universe and that their exact production mechanism is model-dependent (see Section~\ref{sec:production}). 
In the following we discuss the most important observational constraints on these particles as DM candidates. 
These can broadly be classified into three main categories,
\begin{itemize}
\item the phase space density in DM dense objects, cf. sec.~\ref{sec:PhaseSpace},
\item indirect detection through emission from the decay of sterile neutrino DM, cf. sec.~\ref{sec:decaying-dark-matter},
\item the effect of their free streaming on the formation of structures in the universe, cf. sec.~\ref{sec:structure-formation}.
\end{itemize}
A selection of constraints is summarised in sec.~\ref{OtherConstraints}.

\subsection{Phase Space Considerations}
\label{sec:PhaseSpace}
Being out of equilibrium implies $n_N\ll n_\nu$ for the number density $n_N$ of sterile neutrinos, where the number of relic neutrinos, $n_\nu$, is given by Eq.~\eqref{eq:1}.  
As a result, the observed DM density $\rho_{DM} = M n_N$ can be achieved with the mass of DM particles that satisfy (even saturate) the lower bound for the fermionic DM, the \emph{Tremaine-Gunn bound}~\cite{Tremaine:1979we}, assuming that the phase-space density of DM particles in a galaxy does not exceed that of the degenerate Fermi gas.
 The strongest lower bound comes from the dwarf spheroidal satellites of the Milky way (dSph) and is $\mathcal{O}(100)$~eV. We do not quote the exact numerical value, as it depends on the set of dSphs used, and on the way to estimate the phase-space density of DM particles based on the observational data (for the discussion see~\cite{Boyarsky:2008ju,Gorbunov:2008ka,Domcke:2014kla,DiPaolo:2017geq}).
Assuming a particular primordial phase-space distribution of DM particles that
obeys the Liouville theorem in the process of evolution, one may additionally
strengthen the bounds~\cite{Boyarsky:2008ju} (see also \cite{Wang:2017hof}
where the several newly-discovered satellite galaxies from wide-field optical
surveys such as the Dark Energy Survey (DES) and Pan-STARRS  were included). A
pedagogical review of this argument is given e.g.\ in~\cite{Boyarsky:2008ju} or in~\cite[Section 4.1]{Adhikari:2016bei}.\footnote{In ref.~\cite{Destri:2012yn}  a stronger (but disputed) claim has been made, as it was argued that phase space considerations and quantum statistics actually favour a keV mass for the DM, cf. also \cite{Boyanovsky:2007ay,deVega:2009ku}.}


\subsection{Decaying Dark Matter}
\label{sec:decaying-dark-matter}

\subsubsection{Existing Constraints}
For sterile neutrino masses below twice the value of the electron mass the dominant decay channel is $N \to \nu_\alpha\nu_\beta\bar\nu_\beta$  (all possible combinations of flavours are allowed, including charge conjugated decay channels as $N$ is a Majorana particle). 
The total decay width $N\to 3\nu$ is given by~\cite{Pal:1981rm,Barger:1995ty} 
\begin{equation}
  \label{eq:11}
  \Gamma_{N\to 3\nu} = \frac{G_F^2 \M^5}{96\pi^3} \sum_\alpha  |\theta_\alpha|^2 \approx
  \frac1{\unit[1.5\times
    10^{14}]{sec}}\left(\frac{\M}{\unit[10]{keV}}\right)^5 \sum_\alpha  |\theta_\alpha|^2
\end{equation}
From now on we will discuss the total mixing angle and use the simplified notation
\begin{equation}
  \label{eq:4}
  \theta^2 \equiv \sum_{\alpha=e,\mu,\tau} |\theta_\alpha|^2,
\end{equation}
within this section, not to be confused with the full mixing matrix $\theta$ in eq.~(\ref{thetaDef}) and following.
If one requires that the lifetime corresponding to the process~\eqref{eq:11} should be (much) longer than
the age of the Universe, $t_{\text{Universe}} = 4.4\times
10^{17}\unit{sec}$~\cite{Ade:2015xua}
the  bound on the sum of the mixing angles $\theta_\alpha^2$ becomes (see Fig.~\ref{fig:SummaryPlot})
\begin{equation}
  \label{eq:5}
  \theta^2<3.3\times 10^{-4}\left(\frac{\unit[10]{keV}}{\M}\right)^5\quad\text{---
    lifetime longer than the age of the Universe}
\end{equation}
Already the requirement that the sterile neutrino lifetime is longer than the age of the Universe (bound~\eqref{eq:5}) implies that the contribution of DM sterile neutrino to the neutrino masses, $\delta m_\nu \sim \M \theta^2$, is smaller than the solar neutrino mass difference $m_{\rm sol} = 0.0086$~eV)~\cite{Asaka:2005an,Boyarsky:2006jm}.  Therefore, at
least two more sterile neutrinos are required to explain to observed mass
differences, as discussed already in~\ref{sec:SeeSawwMechanism}.
If the neutrino masses are due to exactly two sterile neutrinos,
the lightest active mass eigenstate, $m_1$ is essentially zero, which bounds
the total sum of neutrino masses to be
\begin{equation}
  \label{eq:6}
  \sum_i m_i \simeq \kappa m_\atm
\end{equation}
where $\kappa = 1$ for the case of \emph{normal hierarchy} and $\kappa=2$ for
the case of \emph{inverted hierarchy}. This is a non-trivial
prediction that can be checked by ESA's Euclid space
mission~\cite{Audren:2012vy}.

\begin{figure}[!t]
  \centering
  \includegraphics[width=0.5\textwidth]{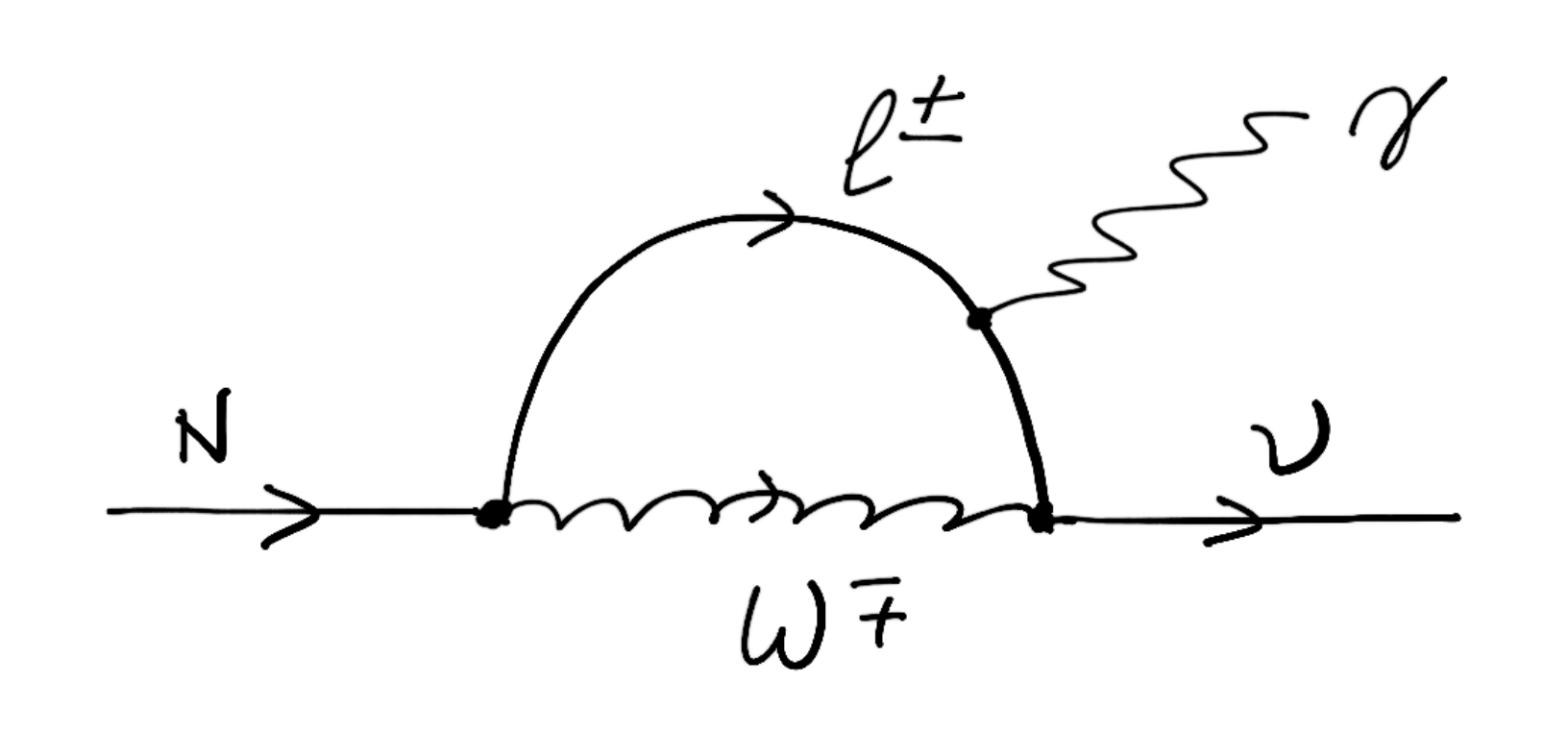}
  \caption[Example for a Feynman diagram that contributes to the
  radiative decay of sterile neutrino]{Radiative decay of sterile
    neutrino $N \to \gamma + \nu_\alpha$. A similar diagram where a photon
    couples to the $W$-boson is not shown. The $N$ coupling to weak gauge bosons exists for each
    flavour and is proportional to $\theta_\alpha$, cf. (\ref{NWW}).}
  \label{fig:rad-decay}
\end{figure}

Along with the dominant decay channel, $N\to 3\nu$ sterile neutrino also
possesses a loop mediated \emph{radiative decay} $N\to \nu + \gamma$~\cite{Pal:1981rm}
(Fig.~\ref{fig:rad-decay}). The decay width of this process is~\cite{Pal:1981rm,Barger:1995ty}
\begin{equation}
\label{gamma}
  \Gamma_{N\to\gamma\nu} = \frac{9\, \alpha\, G_F^2}
  {256\pi^4}\theta^2 \M^5 =
  5.5\times10^{-22}\theta^2
  \left[\frac{\M}{1\,\mathrm{keV}}\right]^5\;\mathrm{sec}^{-1}\:.
\end{equation}
While it is suppressed by $\frac{27\alpha}{8\pi} \approx \frac1{128}$ as
compared to the
main decay channel, such a decay produces a
photon with energy $E=\frac12 \M$. 
If the sterile neutrino is a main ingredient of the DM, then such a mono-chromatic
signal is potentially detectable from spots on the sky with large DM overdensities
\cite{Dolgov:2002wy,Abazajian:2001vt,Herder:2009im}.
This opens an exciting possibility of astrophysical searches of sterile
neutrino DM.

\begin{figure}[!t]
  \centering
  \includegraphics[width=0.75\textwidth]{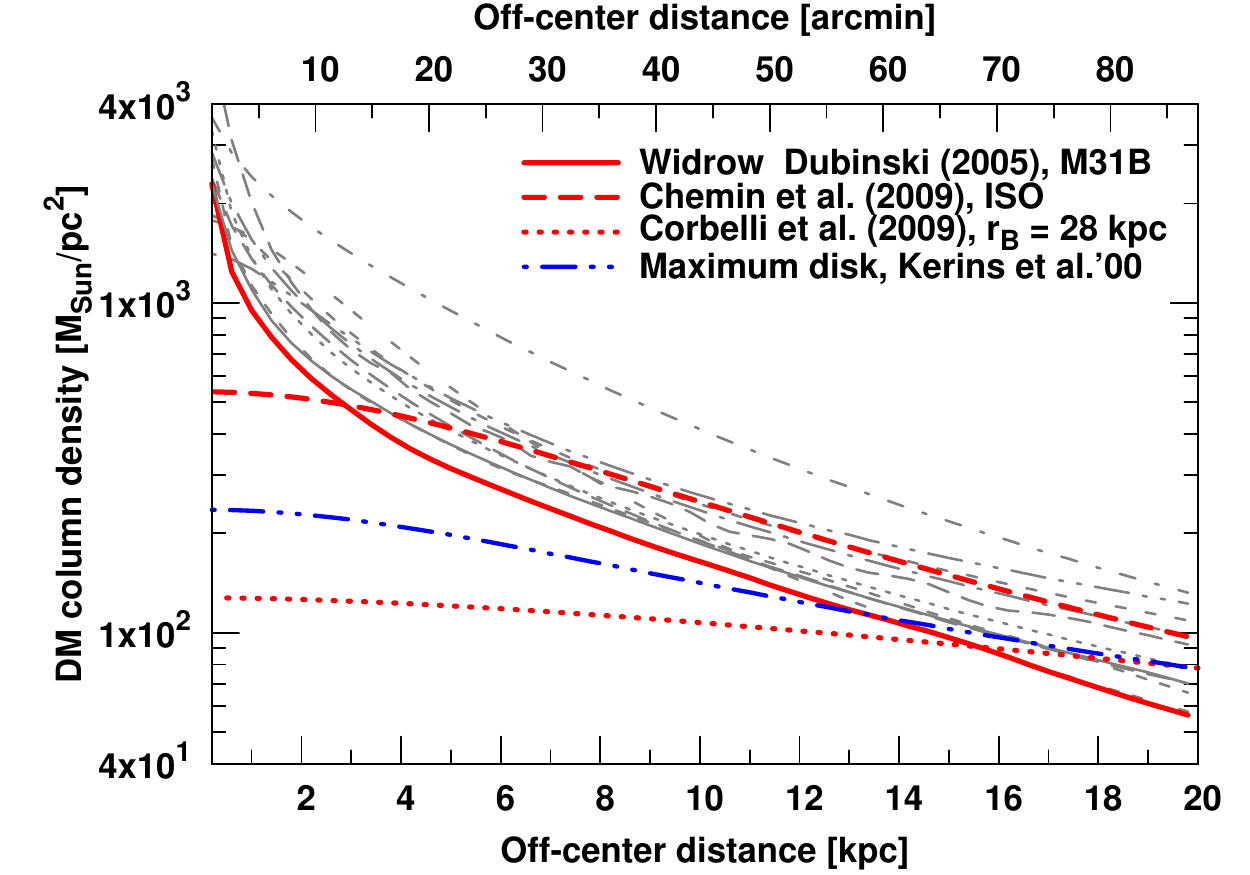}
  \caption[DM column density for M31]{DM column density
    (DM density distributions, integrated along the line of sight)
    for different DM density profiles of Andromeda galaxy (M31). The
    profiles used are \protect\cite{Geehan:06,Tempel:07,Klypin:02,
      Chemin:09,Corbelli:09,Widrow:05,Kerins:03}. The extreme profiles
    of \emph{Corbelli et al.}~\cite{Corbelli:09} and \emph{Kerins et al.}~\cite{Kerins:03}
    are unphysical but demonstrate how much matter can be ``pushed'' into
    baryonic, rather than DM mass.}
  \label{fig:M31}
\end{figure}

The number of photons from DM decay is
proportional to the \emph{DM column density} -- integral of the DM distribution along the line of sight (l.o.s.):
\begin{equation}
  \label{eq:7}
  \CS = \int_{\text{l.o.s.}}\rho_{DM}(r)dr
\end{equation}
Averaged over a sufficiently large field-of-view (of the order of $10'$ or more) this
integral becomes only weakly sensitive to the underlying DM
distributions (see, e.g.\ Fig.~\ref{fig:M31} where this is illustrated for the case of Andromeda galaxy,
based on~\cite{Boyarsky:2010ci}).\footnote{The same scatter in the DM density profiles gives a much larger
uncertainty in the annihilation $J$-factor $\mathcal{J} =
\int\rho^2_{DM}(r)dr$ to which annihilating DM signal is proportional.}

As a result a vast variety of astrophysical objects of
different nature would produce a comparable decay
signal~\cite{Boyarsky:2006fg,Boyarsky:2009rb,Boyarsky:2009af}.  Therefore, 
\begin{itemize}\item one has a  freedom of choosing the
  observational targets, avoiding complicated astrophysical backgrounds;
\item if a candidate line is found, its surface brightness profile may be
  measured, distinguished from astrophysical lines (which usually decay in
  outskirts) and compared among several objects with the same expected signal.
 
\end{itemize}
\emph{In this way one can efficiently distinguish the decaying DM line from
  astrophysical backgrounds.}

Searches of this decaying DM signal in the keV--MeV mass range have been conducted using a wide range of X-ray telescopes:
\textit{XMM-Newton}~\cite{Boyarsky:2005us,Boyarsky:2006zi,Boyarsky:2006fg,Watson:2006qb,Boyarsky:2006ag,Boyarsky:2007ay,Loewenstein:2012px,Bulbul:2014sua,Ruchayskiy:2015onc,Franse:2016dln,Boyarsky:2014ska,Boyarsky:2014jta,Iakubovskyi:2015kwa,Gewering-Peine:2016yoj},
\textit{Chandra}~\cite{RiemerSorensen:2006fh,RiemerSorensen:2006pi,Boyarsky:2006kc,RiemerSorensen:2009jp,Loewenstein:2009cm,Boyarsky:2010ci,Mirabal:2010jj,Watson:2011dw,Riemer-Sorensen:2014yda,Hofmann:2016urz},
\textit{Suzaku}~\cite{Loewenstein:2012px,Kusenko:2012ch,Urban:2014yda,Tamura:2014mta,Sekiya:2015jsa},
\textit{Swift}~\cite{Mirabal:2010an},
\textit{INTEGRAL}~\cite{Yuksel:2007xh,Boyarsky:2007ge}, HEAO-1~\cite{Boyarsky:2005us} and
Fermi/GBM \cite{Horiuchi:2015pda,Ng:2015gfa}, as well as a rocket-borne X-ray
microcalorimeter~\cite{Abazajian:2006jc,Boyarsky:2006hr,Figueroa-Feliciano:2015gwa}, \nustar~\cite{Riemer-Sorensen:2015kqa,Neronov:2016wdd,Perez:2016tcq} (we use a more conservative of two \nustar bounds \cite{Neronov:2016wdd} and \cite{Perez:2016tcq} for our final summary plot in Section~\ref{Conclusion}).
The bounds from the non-observation of DM decay
line in X-rays discussed in more detail in Refs.\cite {Boyarsky:2012rt,Adhikari:2016bei}.

\subsubsection{Status of the 3.5~keV Line}
\label{sec:3_5kev}

Recently, an unidentified feature in the X-ray spectra of galaxy clusters~\cite{Bulbul:2014sua,Boyarsky:2014jta} as well as Andromeda~\cite{Boyarsky:2014jta} and the Milky Way galaxies~\cite{Boyarsky:2014ska} have been reported by two independent groups.
The signal can be interpreted as coming from the decay of a DM particle with the mass $\sim 7$~keV (in particular of a sterile neutrino with the mixing angle $\sin^22\theta \simeq (0.2 - 2)\times 10^{-10}$).
The signal sparked a great deal of interest.
A number of works have subsequently found the line in the spectra of galaxies clusters~\cite{Urban:2014yda,Iakubovskyi:2015dna,Franse:2016dln} or galaxies~\cite{Ruchayskiy:2015onc,Neronov:2016wdd,Perez:2016tcq,Cappelluti:2017ywp}).
Other DM-dominated objects did not reveal the presence of the line~\cite{Riemer-Sorensen:2014yda,Malyshev:2014xqa,Riemer-Sorensen:2015kqa,Anderson:2014tza,Figueroa-Feliciano:2015gwa,Tamura:2014mta,Sekiya:2015jsa}.
Possible explanations of the origin of the line also included: statistical fluctuation, unknown astrophysical line; an instrumental feature.

\textbf{Statistical fluctuation?}
The original detections were at the level $3-4\sigma$.
By now the line has been observed in the spectra of many objects, with its formal statistical significance exceeding $5\sigma$~\cite{Bulbul:2014sua,Boyarsky:2014jta,Boyarsky:2014ska} (even taking into account the ``look elsewhere effect''~).
The intensity of the line is consistent with the estimates of DM column density~\cite{Boyarsky:2014ska,Ruchayskiy:2015onc,Iakubovskyi:2015dna}.
So, while in some object the line can be just a fluctuation ``in the right position'', this cannot be the only explanation of its origin.
Recent observation of the line at $11\sigma$ in the blank sky dataset with the \nustar satellite~\cite{Neronov:2016wdd} made the ``statistical fluctuation'' even less likely.\footnote{The $3.5$~keV line has been previously marked as ``instrumental'' in \cite{Wik:2014boa}, ``possibly related to the solar activity''. However, the analysis of \cite{Neronov:2016wdd} did not find the variation
of the line flux between the ``sun-illuminated'' and ``no-
sun''  parts  of  the  data  set beyond the mere change of exposure.}

\textbf{An unknown instrumental effect (systematics)?}
Given high statistical power of the datasets (with the statistical errors being $1\%$ or less) a natural ``suspect'' would be a systematics of the instrument (for example, a per-cent level unknown feature in the effective area of the telescope would explain such a ``bump'' in the spectrum).
However, many works have demonstrated that such an explanation does not hold.
First of all, the line was shown to scale correctly with the redshift of the source (see~\cite{Bulbul:2014sua,Boyarsky:2014jta,Iakubovskyi:2015dna}) which is impossible to explain with any feature related to the instrument.
In addition, the line has been observed by 4 instruments.
Therefore any explanation should appeal to instrumental feature \emph{common} in all 4 telescopes.
In particular.
the presence of the \textsc{Au} M-absorption edge at energy $\sim 3.4249$~keV \citep{HENKE1993181,Kurashima:16}, invoked as a reason for explaining the feature in \textit{XMM-Newton} and \textit{Suzaku} would not explain the feature in the spectra observed with \textit{Chandra}~ \cite{Cappelluti:2017ywp} (where Iridium is used instead of Gold~\cite{Iridium_Chandra}) or in \nustar~\cite{Neronov:2016wdd,Perez:2016tcq} (where observations are based on ``bounce-0'' (stray) photons not collected by the mirrors).
In addition, the systematics related to the properties of CCD detector (common in \textit{XMM-Newton}, \textit{Chandra}, and \textit{Suzaku}) would not explain the feature in the spectrum of \nustar with its Cadmium-Zinc-Telluride detectors.
Finally, an $11\sigma$ detection of the 3.5 keV signal by \nustar from a quiescent region of the sky~\cite{Neronov:2016wdd} and a subsequent confirmation of this result with the Chandra X-ray observatory \cite{Cappelluti:2017ywp} suggest that the line is not a statistical fluctuation and nor an artefact of the XMM-Newton's instrumentation.

\textbf{Atomic transition as the origin of the line?}
It was argued that the line could originated from atomic transition (e.g.\
Potassium~XVIII with lines at $3.47$ and  $3.51$~keV~\cite{Riemer-Sorensen:2014yda,Jeltema:2014qfa,Carlson:2014lla,Phillips:15} or Charge Exchange between neutral Hydrogen and bare sulphur ions leading to the enhanced transitions between highly excited states and ground states of \textsc{S XVI}~\cite{Gu:2015gqm,Shah:2016efh,Gu:2017pjy}.
The corresponding sulphur transitions have energies in the range $3.4 - 3.45$~keV -- close to the nominal position of the detected line (for further discussion see~\cite{Boyarsky:2014paa,Bulbul:2014ala,Iakubovskyi:2015kwa,Cappelluti:2017ywp} or~\cite{Adhikari:2016bei}).
It should be noted that the spectral resolution of \xmm (as well as \suzaku, \chandra or \nustar) does not allow to resolve the intrinsic width of the lines (or separate lines of the multiplet of \textsc{K XVIII}).

Recently the centre of the Perseus galaxy cluster has been observed with the \textit{Hitomi} spectrometer~\cite{Aharonian:2016gzq} whose energy resolution at energies of interest is few eV.
The observation did not reveal any atomic lines around $3.5$~keV (and only minor residuals at $3.4 - 3.45$~keV that could be interpreted as \textsc{S XVII} charge exchange, insufficient to account for a strong signal from the centre of the Perseus cluster, as found by \xmm and \suzaku~\cite{Bulbul:2014sua,Urban:2014yda,Franse:2016dln}.

\emph{While these observations rule out atomic transitions as the origin of 3.5 keV line, they leave room for DM interpretation.}
Indeed, typical broadening of the atomic lines in the centre of the Perseus cluster are determined by thermal and turbulent motions and correspond go the width $v_{therm} \sim \unit[180]{km/sec}$~\cite{Aharonian:2016gzq}.
The width of a DM decay line, on the other hand, is determined by the virial velocity that is of the order $v_{vir} \sim \unit[1300]{km/sec}$ \cite{Kent:83}. The corresponding Doppler broadening is wider than the spectral resolution of the \textit{Hitomi} spectrometer and therefore the bounds on the putative DM decay line are much weaker than the limit on the astrophysical line.
\emph{As a result the bounds of \textit{Hitomi} are consistent with \xmm detection in galaxies and galaxy clusters.}

Future progress with our understanding of the nature of 3.5 keV line will come with the next generation of high-resolution X-ray missions, including XARM (\textit{Hitomi} replacement mission)\footnote{\url{https://heasarc.gsfc.nasa.gov/docs/xarm}}, LYNX\footnote{\url{https://wwwastro.msfc.nasa.gov/lynx}} and Athena+~\cite{Neronov:2015kca}. Additionally, microcalorimeters on sounding rockets~\cite{Figueroa-Feliciano:2015gwa} looking into the direction of Galactic Centre may confirm the origin of the line.

\subsection{Sterile Neutrinos and Structure Formation}
\label{sec:structure-formation}

All production mechanisms discussed in Section~\ref{sec:production} tend to predict non-thermal sterile neutrino momentum distributions that potentially exhibit a non-negligible free streaming length in the early universe and will thereby affect the formation of structures in the early universe. 
In the following we review the observational constraints on the free streaming in a mostly model-independent manner.
The conversion of these results into constraints on the mass and mixing of sterile neutrinos is highly model dependent, and there are still considerable uncertainties in different aspects of the problem (including the production in the early universe, the simulation of non-linear structure formation as well as the difficulty to observe small scale structures). 
Here we mostly focus on the (non-resonant or resonant) thermal production mechanism described in Section~\ref{Sec:ThermalProduction} as an example.\footnote{
The other mechanisms have also been studied in detail in the literature, recent results and references can e.g. be found in Refs.~\cite{Merle:2014xpa,Adhikari:2016bei,Schneider:2017qdf}.
}

\subsubsection{Warm Dark Matter}
\label{sec:wdm}

The production of sterile neutrinos occurs at temperatures much higher than
the decoupling temperature of active neutrinos, $T_{dec} \sim
\unit[1]{MeV}$.
Therefore, for  masses  in the keV range sterile
neutrino particles are born \emph{relativistic}.
DM candidates with such properties are commonly referred to as \emph{Warm
  Dark Matter} (or
WDM).\footnote{
Following common convention, we use the term WDM for particles that are relativistic
when they are produced, but become non-relativistic before matter-radiation equality. This is in contrast to "Hot Dark Matter" particles that are relativistic during structure formation and "CDM particles", which are always non-relativistic.} 
The original idea of WDM dates back to 1980s and was  revitalised from
mid-1990s onward (largely due to the idea of sterile neutrino DM~\cite{Dodelson:1993je,Colombi:1995ze}).
It is very difficult to give
proper credits, for an incomplete list of early papers see e.g.\ \cite{Bond:1980ha,Bond:1982uy,Blumenthal:1982mv,Malaney:1995cb,Dodelson:1993je,Colombi:1995ze,Shi:1998km,SommerLarsen:1999jx,Hogan:1999hj,Dalcanton:2000hn,Bode:2000gq,Dolgov:2000ew}. 

Warm DM particles change the formation of structures in the Universe
as compared to the cold DM case.  While DM particles are
still relativistic, they do not cluster, but rather stream freely. This erases
primordial density perturbations at scales below
their \emph{free-streaming horizon} (usually called simply \emph{free
  streaming scale}, see~\cite{Boyarsky:2008xj} for discussion of some
subtleties and confusions associated with the definitions).  This quantity
is defined in the usual way:
\begin{equation}
  \label{eq:3}
  \lambda_{\rm fs}(t) \equiv a(t) \int_{t_i}^t dt'\,\frac{ v(t')}{a(t')} 
  \approx \unit[1]{Mpc}\frac{\unit{keV}}{\M}
  \frac{\langle p_\dm\rangle}{\langle p_\nu\rangle}
\end{equation}
(see~\cite{Boyarsky:2008xj} for details).
Here $v(t)$ is a typical velocity of DM particles, $t_i$ is the
initial time (its particular value plays no role); $a(t)$ is the scale factor
as a function of (physical) time.
In the last equality, $\langle p_\dm\rangle$ and $\langle p_\nu\rangle$ are
the average absolute values of momentum of DM particles and active
neutrinos.  The integral is saturated at early times when DM particles are
relativistic ($v(t) \approx 1$).

The free-streaming horizon indicates at what scales matter power spectrum of
warm DM particles will be distinguishable from that of CDM. \emph{The scales affected by the keV-scale warm DM
  are Mpc or below.}

Historically the first warm DM candidates were heavier analogues of neutrinos -- particles that decoupled from the equilibrium with primordial plasma while being relativistic.
To reconcile their mass with the Tremaine-Gunn bound, discussed above, such particles -- known as \emph{relativistic thermal relics} -- needed to decouple much earlier than the ordinary neutrinos and one had to arrange for a significant entropy dilution that would reduce the number density of the thermal relics, $n_{\tr} \ll n_\nu$, thus reconciling astrophysical and cosmological DM observables.
For thermal relics the suppression of power spectrum at small scales (large comoving $k$) behaves as $k^{-10}$~\cite{Bode:2000gq,Viel:2005qj}.
This scale can be directly related to the free-streaming~\eqref{eq:3} (see~\cite{Boyarsky:2008xj}).

\begin{figure}[!t]
  \centering
  \includegraphics[width=0.5\textwidth]{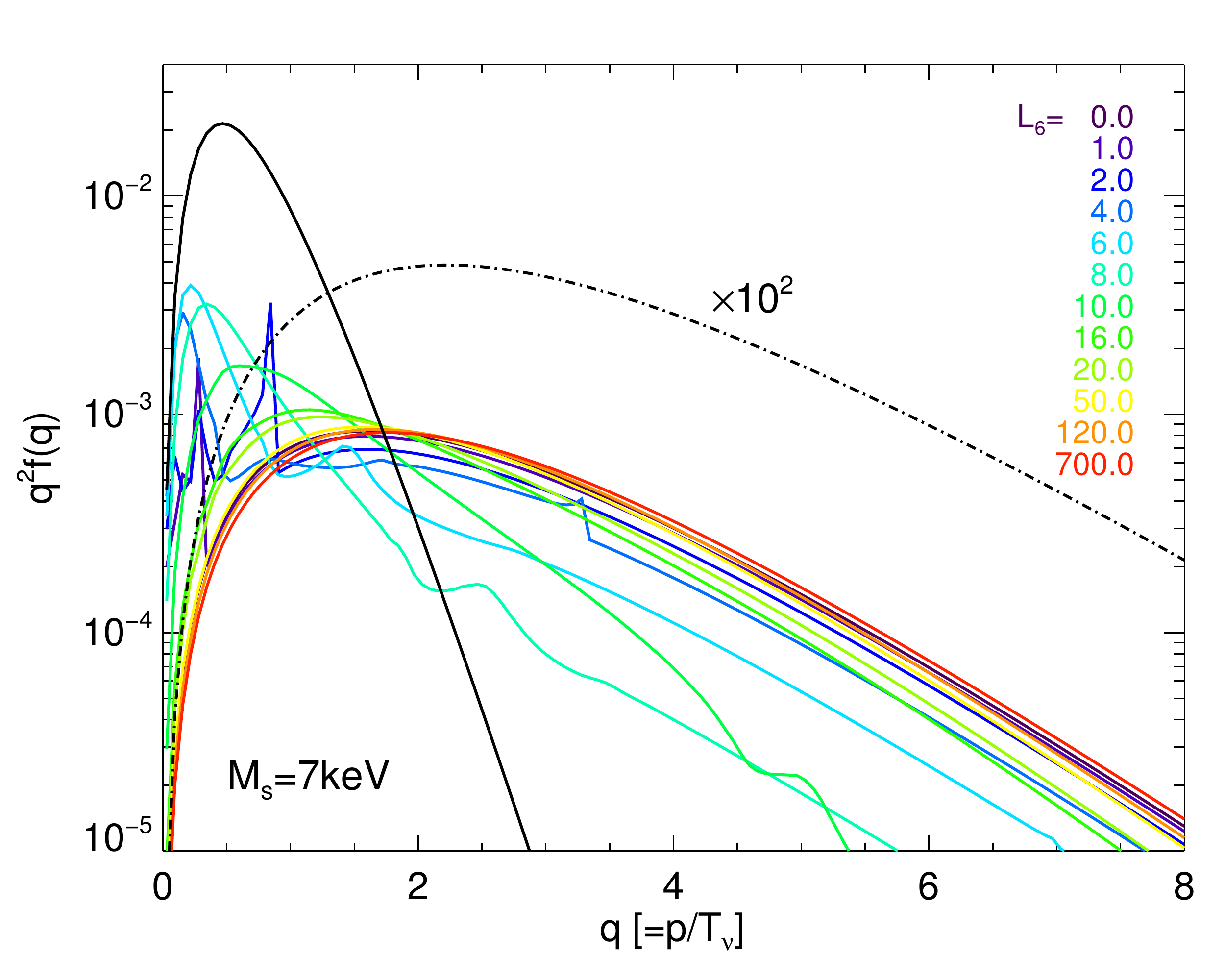}~\includegraphics[width=0.5\textwidth]{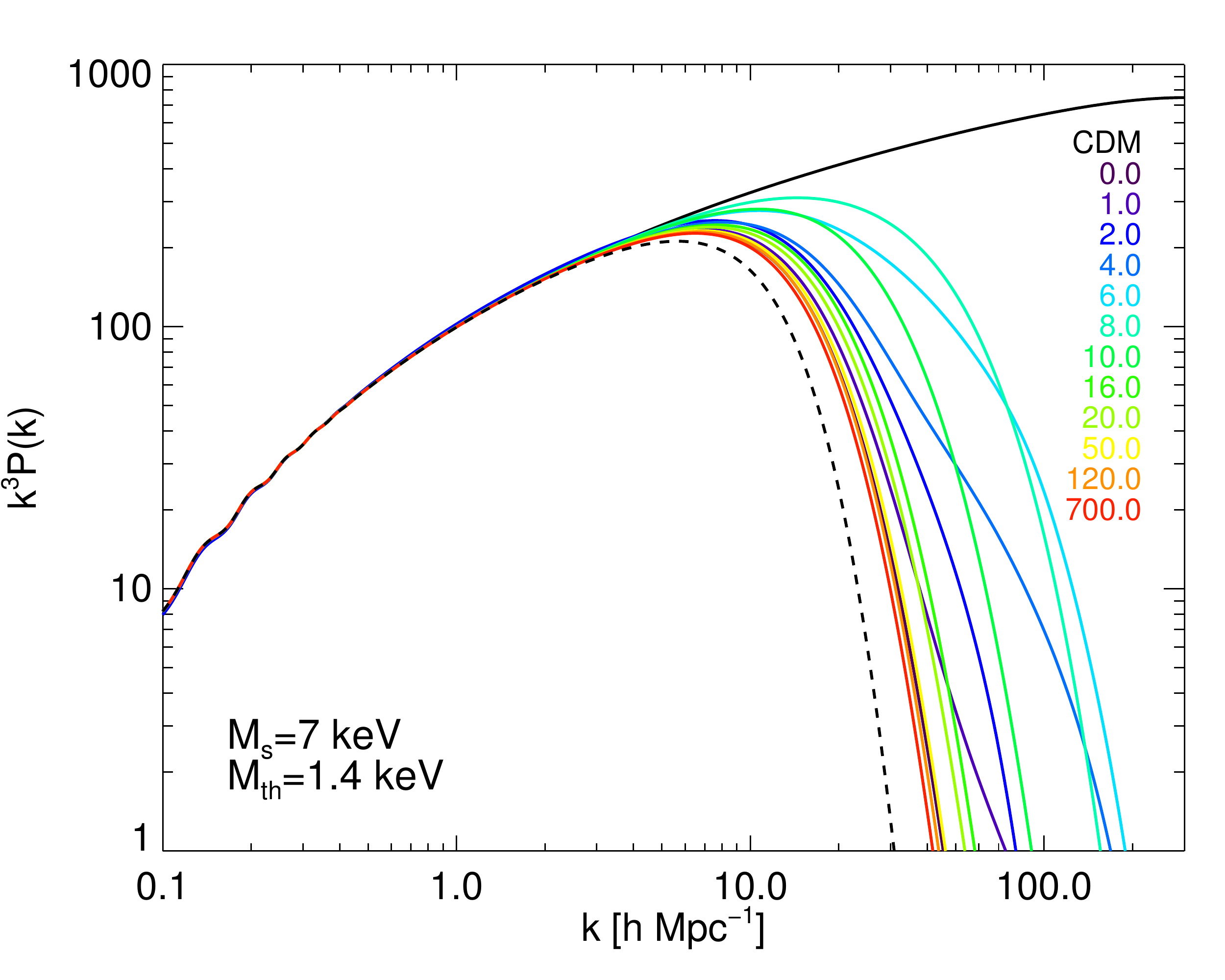}
  \caption[Power spectra for RP sterile neutrinos]{\textbf{Sterile neutrino DM vs.\ thermal relics.}
    \textit{Left:} Sterile neutrino momentum distributions for different values of $L_6$,  the lepton asymmetry in units of $10^{-6}$, with  $L_6=[0,700]$ as indicated by the legend.
    Each distribution $f(q)$ is multiplied by the momentum squared, $q^2 \times f(q)$, to reduce the dynamic range.
    Solid black line -- the Fermi-Dirac distribution of a thermal relic with the mass $M_{\rm th} = 1.4$~keV (the temperature of this distribution is different from $T_\nu$).
    Dashed black line -- Fermi-Dirac distribution with $T=T_\nu$, multiplied by $10^{-2}$ to fit into the plot.
    The momenta are plotted for the thermal plasma at a temperature of 1~MeV.
    The sterile neutrino mass is 7~keV.
    \textit{Right:} Matter power spectra generated from the 7~keV distribution functions shown in the left panel.
    The CDM power spectrum is shown as a solid black line.
    The dashed line corresponds to the power spectrum of a thermal relic of mass 1.4~keV, which is the thermal relic counterpart of the non-resonantly produced 7~keV sterile neutrino.
    From~\cite{Lovell:2015psz}.}
  \label{fig:M7L6PS}
\end{figure}

The situation is different for sterile neutrino DM produced via
non-resonant~\cite{Dodelson:1993je,Abazajian:2001nj,Asaka:2006nq,Asaka:2006rw}
or
resonant~\cite{Shi:1998km,Laine:2008pg,Venumadhav:2015pla,Ghiglieri:2015jua}
mixing with active neutrinos, cf sec.~\ref{Sec:ThermalProduction}.
These particles are produced out of thermal
equilibrium (see Section~\ref{sec:production}) and therefore there is no
universal behaviour of their primordial velocity spectrum,
Fig.~\ref{fig:M7L6PS}. As a result primordial power spectrum depends on the
whole primordial momentum distribution function (see
Fig.~\ref{fig:M7L6PS}). In particular, the small scale behaviour of the matter
power spectra differs drastically even for the particles of the same mass. The
power spectra of sterile neutrino DM can be roughly
\emph{approximated} as a mixture of cold and warm components~\cite{Boyarsky:2008mt}.

To constrain WDM models, including those of sterile neutrinos, there are three main approaches:
\begin{enumerate}
\item Measure the matter power spectrum at relevant scales (Lyman-$\alpha$
  forest; weak lensing; 21 cm line). We discuss the Lyman-$\alpha$ forest in
  Section~\ref{sec:lyman-alpha} and comment very briefly on the other methods.
\item Determine the number of collapsed structures as a
  function of their masses and redshifts (dwarf galaxy counts; reionisation
  history; collapsed objects at high-$z$ Universe). We discuss some of these approaches and their limitations in Section~\ref{sec:counting-dark-matter}.
\item Determine the distribution of matter within individual DM
  dominated objects  (cores in galactic halos). We shall not discuss this question in the current review and refer an interested reader to~\cite[Section 3.3]{Adhikari:2016bei}.
\end{enumerate}
Below we will discuss the most powerful methods, and overview the other ones briefly.

\subsubsection{Lyman-$\alpha$ Forest Method}
\label{sec:lyman-alpha}

The power spectrum of matter at scales $\sim 0.1$--$1$~Mpc and redshifts $z = 2-6$ can be determined by the so-called \emph{Lyman-$\alpha$ forest method}~\cite{Zaldarriaga:2001xs,Gnedin:01,McDonald:1999dt,Croft:1997jf,Croft:2000hs}.

At these redshifts large fraction of primordial hydrogen fills a network of mildly non-linear structures (filaments), forming the so-called \emph{intergalactic medium} (IGM).
The hydrogen makes filaments ``visible'' via absorption features in the spectra of distant quasars.
The statistics of these absorption features allows to reconstruct the distribution of neutral hydrogen and, under the assumption that it traces the matter distribution, the matter power spectrum itself~\cite{Croft:1997jf,McDonald:1999dt,Croft:2000hs,Zaldarriaga:2001xs,Gnedin:01}.
The structure of the filaments is different in cold- and warm DM cosmologies~\cite{Gao:2014yja,Gao:2007yk,Knebe:2003hs} which makes Lyman-$\alpha$ forest method a powerful probe of DM free-streaming~\cite{Hansen:2001zv,Viel:2005qj,Viel:2006kd,Seljak:2006qw,Boyarsky:2008mt,Boyarsky:2008xj,Palanque-Delabrouille:2013gaa,Palanque-Delabrouille:2015pga,Baur:2015jsy,Yeche:2017upn,Baur:2017stq,Irsic:2017ixq,Murgia:2018now}.

The basic observable of the method is the one-dimensional \emph{flux power spectrum} -- a Fourier transform of the transmitted flux correlation function averaged along the lines of sight.
It is not possible to de-project this observable and to reconstruct the underlying matter power spectrum~\cite{Viel:2004bf}.
Therefore the common approach is to compare the data to the mock spectra extracted from hydrodynamical simulations.
Extraction and analysis process attempts to closely replicate the physics of real experiments, in particular, taking into account spectral resolution of the instruments, because the probed scales are directly related to the ability to resolve nearby absorption lines.

The flux power spectra of medium resolution instruments (e.g.\ those coming from SDSS or SDSS-III/BOSS surveys) resembles that of 1D projection of CDM matter power spectrum (grows with the wavenumber $k$ and with time).
Therefore the WDM bounds, based on these datasets~\cite{Viel:2006kd,Seljak:2006qw,Boyarsky:2008mt,Boyarsky:2008xj,Baur:2015jsy,Yeche:2017upn,Baur:2017stq} essentially probe for deviations from the CDM flux power spectrum in excess to the data error bars.
The strongest bounds to-date have been provided by the Lyman-$\alpha$ data based on the BOSS DR9 dataset~\cite{Baur:2015jsy,Yeche:2017upn,Baur:2017stq} due to the large number of quasars it surveyed.
The bounds, based on the BOSS data without two highest redshift bins ($z = 4.2, 4.6$) should be considered as the most reliable ones, as the cross-correlation between Ly-$\alpha$ and \textsc{Si III} (discussed in~\cite{Palanque-Delabrouille:2013gaa,Baur:2015jsy,Palanque-Delabrouille:2015pga,Yeche:2017upn,Baur:2017stq}) affects mostly the highest redshift bins of the BOSS spectrum.
Most of the other relevant systematics is controlled via nuisance parameters (see discussion in~\cite{Palanque-Delabrouille:2015pga,Yeche:2017upn}).
These bounds are shown in Fig.~\ref{fig:SummaryPlot} as dashed light-brown line.

The situation is quite different in the case of high-resolution spectra (e.g.\ \cite{Croft:2000hs,Kim:2003qt,Seljak:2006qw,Viel:2013apy,Irsic:2017ixq,Murgia:2018now}).
The transmitted flux power spectrum reaches a maximum (at $k \sim \unit[0.02]{s/km}$ at $z = 4-6$) and then decreases for larger $k$~\cite{Becker:2006qj,Calverley:2010tz,Becker:2011ee,Viel:2013apy,Garzilli:2015iwa}.
This is a qualitative behaviour expected from the 1D matter power spectrum of  WDM.
And indeed, the work~\cite{Garzilli:2015iwa} demonstrated that the characteristic shape of the suppression of the power spectrum and its evolution with redshifts \emph{can be completely explained by WDM} (cf. also \cite{Garzilli:2018jqh}).

However, free streaming of DM particles is not the only effect that can affect the shape of flux power spectrum.
At redshifts of interest the IGM is in a highly ionised state, being photo-ionized and photo-heated by early sources.
As a result, the amount of flux power spectrum at small scales as compared to the matter power spectrum can be reduced via several distinct astrophysical mechanisms.

First of all, thermal motion of gas leads to the increase of the absorption line width.
This Doppler (temperature) broadening is a one dimensional smoothing effect that originates from observing the hot IGM along the line of sight.
There are numerous measurements of the IGM temperature at $z \le 5$ and only a single temperature measurement at $z \sim 6$~\cite{Bolton:2011ck}.
In order to resolve the degeneracy between the effects of WDM and thermal broadening, the gas temperature at $z\sim 5$ should be measured.
The IGM temperature can be \emph{determined} from the broadening of the Lyman-$\alpha$ absorption lines in QSO spectra \citep{Schaye:1999vr,Ricotti:1999hx,McDonald:2000nn,Theuns:2000va,Zaldarriaga:2000mz,Bolton:2011ck,Viel:2005ha,Lidz:2009ca,Becker:2010cu,Rudie:2012mx,Garzilli:2012gy,Garzilli:2015bha}.
In particular, a new method suitable for high resolution spectra --- \emph{Gaussian optical depth decomposition} --- was presented in \cite{Garzilli:2015bha}.
It is based on the idea that for high-resolution it is possible to identify individual absorption lines and to infer gas temperature directly via measurement of their \emph{broadening}.
Alternatively, it has been proposed to determine the IGM temperature by measuring the level of the transmitted flux \cite{Bolton:2007xi, Viel:2009ak, Calura:2012qq, Garzilli:2012gy}, however there is no agreement between the two methods yet, see \citep{Rollinde:2012vv}.\footnote{Note that, at much higher redshifts $\sim20$ the temperature of the gas can be constrained from $21$ cm observations \cite{Hektor:2018lec}.}

Apart from the Doppler broadening, the growth of neutral hydrogen density fluctuations on small scales is limited by the pressure of the gas that effectively removes the structures smaller than the Jeans scale \cite{Gnedin:1997td,Theuns:1999mz,Desjacques:2004xy,Peeples:2009uj,Kulkarni:2015fga,Garzilli:2015bha,Onorbe:2017ftn} -- the so-called ``pressure smoothing effect''.
Unlike the Doppler broadening that is determined by the instantaneous gas temperature, the pressure smoothing depends on the history of reionisation (the suppression scale due to the pressure effect is determined by the maximal Jeans length in the prior epoch)~\cite{Gnedin:1997td}.
Therefore, to obtain robust bounds on primordial DM velocities with the Lyman-$\alpha$ forest method, \emph{one needs to constrain the thermal history of IGM at redshifts $z \ge 5$}.
Currently, only the measurements of the electron optical depth to reionisation (for review see e.g.~\cite{Reichardt:2015cos}) provide constrains on the reionisation history (for a recent overview of possible reionisation histories see~\cite{Hazra:2017gtx}).

\subsubsection{Measuring Matter Power Spectrum via Weak Gravitational Lensing}
\label{sec:lensing}

An alternative way to constraint the matter power spectrum is provided by weak
lensing~\cite{Markovic:2010te,Smith:2011ev,Casarini:2012qj,Obreschkow:2012yb,Markovic:2013iza,Mahdi:2016las}. However,
the power spectra of CDM and WDM models are distinct mostly at high redshifts
(as WDM structure formation starts \emph{later} and then ``catches up'' with
CDM when entering non-linear stage). Therefore at small redshifts
($z \lesssim 1$), where the weak lensing measurements are performed, the
difference between the CDM and WDM power spectra is minimal unless one goes to
very small (tens of kpc) scales, where the effects of baryonic physics are
important and may even be
dominant~\cite{Semboloni:2011fe,vanDaalen:2011xb,Markovic:2013iza}.

\subsubsection{Counting halos}
\label{sec:counting-dark-matter}

Another way to constrain the primordial velocity of DM particles is to measure the number and properties of virialised DM structures (\emph{halos}).
In CDM structures are formed hierarchically, from the smallest masses (possibly all the way down to the Earth mass halos, see e.g.~\cite{Diemand:2005vz}) upwards.
Away from both smallest and largest mass ends the \emph{halo mass function} $M^2 dn/dM $ is only a slow function of mass (see e.g.~\cite{Tinker:2008ff}).
Similar behaviour holds for \emph{subhalos} -- virialised structures within a massive halo.
The simulations of WDM show that the structures also form hierarchically.\footnote{It is possible that at the smallest scales the fragmentations of the WDM filaments takes place, c.f.~\cite{Knebe:2003hs,Gao:2007yk,Gao:2014yja,Paduroiu:2015jfa}.
  Notice, however, that at such scales WDM simulations are plagued by spurious halo formation and a special care should be taken to disentangle artefacts from physical effects~\cite{Wang:07,Lovell:2011rd,Lovell:2013ola,Angulo:2013sza}.
  Therefore the formation of the WDM structures around the free-streaming is still being debated~\cite{Abel:2011ui,Hahn:2012ma,Lovell:2013ola,Angulo:2013sza}.}

The number of substructures in WDM cosmologies is suppressed (compared to CDM) below a \emph{threshold mass} $M_{\rm fs}$~\cite{Bode:2000gq,Colin:2007bk,Miranda:2007rb,Knebe:2008qy,Polisensky:10,Lovell:2011rd,Dunstan:2011bq,Schneider:2011yu,Benson:2012su,Angulo:2013sza,Schneider:2013ria,Colin:2014sga,Bose:2015mga,Bose:2016irl}.
This threshold mass is related to the free-streaming scale~\eqref{eq:3} via
\begin{equation}
  \label{eq:8}
  M_{\rm fs} \simeq \frac{\pi}6\lambda_{\rm fs}^3
\end{equation}
Therefore by counting the number of observed small mass halos and comparing
this with theoretical predictions, one can constrain primordial velocities of
DM properties.

Many authors have attempted to put bounds on parameters of WDM particles using this
approach~\cite{Zavala:2009ms,Polisensky:2010rw,Bode:2000gq,Colin:2007bk,Miranda:2007rb,Knebe:2008qy,Maccio:2009isa,Polisensky:10,Lovell:2011rd,Dunstan:2011bq,Schneider:2011yu,Anderhalden:2012jc,Lovell:2013ola,Benson:2012su,Angulo:2013sza,Schneider:2013ria,Abazajian:2014gza,Colin:2014sga,Schneider:2014rda,Bose:2015mga,Lovell:2015psz,Lovell:2016nkp,Lovell:2016fec,Bose:2016irl,Nierenberg:2013lqa,Menci:2016eui,Menci:2017nsr,Cherry:2017dwu}.
Below we briefly summarise the main results, explaining why it is difficult to
make robust quantitative predictions, using this method.

Most of the early works compared the number of \emph{observed dwarf satellite galaxies} of the Milky Way (and possibly Andromeda) with theoretical predictions from the DM-only simulations.
Although the exact limits, derived in different works differ significantly, \emph{qualitatively} one may summarise that thermal relic WDM particles with masses $\lesssim 1$~keV are incompatible with the observed Galactic structures, while the thermal relics with the mass $\gtrsim 2$~keV are in broad agreement with observations.

A detailed quantitative comparison, however, is problematic from both theoretical and observational viewpoints.

\textbf{From the observational side} there is a bias between the number of dark halos and of the luminous galaxies, populating these halos.
Indeed, as the mass of the satellite halos drops, it becomes increasingly difficult for them to confine the gas, heated during the reionisation epoch or expelled by stellar feedback (see e.g.~\cite{Bullock:2000wn,Somerville:2001km,Benson:2001au,Benson:2001at,Koposov:2009ru,Maccio:2009aek}).
Therefore, it is possible that the low mass halos exist, but are completely dark.

Additionally, we do not know how typical the Milky Way (and the Local Group) and their satellite distributions are (see e.g.,~\cite{DeRossi:2008vk,Mutch:2011pb}).
Therefore, before one has a more complete census of subhalo population in the different types of galaxies, any quantitative bounds based on the number of dark or luminous satellites is not possible.

Gravitational lensing (and microlensing) is the promising way to probe the presence of substructures directly (c.f.~\cite{Vegetti:2009cz,Vegetti:2012mc,Hezaveh:2014aoa,Xu:2014dda,Li:2015xpc, Li:2016afu,Kamada:2016vsc,Mahdi:2016las,Fedorova:2016gpa,Birrer:2017rpp,Vegetti:2018dly}).
A large sample of lenses together with a robust way to predict the average distribution of sub-halos is required in order to make a statistically robust prediction

\textbf{From the theoretical viewpoint} it is difficult to predict both the number of DM subhalos within a host galaxy and the way they are populated by stars.
One may use semi-analytical models of galaxy formation (SAMs), see~\cite{Baugh:2006pf} for a review.
A number of works used it to constrain the properties of DM and in particular of sterile neutrino DM~\cite{Maccio:2009isa,Kang:2012up,Menci:2012kk,Kennedy:2013uta,Lovell:2015psz,Dayal:2014nva,Dayal:2015vca,Wang:2016rio,Menci:2018lis}.
If one determined observationally a subhalo luminosity functions in a statistically significant sample of galaxies, the SAMs would be a possible way to compare theoretical predictions with observations and therefore to constrain WDM properties.\footnote{SAMs combine theoretical inputs with the results of hydrodynamical simulations and are calibrated on astrophysical observations. Their use for satellite luminosity function of the Milky Way-size galaxies should be taken with caution, being an extrapolation.}
However, at the moment the observational data is limited and therefore median predictions from SAMs do not allow to compare between observations.

At the level of individual simulated halos, the number of satellites depends on a particular realisation in the numerical simulations of an initial density field (even when starting from the same matter power spectrum).
Therefore, subhalo population in numerical simulations exhibits a strong realisation-to-realisation scatter, see the discussion in~\cite{Kennedy:2013uta,Lovell:2015psz,Lovell:2016fec}.
Such a scatter can in particular be larger than the difference between the satellite number in the CDM and WDM cosmologies~\cite{Lovell:2016fec} which explains disagreement between different works on the subject.
Predictions of the luminosity functions based on the cosmological simulations with baryons, such as for example the APOSTLE simulation~\cite{Sawala:2015cdf} or Warm FIRE~\cite{Bozek:2018ekc} are also subject to the realisation scatter and, in addition to that, to uncertainties, related to the modelling baryon physics.

In summary, the satellite count presently does not allow to put robust bounds on sterile neutrino models in a statistically meaningful way. 
One can \emph{check} whether a particular DM model is compatible with observations, but one cannot rule out a model based on such a check.

\subsection{Other Observables }
\label{OtherConstraints}Other observables that rely on the effect of WDM in general and sterile neutrino DM in particular on structure formation to constrain their properties include 
\begin{itemize}
\item the number of dwarf galaxies in the Local Volume~\cite{Klypin:2014ira,Papastergis:2011xe,Horiuchi:2013noa,Bozek:2015bdo,Horiuchi:2015qri,Lovell:2016fec,Lovell:2016nkp},
\item the number of distant (high redshift) galaxies~\cite{Bozek:2015bdo,Nierenberg:2013lqa,Menci:2016eui,Menci:2017nsr,Menci:2018lis},
\item the abundance and structure of cosmic voids~\cite{Tikhonov:2009jq,Reed:2014cta,Yang:2014upa},
\item the abundance of most massive satellites of the Milky Way or the Local group whose apparent lack is known as \emph{too big to fail problem}~\cite{BoylanKolchin:2011de,BoylanKolchin:2011dk,Garrison-Kimmel:2014vqa}.\footnote{Of course, all the above comments regarding basing the prediction on a single object -- the Milky Way -- apply only in this case.}
  Sterile neutrino and other WDM models can alleviate this problem~\cite{Lovell:2011rd,Horiuchi:2015qri,Kennedy:2013uta,Lovell:2016nkp} although the problem can be resolved with astrophysical explanations (see e.g.~\cite{Kennedy:2013uta,Sawala:2012cn,Sawala:2014baa,DiCintio:2013qxa, DiCintio:2014xia, Papastergis:2014aba,Purcell:2012kd}) and
\item WDM may help to solve the tension between the observed shape of DM halos and theoretical predictions \cite{Dubinski:1991bm,Navarro:1995iw,Navarro:1996gj}, known as "core-cusp problem" \cite{deBlok:2009sp}, see e.g. \cite{VillaescusaNavarro:2010qy}.\footnote{
It has been pointed out that self-interactions amongst the sterile neutrinos may improve the situation \cite{Ruffini:2014zfa,Arguelles:2015baa,Mavromatos:2016vbj}.
}
\end{itemize}
The resulting constraints are generally believed to be sub-dominant compared to the bounds discussed above.
Moreover, regarding the last two points it should be emphasised that, at present time, it is not clear whether the apparent tensions between simulations in the standard $\Lambda$CDM model and observation reflect a failure of the model or a problem with the simulations. Present simulations can, for instance, not include baryonic feedback (such as supernova explosions) consistently.
The following (sub-dominant) constraints can be derived from the sterile neutrinos' effect on cosmic history (assuming standard cosmology):
\begin{itemize}
\item Their lifetime must exceed the age of the universe.
\item The history of reionisation~\cite{Barkana:2001gr,Yoshida:03,Hansen:2003yj,Mapelli:05,Biermann:06b,Kusenko:06b,Zhang:07,Yue:2012na,Pullen:2014gna,Dayal:2015vca,Rudakovskiy:2016ngi,Lopez-Honorez:2017csg,Bose:2016hlz,Oldengott:2016yjc} -- delayed structure formation in the WDM universe can affect the reionisation (both via abundance of small halos, their structure and star formation).
  As there are only few detailed constraints on the reionisation history (see the discussion in Section~\ref{sec:lyman-alpha}) it is again only possible to check whether a particular WDM or sterile neutrino model ``fits the data under reasonable assumptions'' (see e.g.~\cite{Rudakovskiy:2016ngi}).
\item The contribution to the number of relativistic degrees of freedom at the time of CMB decoupling should remain within the observational limit \cite{Hernandez:2014fha,Vincent:2014rja}.
\end{itemize}
Finally, a number of astrophysical tests have been proposed in the literature:
\begin{itemize}
\item Possible effects on supernova explosions \cite{Shi:1993ee,Kusenko:1998bk,Fuller:2003gy,Hidaka:2006sg,Hidaka:2007se,Raffelt:2011nc,Wu:2013gxa,Warren:2014qza,Warren:2016slz}.
\item  In refs.~\cite{Kusenko:1998bk,Fuller:2003gy} it has been proposed that sterile neutrinos may help to explain the motion of pulsars. 
\item The authors of ref.~\cite{Kuhnel:2017ofn} argued that compact halos of sterile neutrinos could be detected in the vicinity of (hypothetical) primordial black holes.
\end{itemize}


\section{keV-Scale Sterile Neutrino Dark Matter Production in the Early Universe}
\label{sec:production}

\subsection{Overview}\label{ProductionOverview}
Sterile neutrinos do (by definition) not carry any SM gauge charges, i.e., they do not feel any of the known forces of nature except gravity.\footnote{In principle the Yukawa couplings of course mediate a force. Practically this is not relevant in experiments at energies much below the mass of the Higgs boson.
} 
In order to be viable DM candidates, they must, however, have some interactions with other particles in order to be produced in the early universe. This can be realised in at least three different ways.
\begin{itemize}

\item[I)]  If the sterile states mix with ordinary neutrinos, then they can be produced by the weak interaction through this mixing. In the (type I) seesaw model (\ref{Lseesaw}), this active-sterile mixing occurs generically and is responsible for the light neutrino masses. For given sterile neutrino mass and mixing angle and a fixed thermal history of the universe, this leads an  unavoidable minimal amount of sterile neutrinos that is produced. We discuss production by mixing in Sec.~\ref{Sec:ThermalProduction}.

\item[II)] Neutrino states that appear to be sterile at the energies that are currently experimentally accessible may have new gauge interactions at higher energies. Many extensions of the SM invoke additional gauge symmetries that are ``broken'' at some energy scale above the reach of the LHC. This gives masses of the order of the symmetry breaking scale to the gauge bosons, effectively switching off these gauge interaction at lower energies. If the maximal temperature in the early universe exceeds the mass of the new gauge bosons, the "sterile" neutrinos can be produced thermally by the new gauge interactions. We discuss this possibility in Sec,~\ref{Sec:GaugeProduction}.

\item[III)]  The sterile neutrinos can be produced in the out-of-equilibrium decay of heaver particles in the early universe. We discuss this possibility in Sec.~\ref{Sec:DecayProduction}. Here we only consider the perturbative decay in the radiation dominated era. In principle sterile neutrinos can also produced during (p)reheating after inflation \cite{Bezrukov:2008ut,Bezrukov:2011sz}.
\end{itemize}

Sterile neutrino DM is often associated with keV masses.
Smaller masses are indeed prohibited by phase space considerations, see Sec.~\ref{sec:wdm}.
Larger masses, however, do not contradict any observational data.
The lifetime of sterile neutrinos that decay via active-sterile mixing scales as $\tau\propto \theta^{-2}M^{-5}$ \cite{Pal:1981rm,Barger:1995ty}. Stability on cosmological time scales can be made consistent with arbitrary mass $M$ if one chooses a sufficiently small mixing angle $\theta$, and the lightest singlet fermion may even be stable in extended models \cite{Heeck:2015qra,Dev:2016xcp},  avoiding the bound discussed in sec.~\ref{sec:decaying-dark-matter}.
The keV mass scale is only favoured if the DM is produced via active-sterile mixing and the weak interaction, i.e.,\ option I), which implies a lower bound on $\theta$. 
The production mechanisms II) and III) in principle allow for sterile neutrino DM with masses well above the electroweak scale.

\subsection{Thermal Production via Mixing (``freeze in'')}\label{Sec:ThermalProduction}
In this section we discuss possibility I), i.e., the thermal production of sterile neutrinos via the weak interaction, which is possible due to their mixing $\theta$ with the $SU(2)$ charged SM neutrinos $\nu_L$. 
Active-sterile mixing occurs in most models involving sterile neutrinos, and in particular in the seesaw model (\ref{Lseesaw}), see eqns. (\ref{LightMassEigenstates}) and (\ref{NWW}).
Sterile neutrinos that have no other interactions than the $\theta$-suppressed weak interaction (\ref{WeakWW}) fall into the category of \emph{feebly interacting massive particles} (FIMPs), cf. e.g. \cite{Bernal:2017kxu}.
The $\nu$MSM \cite{Asaka:2005an,Asaka:2005pn} is an example for a specific models in which this possibility is realised. It has also been studied in inverse seesaw scenarios \cite{Abada:2017ieq}.

Let us for a moment ignore the three generations of the SM and consider a simple toy model with only one active neutrino $\nu_a$ and one sterile neutrino $\nu_s$ with vacuum mixing angle $\theta$,
\begin{eqnarray}
\vert \nu_a\rangle & = & \cos\theta\, \vert \nu_1\rangle + \sin\theta\,\vert \nu_2\rangle,\\
\vert \nu_s\rangle & = & -\sin\theta\, \vert \nu_1\rangle + \cos\theta\,\vert \nu_2\rangle.
\end{eqnarray}
Here $\vert \nu_1\rangle$ and $\vert \nu_2\rangle$ are the neutrino energy/mass eigenstates with vacuum mass eigenvalues $m_1$ and $m_2$. 
The mass $M$ of a DM sterile neutrinos is known to be at least a few keV to be consistent with phase space considerations, see Sec.~\ref{sec:wdm}, while the masses $m_i$ of the ``active'' SM neutrinos are smaller than eV. Moreover, observational constraints imply $\theta^2<10^{-6}$. 
We can therefore in good approximation set $m_1=0$, $m_2=M$ an neglect all terms beyond quadratic order in $\theta$. In the seesaw model we may then identify $\nu_1=\upnu$, $\nu_2=N$, $\nu_a=\nu_L$ and $\nu_s=\nu_R$. The generalisation to three flavours is straightforward.

\subsubsection{Non-resonant Production}\label{Subsec:Nonres}  
In vacuum a neutrino that is prepared in any state other than $\vert \nu_1\rangle$ or $\vert \nu_2 \rangle$ would simply undergo coherent oscillations that can be described by an effective Hamiltonian, analogous to the oscillations of the SM neutrinos into each other.\footnote{The vacuum decay of the heavy neutrinos can be neglected on the relevant time scale.}
The presence of the primordial plasma has two effects. 

On one hand, forward scatterings with particles from the plasma modify the effective dispersion relations of the neutrinos \cite{Klimov:1981ka,Weldon:1982bn}.
This effect is similar to the refraction 
of photons in glass or water 
or the modified dispersion relations ("bands") of electrons in a solid state.
It can effectively be described by introducing effective in-medium masses and an effective in-medium mixing angle,
\begin{eqnarray}\label{mixing:states}
\vert \nu_a\rangle & = & \cos\theta_m\left( t\right)\, \vert \nu_1\left( t\right)\rangle + \sin\theta_m\left( t\right)\,\vert \nu_2\left( t\right)\rangle,\\
\vert \nu_s\rangle & = & -\sin\theta_m\left( t\right)\, \vert \nu_1\left( t\right)\rangle + \cos\theta_m\left( t\right)\,\vert \nu_2\left( t\right)\rangle,
\end{eqnarray}
The effective mixing angle $\theta_m$ and the definition of the thermal mass eigenstates are time dependent because the temperature of the primordial plasma changes due to Hubble expansion. 
Since the change in temperature is adiabatically slow ($\dot{T}/T^2\ll 1$), the cooling does not disturb the coherent evolution of the quantum system.
A neutrino is produced by the weak interaction in the state $|\nu_a\rangle$.
At any later point in time, the quantum state has an ``active'' and a ``sterile'' component \cite{Pontecorvo:1968fk}. The probability to find an active or sterile neutrino in a measurement can be computed by solving Schr\"odinger's equation.
Hence, $|\nu_s\rangle$ gets populated through the coherent oscillations.

On the other hand, sterile neutrinos can also be produced or annihilated in scatterings mediated by the weak interaction.
These scatterings can be viewed as a measurement, they destroy the coherence of the quantum state and force it into a weak interaction (flavour) eigenstate.
This phenomenon, which was initially discovered in the context of light neutrinos~\cite{Barbieri:1989ti,Kainulainen:1990ds}, has been applied to the DM problem by Dodelson and Widrow~\cite{Dodelson:1993je}. 
The mixing therefore leads to the production of sterile neutrinos in the early universe two ways:
\begin{itemize}
\item There are coherent oscillations between active and sterile states. 
\item Sterile neutrinos are produced in decoherent scatterings.
\end{itemize}
The production of sterile neutrino DM occurs mainly via decoherent scatterings.
The active neutrino scattering cross section is $\sigma \sim G_{\rm F}^2E_\nu^2$, where $E_\nu$ is the neutrino energy, leading to an overall scattering rate 
$\Gamma_\nu \sim \sigma \cdot {\rm flux} \sim G_{\rm F}^2\, T^5$, where $G_{\rm F}$ is the Fermi constant and we used $E_\nu\sim T$. 
The probability that a neutrino is ``measured'' in a sterile state is $\sim \sin^2 (2\theta_m)$. This gives a sterile production rate $\Gamma_N \sim G_F^2 T^5 \sin^2 (2\theta_m)$. 

The effective in-medium mixing angle $\theta_m$ for a neutrino state with momentum $p$  is given by
\begin{equation}\label{eq:mixing}
\sin^2(2\theta_m) = \frac{\Delta^2(p) \sin^2(2\theta)}{\Delta^2(p)\sin^2(2\theta) + \left[\Delta(p)\cos(2\theta) - V_D - V_T\right]^2}.
\end{equation}
Here$\Delta(p) = \Delta m^2/(2p)$ is  given by the splitting $\Delta m^2=m_2^2-m_1^2\simeq M^2$ between the two mass eigenvalues  in vacuum. 
$V_D$ and $V_T$ are the finite density and finite temperature \emph{matter potentials}, which are generated by forward scatterings in the plasma and are only felt by the active neutrino. 
The finite density potential $V_D$ dominates in highly matter-antimatter asymmetric environments, such as stars. In the early universe it is negligible if all matter-antimatter asymmetries are of the order of the baryon asymmetry of the universe $\eta_B\sim 10^{-10}$. 
The finite temperature potential $V_T$ in this approximation is simply a temperature dependent shift in the effective in-medium mass.\footnote{In principle the dispersion relations of fermions in a plasma can have a rather complicated momentum dependence \cite{Klimov:1981ka,Klimov:1982bv,Weldon:1989ys,Quimbay:1995jn}, but for the present purpose the approximation of a momentum independent thermal mass is sufficient.}
In the simple single flavour picture they can be estimated as 
\begin{eqnarray}
V_T&\simeq& - \frac{8}{3}\sqrt{2} G_F\left[\frac{\rho_\nu}{m_Z^2} + \frac{\rho_\ell}{m_W^2}\right] E_\nu 
\sim 2 G_F T^4 \left[
\frac{1}{m_Z^2} + r_\ell  \frac{2}{m_W^2}
\right]p \equiv G_{\rm eff}^2 T^4 p
\label{VTdef}\\
V_D&\simeq& 2\sqrt{2} G_F n_\gamma \Lnu_\nu = 
2\sqrt{2} G_F \frac{2\zeta(3)}{\pi^2} T^3 \Lnu_\nu
,\label{VDdef}
\end{eqnarray}
complete expressions for three flavours can e.g. be found in ref.~\cite{Venumadhav:2015pla}.
Here $r_\ell$ is a numerical factor that takes account of the deviation of the charged lepton occupation numbers from an equilibrium distribution for massless Dirac fermions, i.e., $\rho_\ell = r_\ell (7/8)(\pi^2/30)4T^4$.
For $r_\ell \sim 1$ one can approximate $ G_{\rm eff}^2 \sim 10^2 G_F^2$.
$L_\nu$ is the difference between neutrino and antineutrino densities normalised to the photon density $n_\gamma = 2\,\zeta\left( 3 \right) T^3 /\pi^2$, i.e.,   $\Lnu_\nu \equiv {\left( n_{\nu} - n_{\bar\nu} \right)}/n_\gamma$.
The negative sign in eq.~(\ref{VTdef}) is crucial, it implies that $V_T$ can only reduce $\theta_m$.
The physical reason is that finite temperature corrections at $T\ll m_W$ modify the neutrino dispersion relations in matter in a way that $E_\nu^2<(m_i^2 + p^2)^{1/2}$ \cite{Notzold:1987ik}, i.e., they increase the effective mass splitting.\footnote{Note that the sign changes at temperatures above the W mass \cite{Lello:2016rvl}.} 
Another way to say this is that finite temperature effects give a positive contribution to the neutrino refractive index.

At low temperatures the sterile neutrino production is suppressed because scatterings become too rare in a dilute plasma, which can be seen from the factor $T^5$ in $\Gamma_N$. At high temperatures, however, $\theta_m$ is suppressed due to $V_T$. This suppression prevents sterile neutrinos with mixing angles smaller than $\sin^2 (2\theta) \sim 10^{-6} \, (10\text{ keV}/M)$ to reach thermal equilibrium in the early universe. Sterile neutrino DM is therefore produced via incomplete \emph{freeze in} (rather than \emph{freeze out}), and the particles' momentum distribution is non-thermal, which is important in the context of cosmic structure formation, see Sec.~\ref{fig:SummaryPlot}.
The production peaks at an intermediate temperature, which roughly corresponds to the temperature when the suppression of $\theta_m$ becomes inefficient ($| V_T| \sim \Delta(p)$). For keV mass sterile neutrinos this occurs at $T\sim 0.1 - 1$ GeV. 

For any $\theta\neq 0$, there is an unavoidable contribution of sterile neutrinos to $\Omega_{\rm DM}$ due to this so-called non-resonant mechanism (or Dodelson-Widrow mechanism), which is uniquely determined by $M$ and $\theta$. If one requires this contribution to compose the entire observed $\Omega_{\rm DM}$, then $M$ and $\theta$ must lie along the solid black line in Fig.~\ref{fig:SummaryPlot}. This scenario is already excluded (or at least strongly disfavoured) by a combination structure formation bounds or X-ray searches: For values of $M>2$ keV, the required mixing angle is ruled out by the non-observation of an X-ray emission line from DM dense regions, see Sec.~\ref{sec:decaying-dark-matter}. For smaller masses, the free streaming of the DM particles is in conflict with the observed small scale structure in the universe, see Sec.~\ref{sec:structure-formation}. 
While a rough estimate indicates that the resulting sterile neutrino distribution function is proportional to a Fermi-Dirac distribution \cite{Dodelson:1993je}, the actual spectrum is in fact ``colder'', see Fig.~6 in Ref.~\cite{Asaka:2006nq}. 
Hence, though this production mechanism is ``thermal'' in the sense that the $N$ are produced in scatterings in a thermal plasma, it does not lead to a thermal spectrum. This is typical for freeze in production.\footnote{The $N$ may of course thermalise if they have additional interactions at low energies \cite{Hansen:2017rxr}.}

\subsubsection{Resonant Production}\label{Subsec:Res}
So far we have assumed that $V_D$ can be neglected, which is justified if the lepton asymmetries in the primordial plasma are as small as the baryon asymmetry. 
However, the observational constraints on the lepton asymmetry in the early universe 
are much weaker than those on the baryon asymmetry \cite{Oldengott:2017tzj}, (cf. also e.g. \cite{Canetti:2012zc} and references therein).
In the following we illustrate the effect of lepton asymmetries within the simplified single flavour picture introduced in the previous paragraph \ref{Subsec:Nonres}, in which case there is only one single relevant asymmetry $\Lnu_\nu$.
In reality there are of course three independent asymmetries $\Lnu_{\nu_\alpha} \equiv {\left( n_{\nu_\alpha} - n_{\bar\nu_\alpha} \right)}/n_\gamma$, where $n_{\nu_\alpha}, n_{\bar\nu_\alpha}$ are the thermal neutrino and antineutrino densities in flavour $\alpha$.
We continue to neglect the asymmetries in quarks and charged leptons, which are of the order of the BAU, i.e., we focus on the regime where $\Lnu_\nu$ is much larger than the BAU at the time of DM production.
This situation can e.g. be realised in the $\nu$MSM , where the required $L_\alpha$ \cite{Laine:2008pg} can be generated by the $CP$ violating interactions of heavier sterile neutrinos that also generate the observed BAU   \cite{Shaposhnikov:2008pf,Canetti:2012vf,Canetti:2012kh}.

It is well-known that the Mikheyev-Smirnov-Wolfenstein (MSW) effect~\cite{Mikheev:1986gs,Wolfenstein:1977ue} can resonantly enhance the transition between two kinds of neutrinos when finite density corrections cause a ``level crossing'' in the neutrino dispersion relations. A similar effect can occur between active and sterile neutrinos. The dispersion relation of the predominantly sterile mass state is in good approximation described by 
$E_N^2\simeq p^2 + M^2$,
as corrections $\propto\theta^2 T^2$ are negligible compared to the sterile neutrino vacuum mass $M$. The active neutrino dispersion relations, however, are significantly modified by $V_T$ and $V_D$. 
As emphasised in the previous subsection \ref{Subsec:Nonres}, $V_T$  
at $T<m_W$ gives a positive contribution to the neutrino refractive index that can be viewed as the effect of a momentum dependent ``negative thermal mass'' for the active neutrinos. 
Hence, $V_T$ increases the effective mass splitting and cannot lead to a resonant enhancement of the mixing angle (\ref{eq:mixing}).
 The contributions to $V_D$, on the other hand, are proportional to the matter-antimatter asymmetries in different species in the primordial plasma, which can have either sign.
 The effective mixing angle (\ref{eq:mixing}) is of order one when the effective mass splitting vanishes, which requires  $\Delta(p)\cos(2\theta) - V_D -V_T=0$ or 
\begin{equation}
M^2 -  2{\frac{4 \sqrt{2} \zeta\left(  3\right)}{\pi^2}}G_F \Lnu_\nu p  T^3+ 2 G_{\rm eff}^2 p ^2 T^4 =0,
\label{res}
\end{equation}
where we have neglected terms $\mathcal{O}[\theta^2]$ and the light neutrino mass.  
Solving (\ref{res}) for $\x\equiv p/T$ with fixed $T$ allows one to identify at which temperature a mode $p$ enters the resonance. 
This can happen twice per mode for each active flavour,

\begin{eqnarray}
\x _{res} 
&=& \frac{G_F}{G_{\rm eff}^2 T^2} \frac{4\zeta(3)}{\sqrt{2}\pi^2}\Lnu_\nu\left[
1 \pm
\sqrt{
1-
\frac12 \frac{M^2}{T^2}\frac{G_{\rm eff}^2}{G_F^2}
\frac{\pi^4}{8\zeta(3)^2}\frac{1}{\Lnu_\nu^2}
}
\right]
\label{epsilonres}
\end{eqnarray}
From the square root it is clear that the resonance requires a minimal $\Lnu_\nu$,\footnote{
Equation (\ref{res}) suggests that the resonance only occurs if $\Lnu_\nu$ has the right sign. This, however, is a result of the simplified quantum mechanical treatment following (\ref{mixing:states}), which neglects the internal neutrino degrees of freedom. 
If $|\Lnu_\nu|$ fulfils the condition (\ref{cond}), then the resonance always occurs, but depending on the sign of $\Lnu_\nu$, the produced sterile neutrinos have either positive or negative helicity. 
}
\begin{equation}
|\Lnu_\nu|> \frac12 \frac{M}{T}\frac{G_{\rm eff}}{G_F}
\frac{\pi^2}{2\zeta(3)}
\label{cond},
\end{equation}
which is nothing but the condition that $V_D$ must be large enough to overcome the vacuum mass splitting. Plugging in keV values for $M$ shows that the required lepton asymmetry exceeds $\eta_B$ by several orders of magnitude.

\begin{figure}
	\centering
 	\includegraphics[width=0.6\textwidth]{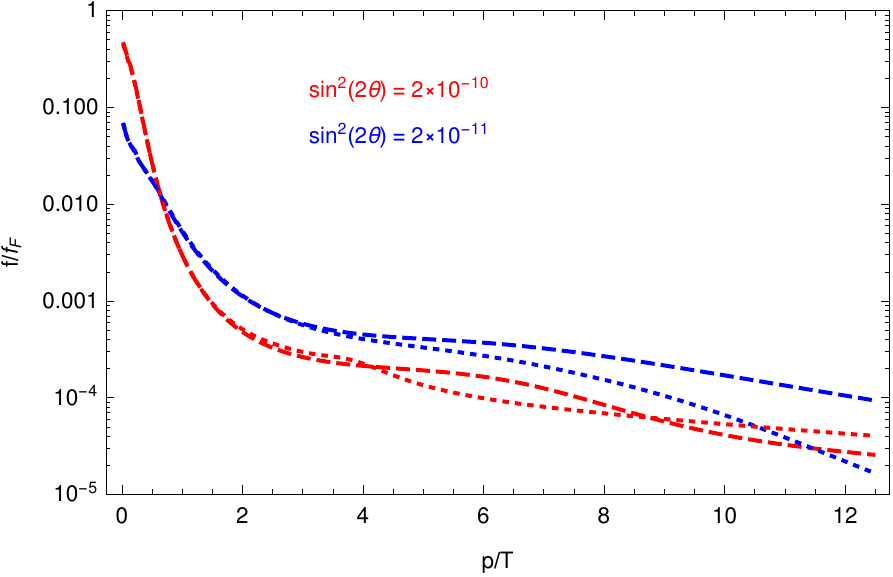}
	\caption{
The momentum distributions of resonantly produced sterile neutrino DM with $M=7.1$ keV for two different choices of $\theta$. For the dotted line it was assumed that sterile neutrinos only mix with electron neutrinos ($\theta=\theta_e$), for the dashed line they only mix with tau neutrinos ($\theta=\theta_\tau$). In all cases it was assumed that $L_e=L_\mu=0$ at $T>4$ GeV, and the initial value of $L_\tau$ was adjusted to explain the observed $\Omega_{\rm DM}$. The data was kindly provided by the authors of Ref.~\cite{Ghiglieri:2015jua} via \href{http://www.laine.itp.unibe.ch/dmpheno/}{http://www.laine.itp.unibe.ch/dmpheno/}.
}
	\label{ResonantSpectra}
\end{figure}

\begin{figure}
	\centering
	\includegraphics[width=0.6\textwidth]{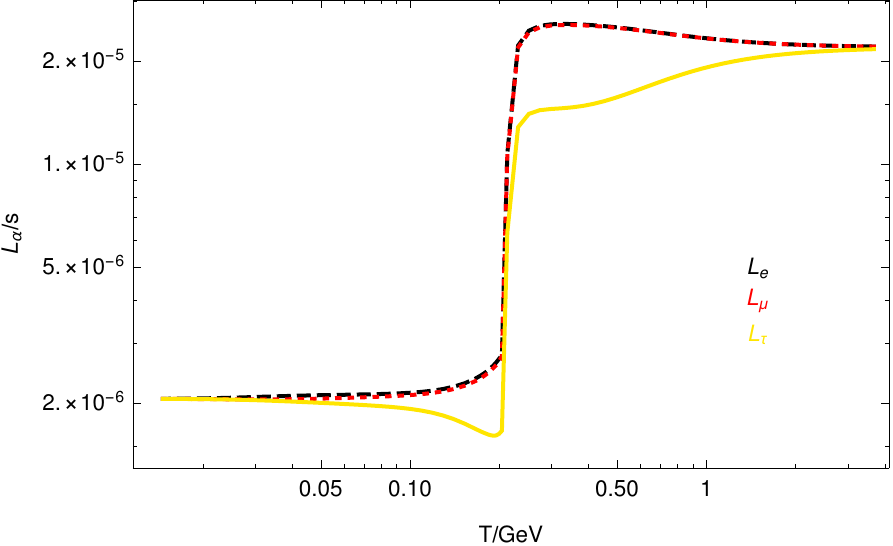}
	\caption{
Time evolution of the lepton asymmetries with $M=7.1$ keV for $\theta_\tau^2=2\times 10^{-10}$, $\theta_e=\theta_\mu=0$. 
In all cases it was assumed that $L_e=L_\mu=L_\tau$ at $T>4$ GeV, and the initial value was adjusted to explain the observed $\Omega_{\rm DM}$. 
The jump clearly shows the depletion at the moment of resonant DM production.
The data was kindly provided by the authors of Ref.~\cite{Ghiglieri:2015jua} via \href{http://www.laine.itp.unibe.ch/dmpheno/}{http://www.laine.itp.unibe.ch/dmpheno/}.
	\label{Dilution}
    }
\end{figure}

In the seesaw model, both, heavy and light neutrinos are fundamentally Majorana fermions. 
However, at the relevant temperatures $T\gg M \gg m_i$, lepton number violating processes are strongly suppressed and the two helicity states of the neutrinos act as ``particle'' and ``antiparticle''. This allows to introduce a generalised total lepton number $\tilde{L}$, to which those neutrinos with left handed helicity contribute $+1$ and those with right handed helicity contribute $-1$.\footnote{
The helicity based lepton number $\tilde{L}$ is not to be confused with the generalised lepton number $\bar{L}$ defined in eq.~(\ref{LbarDef}), cf. \cite{Antusch:2017pkq} for a discussion.} 
The conservation of $\tilde{L}$ implies that the production of sterile neutrinos depletes $\Lnu_\nu$, and the process of resonant production can be pictured as a conversion of SM lepton number $\Lnu_\nu$ into DM abundance via the MSW effect.
In a realistic scenario this happens in a flavour dependent manner and one has to track the different asymmetries  $L_\alpha$ individually, see Fig.~\ref{Dilution}.
The lepton asymmetries $L_\alpha$ then play a double role: They act as a reservoir for the production and as a catalyst that causes the resonant enhancement at the same time. 
The total amount of resonantly produced DM particles and their momentum distribution depend on a number of factors. 
As the universe cools down, different modes $p$ enter the resonance at different times. The duration for which the production in that mode remains resonant is given by the time for which $\theta_m(t)$ for that mode remains to be of order unity. The number of produced particles is governed by the value of the scattering rate $\sin^2(2\theta_m)\Gamma_\nu$ during that time.
The time dependence does not only come from the change in $T(t)$ due to Hubble expansion, but also from the depletion of the $L_\alpha$ due to the DM production. If the $L_\alpha$ are depleted too rapidly, the resonant production terminates itself because the rapid change in the asymmetry violates the adiabaticity that is required for the MSW effect. The resonant production for sure comes to an end when the $L_\alpha$ have been reduced so much that condition (\ref{cond}) cannot be fulfilled any more. From Eq.~(\ref{res}) it is obvious that low momentum modes enter the resonance first, meaning that the resonantly produced DM is comparably cold \cite{Abazajian:2001nj}, see Fig.~\ref{ResonantSpectra}. 
This is important because it allows to evade the constraints from structure formation that rule out non-resonant thermal production as the sole source of DM.
A quantitative determination of the DM spectra requires a detailed analysis of the evolution of the different neutrino flavour and helicity states at temperatures below $1$ GeV, which is complicated by the fact that the production peaks at temperatures when quarks and gluons begin to hadronise \cite{Abazajian:2002yz}.

\subsubsection{Treatment in Quantum Field Theory}

While the simple quantum mechanical description employed in the previous section is captures most aspects of the thermal production qualitatively, it is not sufficient to make precise predictions for the formation of structures in the early universe.
A more accurate computation can be performed in the framework of quantum field theory. Here we only briefly sketch the approach, a more detailed description can e.g. be found in refs.~\cite{Venumadhav:2015pla,Ghiglieri:2015jua}. 
The tiny mixing angles $\theta_\alpha$ justify to distinguish two vastly different sets of time scales, which makes the problem calculable. At the time when the production of sterile neutrino DM peaks, all SM degrees of freedom are kept in kinetic equilibrium by their fast gauge interactions. In contrast to that, processes that change the number of sterile neutrinos and/or the lepton charges in the plasma are slow (compared to the Hubble rate $H$) due to a suppression $\sim \theta^2$. 
The goal is then to formulate and solve a set of equations that describes these slow degrees of freedom. 
The most common one is to derive so-called \emph{density matrix equations} \cite{Sigl:1992fn}.
These are typically of the form 
\begin{equation}\label{DensityMatrixEquation}
\mathcal{H}X
\frac{d}{dX}
\uprho = -i[ H_{\rm eff},\uprho] - \frac{1}{2}\{\Gamma_{\rm eff}, \uprho-\uprho^{\rm eq}\},
\end{equation} 
where the density matrix 
\begin{eqnarray}
\uprho=\left(
\begin{tabular}{c c}
$\uprho_{\nu\nu}$ & $\uprho_{\nu N}$\\
$\uprho_{N \nu}$ & $\uprho_{N N}$
\end{tabular}
\right)
\end{eqnarray}
is composed of the expectation values
\begin{equation}\label{rhoNdef}
(\uprho_{NN})_{IJ}^{h h'}\propto
\frac{\langle a_I^h(p)^\dagger a_J^{h'}(p)\rangle}{Vs} \ , \ (\uprho_{\nu\nu})_{ij}^{h h'}\propto
\frac{\langle b_i^h(p)^\dagger b_j^{h'}(p)\rangle}{Vs} \ , \ 
(\uprho_{\nu N})_{iJ}^{h h'}\propto\frac{\langle b_i^h(p)^\dagger a_J^{h'}(p)\rangle}{Vs}
\end{equation} 
of the the ladder operators that appear in the plane wave expansion of the neutrino fields
\begin{eqnarray}
N_I=\sum_h\int\frac{d^3\textbf{p}}{(2\pi)^3}\frac{1}{2\sqrt{\textbf{p}^2+M_I^2}}\left(u_{I,\textbf{p}}^h e^{-i\textbf{px}}a_{I,h}(\textbf{p,t})
+ v_{I,\textbf{p}}^h e^{i\textbf{px}}a_{I,h}^\dagger(\textbf{p,t})\right),\nonumber\\
\upnu_i=\sum_h\int\frac{d^3\textbf{p}}{(2\pi)^3}\frac{1}{2\sqrt{\textbf{p}^2+m_i^2}}\left(u_{i,\textbf{p}}^h e^{-i\textbf{px}}b_{i,h}(\textbf{p,t})
+ v_{i,\textbf{p}}^h e^{i\textbf{px}}b_{i,h}^\dagger(\textbf{p,t})\right)\nonumber.\end{eqnarray}
Here $V$ is an arbitrarily chosen unit volume and $s$ is the entropy density of the universe.
The expectation values $\langle\ldots\rangle={\rm Tr}(\hat{\varrho}\ldots)/{\rm Tr}(\hat{\varrho})$ are defined as usual in quantum statistics, where $\hat{\varrho}$ is the quantum statistical density operator.
$\uprho^{\rm eq}$ is the equilibrium density matrix and $X=M/T$ a dimensionless time variable, where $M$ is an arbitrarily chosen mass scale. It is convenient to identify $M$ with the DM particle's mass.
For $I=J$ or $i=j$ the entries of $\uprho$ simply give the occupation number of particles of flavour $I$, helicity $h$ and momentum $p$.
The off-diagonal elements characterise correlations between different flavours.
The function
\begin{equation}\label{Hdef}
\mathcal{H}\equiv  
\frac{1}{X}\left(\frac{d}{d X} \frac{1}{2H}\right)^{-1}=
\frac{1}{X}\left(\frac{d}{d X}\frac{1}{2}\sqrt{\frac{45}{4\pi^3 g_*}}\frac{m_P}{M^2} X^2\right)^{-1}
\end{equation}
corresponds to the Hubble parameter if the number of degrees of freedom $g_*$ is constant during the evolution. 
The effective Hamiltonian $H_{\rm eff}$ describes coherent flavour oscillations.
In vacuum this term is simply given by the neutrino mass matrix, but at finite temperature and density, it receives contributions from $V_T$ and $V_D$. The dissipation term with $\Gamma_{\rm eff}$ comes from decays, inverse decays and scatterings, i.e.\, processes that change the number of particles of a given species in a given mode. It involves decoherence and leads to scattering-induced sterile neutrino production. 
Due to the separation of different time scales, it is usually sufficient to track $\uprho_{NN}$ and the momentum integrated asymmetries 
\begin{eqnarray}
 L_\alpha&=&\frac{s}{\rho_\gamma}\int\frac{d^3\textbf{p}}{(2\pi)^3}
 \left[(\uprho_{\nu\nu})_{\alpha\alpha}^{++} - (\uprho_{\nu\nu})_{\alpha\alpha}^{--}\right].
\end{eqnarray}
  A consistent set of density matrix equations for sterile neutrinos and lepton asymmetries has first been presented in Ref.~\cite{Asaka:2005pn}.
Similar equations have later been derived from first principles by a number of authors \cite{Cirigliano:2009yt,Beneke:2010wd,Garbrecht:2011aw,Fidler:2011yq,Canetti:2012kh,Dev:2014wsa,Dev:2014laa,Drewes:2016gmt,Antusch:2017pkq,Ghiglieri:2017gjz}. A brief summary of the underlying physical assumptions is e.g.\ given in Sec.~5 of Ref.~\cite{Adhikari:2016bei}.
An alternative approach that is based on a perturbative solution of the quantum kinetic equation for $\hat{\varrho}$ itself (known as \emph{von Neumann equation}) has been presented in \cite{Asaka:2006nq,Laine:2008pg} (cf. also \cite{Gagnon:2010kt}) and was further developed in ref.~\cite{Ghiglieri:2015jua}.

\subsubsection{Uncertainties and Open Questions}

In recent years, there has been considerable progress in the computation of sterile neutrino spectra. State of the art calculations can be found in Refs.~\cite{Ghiglieri:2015jua,Venumadhav:2015pla}. 
There are, however, a number of unsolved problems.
The biggest uncertainties probably come from hadronic contributions in the transition region between the quark and hadron phases of the primordial plasma at $T \gtrsim 100$~MeV. 
During this phase, $g_*$ roughly changes by a factor three, which affects the equation of state of the plasma and thereby the expansion of the universe in (\ref{Hdef}). 
Moreover, neutrinos also scatter with quarks/hadrons, which affects both, $V_D$ and $\Gamma_\nu$. 
The calculations in Refs.~\cite{Ghiglieri:2015jua,Venumadhav:2015pla} include the effects on the equation of state as modelled in Ref.~\cite{Laine:2006cp}.
Ref.~\cite{Venumadhav:2015pla} uses lattice results for the quark number susceptibilities to estimate the effect on $V_D$.
It is hard to estimate the effect of hadronic thermal scatterings on $\Gamma_\nu$. In principle it can be related to well-defined mesonic spectral functions~\cite{Asaka:2006rw}, but it is practically very hard to determine these from lattice calculations. 
Refs.~\cite{Venumadhav:2015pla,Lello:2015uma} have used a chiral effective theory approach,
while Ref.~\cite{Ghiglieri:2015jua} simply used a smooth interpolation between the confined and deconfined phases.
In addition to the hadronic contributions, there are also a number of questions involving the weak interaction. 
For instance, resummations are necessary to obtain gauge invariant results for the quantities $V_T, V_D, \Gamma_\nu$.
In the context of leptogenesis, which occurs at temperatures above the electroweak scale, correct leading-order results require a resummation to account for the so-called Landau-Pomeranchuk-Migdal (LPM) effect~\cite{Anisimov:2010gy,Garbrecht:2013urw}. These calculations have been extended to temperatures below the electroweak scale in Ref.~\cite{Ghiglieri:2016xye}, but have not yet been included in the computation of sterile neutrino DM spectra. Finally, the treatment in terms of the on-shell occupation numbers (\ref{rhoNdef}) does not take into account the full quasiparticle spectrum at high temperature, which includes a collective ``hole'' excitation \cite{Klimov:1981ka,Klimov:1982bv,Weldon:1989ys,Quimbay:1995jn}. 
All these corrections cannot be quantitatively understood in terms of the simple picture of active-sterile oscillations  and require a fully field theoretical approach. However, the basic physical idea remains the same, and the considerations in sections \ref{Subsec:Nonres} and \ref{Subsec:Res} are sufficient for a qualitative understanding.

In addition to these technical issues, the question arises how the large lepton asymmetries that are required to fulfil the condition (\ref{cond}) can be generated. Since electroweak sphalerons keep the baryon and lepton numbers in equilibrium \cite{Kuzmin:1985mm} at $T>130$ GeV \cite{D'Onofrio:2014kta} and $\eta_B$ is known to be $\sim 10^{-10}$ \cite{Ade:2015xua}, this lepton asymmetry should be generated at lower temperatures. A late time lepton asymmetry is e.g.\ necessarily produced if the BAU is generated in heavy neutrino oscillations \cite{Akhmedov:1998qx,Asaka:2005pn}, as in the $\nu$MSM \cite{Asaka:2005an}. Pushing the asymmetry that is produced at $T\ll v$  to the large values required for resonant DM production requires some tuning \cite{Canetti:2012vf,Canetti:2012kh}.
It has been suggested that may be avoided if some of the asymmetry that is produced before the heavier neutrinos come into equilibrium is protected from the washout due to an almost conserved quantum number \cite{Eijima:2017anv} or because it can be stored in magnetic fields \cite{Boyarsky:2011uy}.
The study of these (and possibly other) possibilities is a topic of ongoing research.

\subsection{Thermal Production via New Gauge Interactions (``freeze out'')}\label{Sec:GaugeProduction}
Sterile neutrinos  are by definition not charged under any SM gauge group. In the minimal seesaw model (\ref{Lseesaw}), the $\nu_R$ only couple to the SM via their Yukawa interactions $F$, and the low energy mass eigenstates $N$ can only be produced via their $\theta$-suppressed weak interaction (\ref{WeakWW}), as discussed in the previous section. 
There is, however, no reason why the $\nu_R$ should not be charged under new gauge interactions if the model (\ref{Lseesaw}) is embedded in a more general framework of particle physics. For instance, the existence of $\nu_R$ with gauge interactions is a generic prediction of \emph{Grand Unified Theories} based on the gauge group $SO(10)$.
The probably most studied scenario with $\nu_R$ gauge interactions are left-right symmetric models \cite{Pati:1974yy,Mohapatra:1974hk,Senjanovic:1975rk,Wyler:1982dd}, which can effectively describe $SO(10)$ theories or other models at intermediate energies. We shall use this scenario as an example in the following.
Moreover, we for simplicity assume that the gauge coupling constant of the right handed $SU(2)$ has the same value as the left handed one.
The conclusions apply to other setups in a very similar manner,\footnote{A more general discussion can e.g.~be found in Ref.~\cite{Patwardhan:2015kga}.}
though details related to the breaking of gauge symmetry \cite{Kusenko:2010ik} or a non-minimal SM-singlet fermion sector 
\cite{Escudero:2016tzx} can affect the final DM abundance.

Data from the LHC implies that the masses $m_{W_R}$ of the new gauge bosons should be significantly larger than a TeV \cite{Aad:2015xaa,Khachatryan:2016jqo,Sirunyan:2018pom}.\footnote{Note that the bounds from X-ray searches discussed in Sec.~\ref{sec:decaying-dark-matter} also impose constraints on the additional gauge interactions, which can generate low energy effective operators that mediate DM decays. This may lead to stronger bounds on $\theta$ from given observational data.} 
In comparison to this, the keV masses of the sterile neutrinos can be neglected, and they are relativistic at the time of their freezeout. The $N$-abundance is simply given by the equilibrium abundance of massless fermions at the time of freezeout. At temperatures $T\ll m_{W_R}$ the sterile neutrinos are kept in thermal equilibrium by a Fermi-theory like interaction, but with the W mass $m_W$ replaced by the RH W mass $m_{W_R}$. The calculation therefore goes along the same lines as for light neutrinos.
The cross section can be estimated as $\sigma\sim G_F^2 T^2 (m_W/m_{W_R})^4$ and the interaction rate as $\Gamma_N\sim G_F^2 T^5 (m_W/m_{W_R})^4$. The temperature $T_f$ of freezeout can be estimated from the requirement $\Gamma_N=H$, where $H=T^2/m_{\rm Pl}\times \sqrt{(4\pi^3 g_*)/45}$ is the Hubble rate,
\begin{equation}
T_f\sim  g_{*}^{1/6}(T_f)\left(\frac{m_{W_R}}{m_W}\right)^{4/3} {\rm MeV}.
\end{equation}
The relic abundance of a fermionic hot relic with two internal degrees of freedom right after freezeout is given by
\begin{equation}
\frac{n_N}{s}(T_f) = \frac{135\zeta(3)}{4\pi^4}\frac{1}{g_{*s}(T_f)}.
\end{equation} 
The present day abundance can then be estimated by
\begin{equation}
\frac{n_N}{s}(t_0) = \frac{135\zeta(3)}{4\pi^4}\frac{1}{g_{*s}(T_f)}\frac{g_{*s}(T_0)}{g_{*s}(T_f)}\frac{1}{S},\label{noversnow}
\end{equation} 
where the factor $1/S$ takes account of a possible injection of entropy from some nonequilibrium process after the freezeout, e.g. due to the decay of a heavy particle \cite{Scherrer:1984fd}.
This leads to a contribution $\Omega_N$ to the energy density of the universe that compares to $\Omega_{\rm DM}$ as
\begin{eqnarray}\label{dilutionabundance}
\frac{\Omega_N}{\Omega_{\rm DM}}=\frac{1}{S}\left(\frac{10.75}{g_{*s}(t_f)}\right)\left(\frac{M}{\rm keV}\right)\times 100.
\end{eqnarray}
For keV masses, this number is much larger than one, i.e., in strong contradiction with observation, if $S\sim 1$ and $g_{*s}(t_f)\sim 10^2$. 
$\Omega_N$ can be made consistent with the observed $\Omega_{\rm DM}$ in different ways. 
One possibility is that the number $g_{*s}$ of degrees of freedom could change by three orders of magnitude. The existence of hundreds of new particles may seem far fetched, but is in principle not impossible. 
Another possibility is that the $N$ have some new interactions that remain efficient after they decoupled from the SM and allow them to reduce their number densities by self-annihilation \cite{Herms:2018ajr}.
Finally, it could be that the $N$ particles are diluted by an entropy injection into the SM plasma after their freezeout. This has the nice side effect that the momentum distribution of the DM particles is cooled by a factor $S^{-1/3}$, which makes it easier to fulfil constraints from structure formation discussed in Sec.~\ref{sec:structure-formation}.
The entropy could, for example, be produced in the decay of heavier sterile neutrinos \cite{Asaka:2006ek,Bezrukov:2009th,Bezrukov:2012as,Nemevsek:2012cd} or other singlets \cite{King:2012wg}. This decay may at the same time generate the baryon asymmetry of the universe via thermal leptogenesis~\cite{Bezrukov:2012as}. We use the decay of heavier neutrinos $N_2$ as an example in the following. Of course, the entropy could also be produced by the decay of any other heavy particle.\footnote{Similar scenarios have also been studied in theories other than the seesaw model, including radiative neutrino mass generation \cite{Ma:2012if,Hu:2012az,Baumholzer:2018sfb} and composite neutrinos \cite{Robinson:2012wu}.}
If $N_1$ is the DM candidate and $N_2$ a heavier neutrino, then the $N_1$ density is given by \cite{Nemevsek:2012cd}
\begin{equation}
  \label{eq54:omegaNfull}
  \Omega_{N_1} \simeq
  0.265 \left(\frac{M_1}{1{\rm keV}}\right)
  \left(\frac{1.6 {\rm GeV}}{M_{N_2}}\right)
  \left(\frac{1\,\text{sec}}{\tau_{N_2}}\right)^{1/2}
  \frac{g_*(T_{f,2})}{g_*(T_{f,1})}.
\end{equation}
Here $\tau_{N_2}$ is the $N_2$ lifetime and $T_{f,1}, T_{f,2}$  are the freezeout temperatures of $N_1$ and $N_2$.
There are two main constraints on the allowed values of these parameters:
\begin{itemize}
\item  $\tau_{N_2}\lesssim 1\,\text{sec}$ in order to be consistent with the production of the observed abundances of light elements in big bang nucleosynthesis. With \eqref{eq54:omegaNfull} this gives a lower bound on the $N_2$ mass $M_2$.
\item $N_2$ should be relativistic when it freezes out, $T_{f,2}>M_2$, as otherwise the entropy release is less efficient. 
This implies a lower bound on the scale of the gauge interactions that keep $N_2$ in thermal equilibrium.
\end{itemize}
A careful analysis for a minimal left-right symmetric theory with two heavier neutrinos $N_2$ and $N_3$ has been performed in~\cite{Nemevsek:2012cd}.
The above bounds generally require  $M_{2,3}\gtrsim (M_1/1\keV)\times1.6\GeV$ and $m_{W_R}\gtrsim(M_1/1\GeV)^{3/4}\times10\TeV$. Tuning the flavour structure to separate the freeze-out temperatures of the diluting particles and the DM sterile neutrinos, the constraints can be relaxed for specific parameter choices.

Finally, it should be pointed out that the above considerations crucially rely on the assumption that the sterile neutrinos are in thermal equilibrium initially. This implicitly assumes that the symmetry breaking scale of the new gauge group under which $\nu_R$ is charged
is smaller than the maximal temperature in the radiation dominated cosmological epoch.
This temperature is unknown.
The only observational constraint is that it should be larger than a few MeV to produce the observed amounts of light elements in the intergalactic medium in big bang nucleosynthesis \cite{Barenboim:2017ynv}. 
In inflationary cosmology it associated with the reheating temperature, which can in principle be constrained from CMB observations \cite{Martin:2014nya,Drewes:2015coa}, though this is somewhat model dependent. For a specific inflationary model it has e.g. been studied in ref.~\cite{Drewes:2017fmn}.

\subsection{Non-thermal Production in the Decay of Heavier Particles}\label{Sec:DecayProduction}

The production mechanisms I) and II) discussed in the previous subsections both involve  scatterings of particles in the primordial plasma, which are in good approximation is in thermal equilibrium. 
As such, they can both be classified as ``thermal'' production, though one should keep in mind that the sterile neutrinos themselves do not reach thermal equilibrium if they are solely produced via the mixing $\theta$, and their momentum distribution can be highly non-thermal.
A physically rather different picture emerges when the heavy neutrinos are produced in the decay of other massive particles. If this decay occurs far from thermal equilibrium (e.g., long after the decaying particle has frozen out or before it even reaches thermal equilibrium), then this is a truly non-thermal production mechanism. 

Heavy neutrinos can be produced in the decay of practically any heavier particle that the interact with.
In the minimal seesaw model (\ref{Lseesaw}) they are e.g. produced in the decay of pions \cite{Asaka:2006nq,Lello:2014yha,Lello:2015uma}, the  Higgs boson \cite{Bezrukov:2008ut} and W bosons \cite{Lello:2016rvl}. All these processes are subdominant in comparison to the production in decoherent scatterings described in Sec.~\ref{Sec:ThermalProduction}. 
However, in  extensions of the SM the production from the decay of new particles may dominate.
There are countless possibilities to implement this idea in specific models. The simplest scenarios involve a scalar singlet $\phi$ \cite{Shaposhnikov:2006xi,Petraki:2007gq,Boyanovsky:2008nc} that can couple to the heavy neutrinos $N$ via a Yukawa interaction $y\phi\bar{N}N$ (c.f. e.g. also \cite{Roland:2014vba,Matsui:2015maa,Roland:2016gli,Shakya:2018qzg}). If $\phi$ has a non-vanishing vacuum expectation value $\langle\phi\rangle$, then this term could in principle also generate the Majorana mass for $\phi$. Other possibilities invoke the decay of a charged scalar \cite{Frigerio:2014ifa,Drewes:2015eoa},\footnote{In this case the sterile neutrinos may decay into light Majorons \cite{Chikashige:1980ui,Schechter:1981cv}, which in principle could then form the DM \cite{Frigerio:2011in,Queiroz:2014yna}.
} 
an additional Higgs doublet \cite{Adulpravitchai:2015mna,Drewes:2015eoa},  its own SUSY partner \cite{Shakya:2016oxf}, 
vector bosons \cite{Shuve:2014doa,Caputo:2018zky} or fermions~\cite{Abada:2014zra}.
We here restrict ourselves to the simplest case of a scalar singlet $\phi$. 

For heavy neutrinos with keV masses the production occurs when they are highly relativistic, so that we can neglect their mass and identify the occupation number $f_{N\p}$ for the heavy neutrino momentum mode $\p$ with the phase space distribution function $f_N(\p)$ at energy $|\p|$.
The production in the decay of $\phi$-particles\footnote{
Heavy neutrinos can either be produced in the decay of non-coherent $\phi$-particles that form a gas or by the decay of the coherent condensate $\langle\phi\rangle$. Near the minimum of the effective potential, the decay rates for both processes are identical, see e.g. Ref.~\cite{Cheung:2015iqa} and references therein.
} in a thermal plasma is governed by the quantum kinetic equation
\begin{eqnarray}
\partial_t f_N(\p)
&=&2\frac{\tilde{\Gamma}_0}{m_\phi}\frac{M_\phi^2}{\p^2}
\int_{|M_\phi^2/(4\pp)-\pp|}^{\infty} d\qq \frac{\qq}{\Omega_{\phi\q}}
\left[1 -f_N(\pp) - f_N(\Omega_{\phi\q}-\pp)\right] 
\left[f_{\phi\q}-\tilde{f}_{\phi\q}\right]\nonumber\\
&=&2\frac{\tilde{\Gamma}_0}{m_\phi}\frac{M_\phi^2}{\p^2}
\int_{M_\phi^2/(4\pp)+\pp}^{\infty} d\Omega_{\phi\q}
\left[1 -f_N(\pp) - f_N(\Omega_{\phi\q}-\pp)\right] 
\left[f_{\phi\q}-\tilde{f}_{\phi\q} \right]\nonumber,\label{sigletmomentutmdependentproduction}
\end{eqnarray}
which is derived in detail in Ref.~\cite{Drewes:2015eoa}.
Here $\tilde{\Gamma}_0=y^2 m_\phi/(16 \pi)$ is the vacuum decay rate for $\phi$-particles at rest into $N$-particles, $M_\phi$ is the thermal $\phi$-mass in the plasma and $\Omega_{\phi\q}\simeq(\q^2 + M_\phi^2)^{1/2}$ the dispersion relation. The value of $M_\phi$ depends on the interactions that $\phi$ has with other particles in the plasma. $f_{\phi\q}$ is the occupation number for $\phi$ particles with momentum $\q$ and $\tilde{f}_{\phi\q}=\tilde{\Gamma}_{\phi\q}^</(\tilde{\Gamma}_{\phi\q}^>-\tilde{\Gamma}_{\phi\q}^<)$
\cite{Drewes:2012qw}. We use $\Gamma_{\phi\q}^\gtrless$ to refer to the total gain and loss rates for $\phi$-particles  that drive $\phi$ towards thermal equilibrium (including interactions with particles other than $N$)\footnote{ In equilibrium they are given by $\Gamma_{\phi\q}^<=f_B(\Omega_{\phi\q})\Gamma_{\phi\q}$ and $\Gamma_{\phi\q}^>=[1 + f_B(\Omega_{\phi\q})]\Gamma_{\phi\q}$, where $f_B$ is the Bose-Einstein distribution and $\Gamma_{\phi\q}=\Gamma_{\phi\q}^>-\Gamma_{\phi\q}^<$ is the full thermal damping rate for the $\phi$-mode $\q$ (i.e. the quasiparticle width), leading to $\bar f_{\phi\q}=f_B(\Omega_{\phi\q})$. } and $\tilde{\Gamma}_{\phi\q}^\gtrless$ for the contribution to  $\Gamma_{\phi\q}^\gtrless$ from interactions with $N$.
In principle there are additional contributions to the $N$-production from scatterings \cite{Drewes:2015eoa,Konig:2016dzg,Heeck:2017xbu}, but they are negligible in the scenario discussed here. 
Most of the DM is produced at temperatures $T\sim m_\phi$, at which we can safely approximate $M_\phi\simeq m_\phi$. At later times the production is exponentially suppressed due to the Boltzmann factor.
Both, scatterings and corrections to the quasiparticle dispersion relations, may be important in scenarios where the decaying particle is charged and $N$ is always produced along with at least one charged particle \cite{Drewes:2015eoa}.

In the simplest scenarios $\phi$ is in thermal equilibrium when the DM production commences.
For illustrative purposes, we will focus on this case in the following. 
Scenarios in which $\phi$ starts from a nonequilibrium initial state have e.g. been studied in Refs.~\cite{Adulpravitchai:2014xna,Merle:2015oja,Konig:2016dzg}, and it has been shown that this can lead to colder DM spectra \cite{Merle:2014xpa}, but the details are rather model dependent because they depend on the way how $\phi$ couples to the primordial plasma.
With the current approximations, and assuming that the inverse decay $N N \rightarrow \phi$ can be neglected,\footnote{This is justified for small Yukawa couplings $y<10^{-7}$ because the $N$ occupation numbers must remain well below their equilibrium values to explain the observed amount of DM.}
the expression (\ref{sigletmomentutmdependentproduction}) reduces to the well-known form used in the seminal work \cite{Shaposhnikov:2006xi} (see also \cite{Kusenko:2006rh,Petraki:2007gq}):
\begin{equation}
  \frac{\partial f_N(\p)}{\partial t} - H \p \frac{\partial f_{N\p}}{\partial \p}=
  \frac{2 m_\phi\tilde{\Gamma}_0}{\p^2}\int_{\p+m_\phi^2/(4\p)}^\infty f_{\phi\q}d\Omega_{\phi\q},
  \label{eq:kin}.
\end{equation}
If the effective number $g_*$ of degrees of freedom is constant, Eq.~\eqref{eq:kin} can be solved. At late times the solution reads
\begin{equation}
f_N(x) = \frac{16 \Gamma M_0}{3 m_\phi^2} x^2
\int_1^\infty\frac{(z-1)^{3/2} dz}{e^{xz}-1}~,
\label{eq:DistributionSNIE}
\end{equation}
where $x=p/T$ and $M_0 \approx M_{\rm Pl}/(1.66\sqrt{g_*})$.
This leads to a number density of
\begin{equation}
 n_N =\int \frac{d^3\p}{(2\pi)^3} f_N(\p) = \frac{3\Gamma M_0 \zeta(5)}{2\pi m_\phi^2} T^3
 \label{eq:rho}.
\end{equation}
The average momentum of the sterile neutrinos immediately after their production is 
\begin{equation}
\langle \p \rangle = \pi^6/[378\zeta(5)]T = 2.45 T, 
\end{equation}
which is about $20$\% smaller than that for an equilibrium thermal distribution, $p_T=3.15 T$.
At present time, the sterile neutrino contribution to the total energy density of the universe  is~\cite{Shaposhnikov:2006xi}:
\begin{equation}
\Omega_{N} \sim 
\frac{y^2}{S}\frac{g_{*0}}{g_*(T_{\rm prod})}\; \frac{M_{\text{Pl}}}{m_\phi} \;   \frac{M}  { {\text{keV}}}
\simeq\frac{y^2}{S}\frac{3.9}{g_*(T_{\rm prod})}\; \frac{M_{\text{Pl}}}{m_\phi} \;   \frac{M}  { {\text{keV}}},
\label{eq:decays2}
\end{equation}
where $g_{*0}\simeq 3.9$ is the late time value of $g_*$.
The dilution factor $S$ in Eq.~\eqref{eq:decays2} is the same as in \ref{noversnow}, see e.g. Ref.~\cite{Asaka:2006ek}.  If $g_*$ changes significantly at the time when the DM production peaks ($T\sim m_\phi/3$), a more refined analysis is required \cite{Shaposhnikov:2006xi}.
Today, the averaged momentum of sterile neutrinos produced at $T\sim m_\phi/3$ is given by~\cite{Shaposhnikov:2006xi}:
\begin{equation}
\frac{\langle \p \rangle}{T_\gamma}  =  \frac{\pi^6\, }{378\,\zeta (5)} \left(S\frac{g_*(T_{\rm prod})}{3.9}\right)^{-1/3}\approx 2.45\,\left(S\frac{g_*(T_{\rm prod})}{3.9}\right)^{-1/3}.
\label{eq:decays3}
\end{equation}
Comparing this to the average momentum  $\langle p\rangle/T_\gamma=3.15 (4/11)^{1/3}$ in (non-resonant) thermal production via the mixing $\theta$,\footnote{It was pointed out in Ref.~\cite{Abazajian:2005gj} that the above approximation does not hold exactly and $\langle p_\mathrm{NRP}\rangle$ is in fact slightly smaller.}  
the mass bounds from structure formation are weaker for DM produced in decays case~\cite{Bezrukov:2014nza}
\begin{equation}
  M_{\rm decay} = \frac{\langle p\rangle}{\langle p_\mathrm{mixing}\rangle}M_\mathrm{mixing} \simeq 
  \frac{2.45}{3.15}\left[\frac{1}{S}\frac{11}{4}\frac{g_{*0}}{g_*(T_\mathrm{prod})}\right]^{1/3}M_\mathrm{mixing}
  \simeq   1.7\left[S g_*(T_\mathrm{prod})\right]^{-1/3}M_\mathrm{mixing}.
\end{equation}
Here $M_{\rm decay}$ and $M_\mathrm{mixing}$ refer to the constraint that a limit on the free streaming of DM during structure formation imposes on the physical mass $M$ of the sterile neutrinos if they are produced in decays or via mixing (non-resonant), respectively.

\section{Laboratory Searches for keV-scale Sterile Neutrinos}
\label{exp}

\subsection{Overview}
\label{exp-ov}
In analogy to the well-known techniques to search for Cold Dark Matter in the form of Weakly Interacting Massive Particles (WIMPs), two laboratory-based approaches can be distinguished. \footnote{In this section we entirely focus on sterile neutrinos that were produced via their $\theta$-suppressed weak interaction (\ref{NWW}) in the early universe, cf. mechanism I) in sec.~\ref{ProductionOverview}. 
This imposes a lower bound on $|\theta|^2$ which, together with current constraints from X-ray searches, leads to an upper bound of a few tens of keV on the heavy neutrino mass, cf. fig.~\ref{fig:SummaryPlot}. 
If the DM sterile neutrinos are produced via new gauge interactions or in the decay of heavy particles (mechanisms II) and III) in the classification in sec.~\ref{ProductionOverview}, 
then $\theta$ can be arbitrarily small and their masses can be well above the  electroweak scale. In this case the sterile neutrinos themselves and/or the particles involved in their production (new gauge bosons, heavy scalars,$\ldots$) can be searched for in collider experiments, cf. e.g. \cite{Deppisch:2015qwa,Cai:2017mow}.
}
1) A direct detection of the DM particle present in our galaxy by using large-scale detectors, and
2) the production of sterile neutrinos in a radioactive decay and their detection via kinematic considerations. 

In addition to this, one may also look for indirect signatures of sterile neutrinos. It is, for example, well known they can impact the rate of neutrinoless double $\beta$ decay \cite{Bezrukov:2005mx,Blennow:2010th,LopezPavon:2012zg,Drewes:2016lqo,Hernandez:2016kel,Asaka:2016zib}, and the existence of a keV neutrino in principle allows to make predictions for this rate \cite{Abada:2018qok}. However, since heavy neutrinos with $M_I < 100$ MeV themselves cannot give a significant contribution to the decay \cite{Blennow:2010th}, this impact is always indirect and therefore strongly model dependent, i.e., it relies on assumptions about other possible sources of LNV.  The impact of DM sterile neutrinos on most other observables that are commonly used to indirectly test low scale seesaw models (see e.g. \cite{Drewes:2015iva,deGouvea:2015euy} and references therein) is insignificant due to the strong constraints on the mixing angles $\theta$ from X-ray constraints and the requirement that their lifetime is longer than the age of the universe.

Currently, astrophysical bounds are some orders of magnitudes stronger than any laboratory limit. However, these limits rely on underlying cosmological and astrophysical assumptions. A detection through production, in particular, would be completely independent of cosmological and astrophysical input and has the potential to test more exotic scenarios of Physics beyond the SM including sterile neutrinos~\cite{Rodejohann:2014eka, Ludl:2016ane, Barry:2014ika, Bezrukov:2017ike}. Here, new proposals for laboratory-based experiments are presented, that could significantly improve the previous limits and may reach a region of astrophysical interest. 

In the following we assume the existence of a single heavy neutrino $N$ with mass $\M$ and mixing angles $\theta_\alpha$ with the SM flavours $\alpha$ that comprises the entire observed DM density $\Omega_{\rm DM}$. In practice, $N$ may of course be part of a seesaw model with $n>1$ (i.e. $N=N_I$, $M=M_I$ and $\theta_\alpha=\Theta_{\alpha I}$), and the DM may be made of more than one component. 

\subsection{Direct Detection}
\label{exp-int}
keV sterile neutrinos ($\N $) cluster in the gravitational potential wells of galactic halos. If neutrinos of mass ${\M }$ account for the entire local DM, $\rho_{\rm DM} \simeq 0.3\pm0.1~{\rm GeV \cdot cm^{-3}}$ \cite{Bovy:2012tw}, then their local number density is given by $\rm n_{\N } \simeq  (300  \pm100)\cdot \,10^3  / {\M } (\rm{keV}) \, {\rm cm}^{-3}$. This is a factor 1000 larger than the Cosmic (active) Neutrino Background (\cnb) and about 10$^6$ times larger than for WIMPs and therefore motivates the direct search for DM neutrino through their capture or scattering on specific target nuclei.

In what follows we consider an isothermal Milky Way halo and assume the keV relic neutrino background velocity to follow a shifted Maxwellian distribution with an average value of $\rm v_{N}$=220~km/s~\cite{Freese:1987wu}.

\subsubsection{Sterile Neutrino Capture }
\label{exp-int-capt}

The generic electron (anti)neutrino induced $\beta$-capture $\nnbe  + \rm R \rightarrow \rm R' + \rm e^\pm$ on a radioactive nucleus $\rm R$ was addressed in~\cite{Cocco:2007za,Lazauskas:2007da} (cf. also \cite{Li:2010vy}). Due to the positive energy balance $\rm Q_\beta=\rm M_a(\rm R) - \rm M_a(\rm R')>0$~\footnote{$\rm M_a$ denotes the masses of neutral atoms, $\rm M_n$ denotes the nuclear masses.} this exothermic reaction is always allowed independently of the value of the incoming neutrino energy $\rm E_\nu$. This process is therefore appealing for the detection of the Cosmic (light) Neutrino Background (\cnb\ ), which are expected to have today a tiny kinetic energy $\rm T_{C \nu B}\sim 0.5$~meV~\cite{Cocco:2007za,Lazauskas:2007da,Faessler:2014bqa}. 
Considering the case of tritium, a relevant \cnb\,experiment necessitate $\sim100$~g of radioactive target material~\cite{Blennow:2008fh,Li:2010sn,Kaboth:2010kf,Faessler:2013jla,Betts:2013uya,Long:2014zva}. This is a huge technical challenge~\cite{Betts:2013uya} when compared to the already large amount of tritium, $\sim50$~$\mu$g, to be used in the 
Karlsruhe Tritium Neutrino
(KATRIN) experiment~\cite{Angrik:2005ep}. 
The characteristic signal associated with this process as the emerging electron would create a mono-energetic peak at $\rm T_e = \rm E_0 + \rm m_\nu$, where $\rm T_e$ is the electron kinetic energy, $\rm E_0$ is the endpoint energy of the $\beta$-decay, and $\rm m_{\nu}$ is the effective electron neutrino mass. In the case of the \cnb, a sub-eV energy resolution is required to distinguish this signal from the tail distribution of the $\beta$-decay spectrum. 

Likewise the detection of the keV neutrino DM through  $\nu$-capture on $\beta$-decaying nuclei was also considered~\cite{Shaposhnikov:2007cc,Li:2010vy,Long:2014zva}. keV neutrinos induced mono-energetic electron signal would appear at $\rm T_e = {\rm E}_0 + \M $, thus comfortably distinguishable with standard nuclear physics technology. However the expected capture rate suffers from a potentially strong suppression factor $\sin^2\theta_e$ induced by the mixing between the sterile and active neutrino components. 

\begin{figure}[t]
\begin{center}
\begin{tabular}{cc}
\includegraphics*[width=0.46\textwidth]{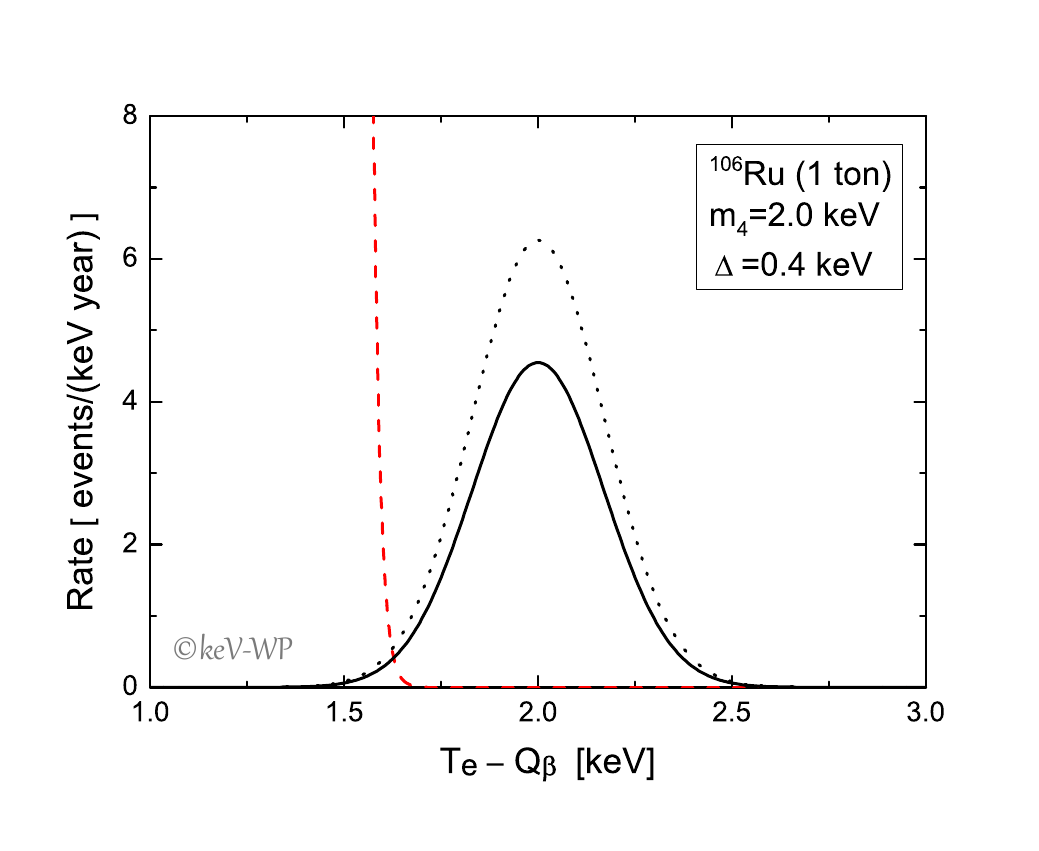}
&
\includegraphics*[width=0.46\textwidth]{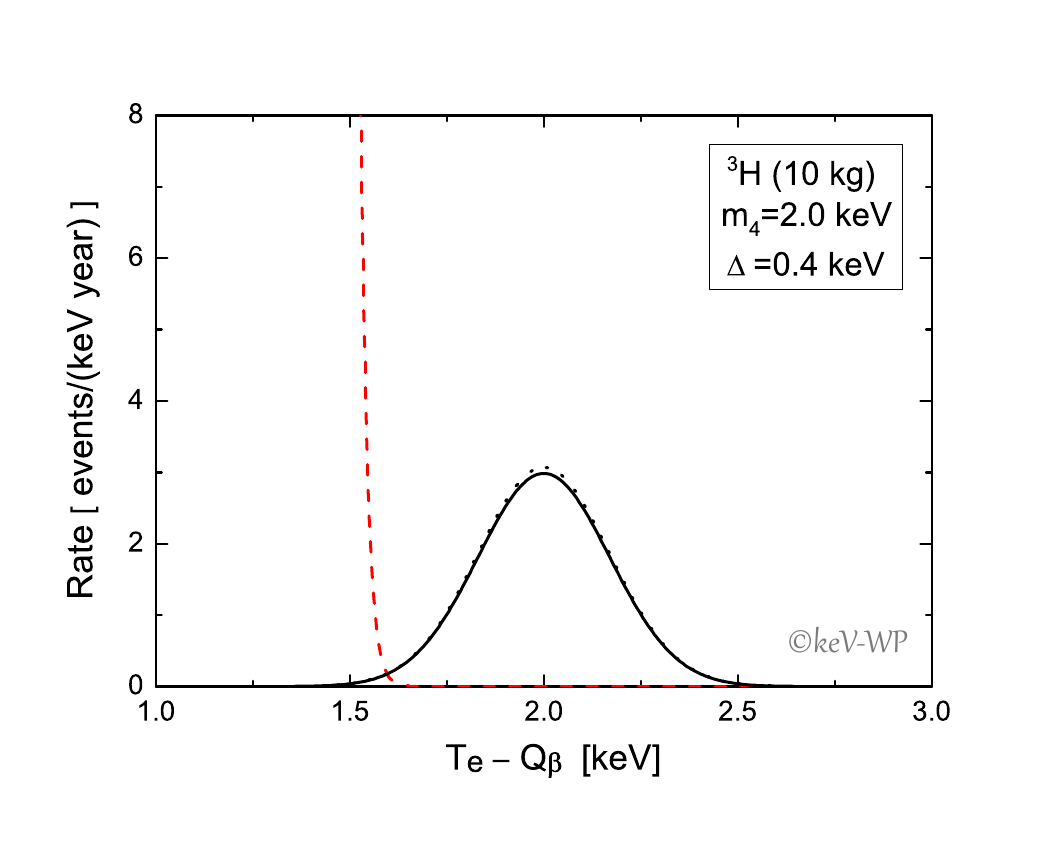}
\end{tabular}
\end{center}
\vspace{-0.5cm}
\caption{The keV sterile neutrino capture rate as a function of the
kinetic energy of electrons with $^{106}{\rm Ru}$ (left panel) and $^3{\rm H}$
(right panel) as the capture targets \cite{Li:2010vy}. The solid (or
dotted) curves denote the signals with (or without) the half-life
effect.
Here $m_4$ is to be identified with the DM particle mass $M$.
}
\label{fig:signature}
\end{figure}

A mass of radioactive material, on the scale of 100~g for tritium~\cite{Long:2014zva}, 10~kg for $^{106}$Ru~\cite{Li:2010vy}, and 600~tons for $^{163}$Ho~\cite{Li:2011mw}, would thus be required to probe a mixing angle of the order of $\sin^2\theta_e \sim10^{-6}$.  As an illustration, Figure~\ref{fig:signature} present the keV sterile neutrino DM capture rate as a function of the kinetic energy of electrons with $^{106}{\rm Ru}$ (left panel) and $^3{\rm H}$ (right panel) \cite{Li:2010vy}. To suppress the $\beta$-decay background, the energy resolution $\Delta$ (i.e., full width at half maximum)
should be smaller $\Delta \lesssim 0.5\rm ~{\rm keV}$, for $\M = 2 ~{\rm keV}$.\\

While procuring this amount of radioactive material is extremely challenging such a mass is in contrast conceivable with a non-radioactive target material~\cite{Lasserre:2016eot}. Indeed keV neutrinos could be capture on a stable nucleus $S$ leading to the radioactive daughter nucleus $D$, produced as a positive ion:
\begin{equation}
\nu \, + \, _{Z}\rm S \rightarrow \rm e^- \, + \, _{(Z+1)} \rm D^+ \, .\label{e:uncap} 
\end{equation} 
This reaction can be stimulated by the mass of a hypothetical sterile neutrino if
\begin{equation}
\M  \geq \Delta(\rm S) - \Delta(\rm D) - \rm E_b(\rm D^+) + \rm E_b(\rm D) \simeq \rm Q_{\beta}^{\rm tab} \, , \label{e:uncap2} 
\end{equation} 
where $\rm E_b$ define the binding energies of the orbital electrons. The difference in the total binding energy of the neutral atom and the single positive ion is  $\sim$100~eV and can be neglected. $\Delta$ are the usual tabulated mass excess and $\rm Q_{\beta}^{\rm tab}$ denotes the tabulated Q-value assuming $\rm m_{\nu}=0$~\cite{Wang:2012aa}. Thus the keV sterile neutrino capture on stable nuclei can be stimulated by the neutrino mass energy for $\M<\rm Q^{tab}_{\beta}\leq 0$. 

The highest negative $\rm Q_{\beta}^{tab}$ value is given for  $^{163}$Dy~\cite{Eliseev:2015pda} that mainly captures neutrinos from its nuclear ground state to the ground state of~$^{163}$Ho 
\begin{equation}
\begin{split}
 ^{163}{\rm Dy} ({\rm gs}, {\rm I}^\pi=5/2^{-}) + \N  ({\M }>2.83 {\rm~keV}) \\
 \rightarrow ^{163}{\rm Ho} ({\rm gs}, {\rm I'}^{\pi '}=7/2^{-}) + e^- .\label{e:DyHo} 
\end{split} 
\end{equation} 
Subsequently the $^{163}{\rm Ho}$ nuclei decays through electron capture (EC) with a half-life of 4570~years. The production rate of $^{163}$Ho is given by
\begin{equation}
\rm R_{^{163}{\rm Ho}}= N_{^{163}{\rm Dy}} \cdot <\sigma_{c} \rm v_{N}> \cdot \rm n_N \times \sin^2\theta_e \, ,
\label{e:DyHoCaptRate} 
\end{equation} 
where $N_{^{163}{\rm Dy}}$ is the number of target~$^{163}$Dy atoms and $<\sigma_{c} v_{\N }>$ denotes the averaged product of the capture cross section and the \rnb\,velocity. 
In the non-relativistic approximation the cross section can be expressed as
\begin{equation}
\sigma_{c}({\M }) \simeq 4.8 \cdot 10^{-43} \cdot {\rm E}_{e}({\M })\cdot {\rm p}_{e}({\M })\cdot {\rm F}({\M }) \,{\rm cm}^{2}\, ,
\end{equation}
where the nuclear size and electron screening effects are taken into account in the Fermi function~\cite{shenter:1983}. It is  appealing to consider an integral experiment where the $^{163}$Dy has been exposed over a geologic time, $\rm t$, enhancing the number of captures given by
\begin{equation}
\begin{split}
N_{^{163}{\rm Ho}}(\rm t,\M ,\sin^2\theta_e,{\rm m}_{^{163}{\rm Dy}})= 
\tfrac{ {m}_{^{163}{\rm Dy}} \cdot \mathcal{N}_A}{A_{^{163}{\rm Dy}} \cdot \lambda^{\rm EC}_{^{163}\rm Ho}} \times \\
\sigma(\M )  <\rm v_{N}>  \rm n_{\N }  (1-e^{-\lambda^{\rm EC}_{^{163}\rm Ho}  \cdot \rm t}) \times \sin^2\theta_e \, ,
\end{split}
\end{equation}
where $A_{^{163}{\rm Dy}}$ is the $^{163}$Dy molar mass and ${m}_{^{163}{\rm Dy}}$ is the target mass.
After an exposure of more than 30000~years an equilibrium is reached between the $^{163}$Ho production and subsequent EC decays. For $\M $=5~keV about $7 \cdot 10^{9} \times \sin^2\theta_e$~atoms of $^{163}$Ho are expected in 1~ton of $^{163}$Dy.

Solar neutrinos (S$\nu$B) are also captured on $^{163}$Dy  and thus constitute a background for the DM search. In a 1 ton $^{163}$Dy target exposed for more than 30000~years $\sim 2.7 \cdot 10^5 $  $^{163}$Ho atoms are expected.

\begin{figure}[ht!]
\begin{center}
\includegraphics[scale=0.6]{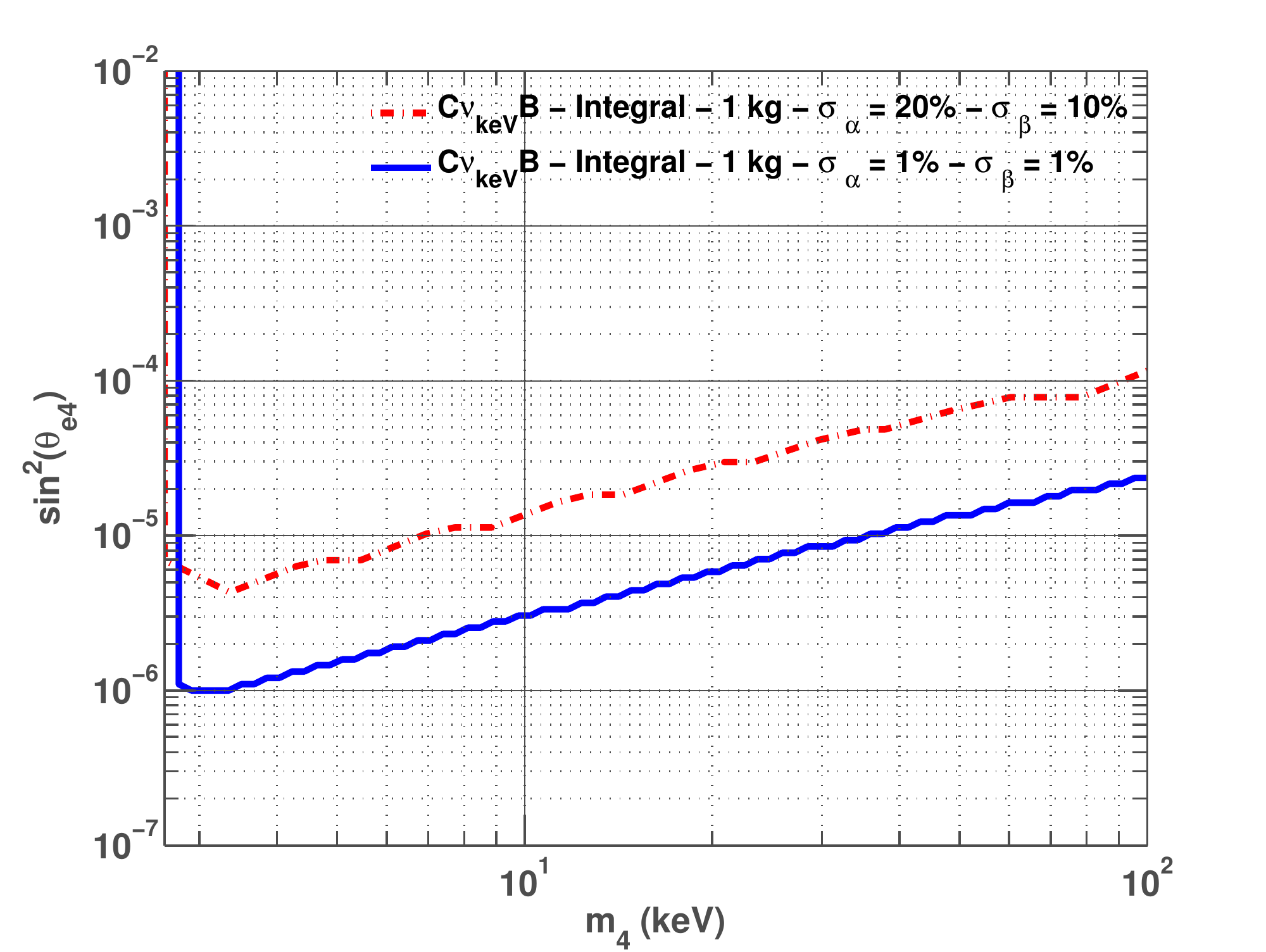}
\caption{\label{f:geosensitivity} 
90\% C.L. sensitivity for \rnb\, integral detection for a target mass of 1~kg assuming the atom counting rate known within 20 \% (1 \% ) and the \snb\,capture rate known within 10\% (1 \%).
Here $m_4$ is to be identified with the DM particle mass $M$ and $\theta_{e4}$ corresponds to $\theta_e$.
}
\end{center}
\end{figure}

A first  detection method consists in counting the number of~$^{163}$Ho atoms in rare-earth ores resulting from the $\nu$-capture on~$^{163}$Dy.  Atoms counting in a magneto-optical trap seems promising~\cite{1367-2630-16-6-063070,Jiang20121}. As for the separation of $^{163}$Ho from other lanthanides, resonance ionisation mass spectrometric (RIMS) techniques with lasers could be considered~\cite{kieck:2015}. 

Results shown in figure~\ref{f:geosensitivity} indicate that a sensitivity of $\sin^2\theta_e \sim 10^{-5}$ is reachable with a kg-scale target mass. This integral approach is limited by the solar neutrino background, however. Indeed, for $\sin^2\theta_e \sim 5 \cdot 10^{-5}$, a similar number of $^{163}$Ho atoms is produced by both the sterile and solar neutrinos. Assuming a one percent uncertainty in the knowledge of the solar neutrinos and the atom counting efficiency the sensitivity could asymptotically tend to $\sin^2\theta_e \sim 10^{-6}$, but not less. \\

Mineral ore containing dysprosium may contain traces of natural uranium and thorium. These radioactive contaminants produce neutrons via spontaneous fission and ($\alpha$,n) reactions on light elements (C, O, Na). The neutron flux induces the capture reaction $^{162}$Er(n,$\gamma$)$^{163}$Er (19~barn) followed by the EC decay of $^{163}$Eu (75~min), leading to $^{163}$Ho. Assuming an thermal neutron flux of $10^{-7}$ n/cm$^2$/s~\cite{Best:2015yma} and taking into account the self-absorption of neutrons on other isotopes, like Gadolinium, we estimate an integral production of less than 1000~$^{163}$Ho atoms for a 1~ton $^{162}$Er target exposed over a geologic time. Assuming a typical dysprosium-rich rock composition, such as the Adamsite~\cite{adamsite}, this production is small compared to the keV neutrino signal for $\sin^2\theta_e > 10^{-6}$. $^{163}$Ho can also be produced by (p,n) or (p,2n) reactions on $^{163}$Dy although these reactions only occur for $>$10~MeV protons. The (p,n) yield is therefore expected to be smaller than for (n,$\gamma$) processes. \\
\begin{figure}[ht!]
\begin{center}
\includegraphics[scale=0.6]{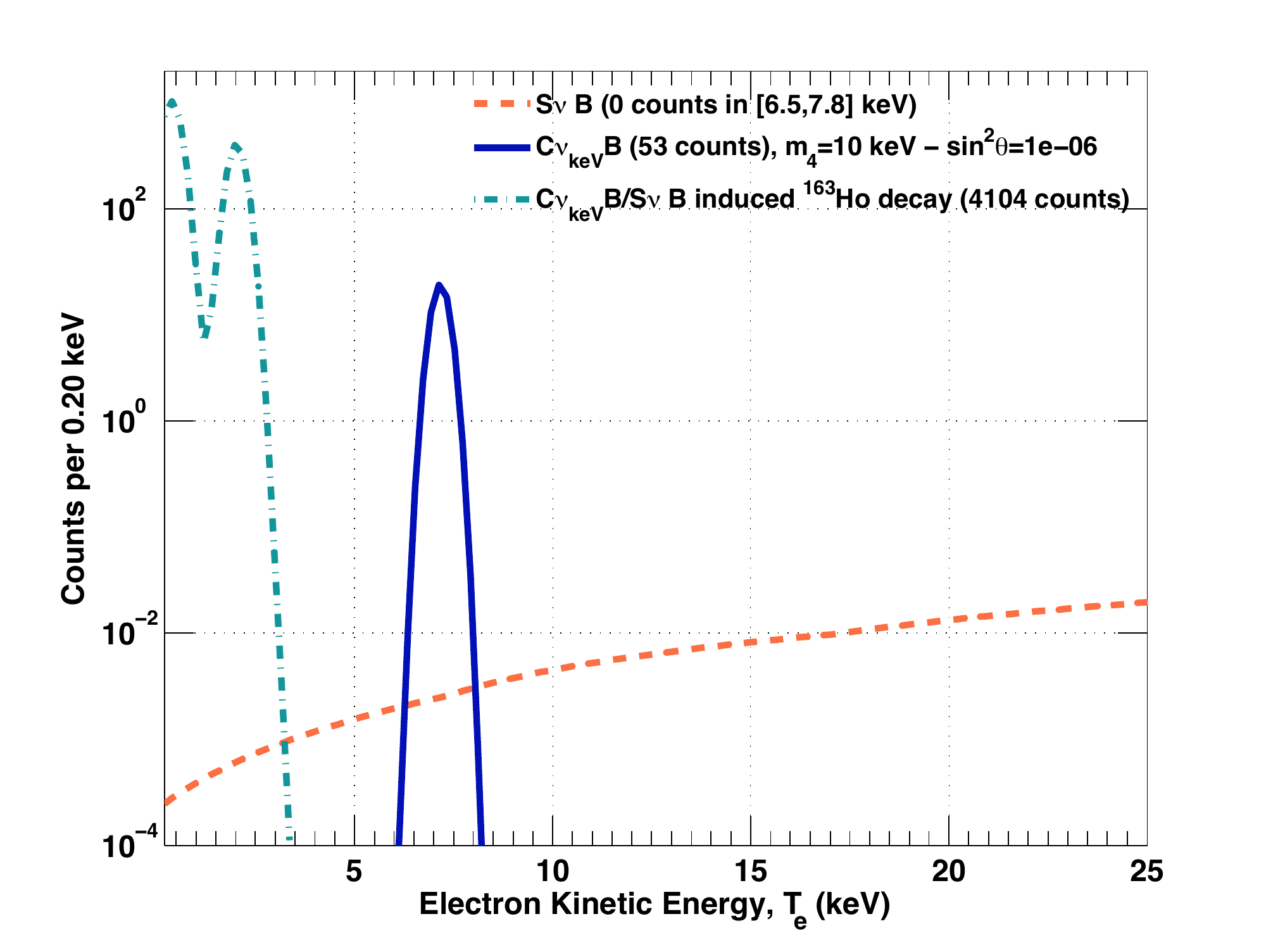}
\caption{\label{f:directexp} 
Electron spectra of sterile and solar neutrino captures on $^{163}$Dy and subsequent $^{163}$Ho decays, for an exposure of 100~ton$\cdot$year and a 0.5~keV resolution (FWHM).}
\end{center}
\end{figure}
To circumvent the limitation due to solar neutrino captures in the integral approach another technique, the real-time detection of keV neutrino captures inside an active $^{163}$Dy-based detector, can be considered. The characteristic signal is provided by the mono-energetic electron peak at ${\rm T}_e = \M  - 2.83$~keV. Let us consider a detector containing 10~tons of $^{163}$Dy exposed for 10~years (see figure~\ref{f:directexp}). For $\sin^2\theta_e = 10^{-6}$ and $\M $=10~keV the keV neutrino captures would induce a peak at T$_{\rm e,\rm peak}$=7.2~keV containing 53 electrons, to be discriminated against backgrounds. Compared to the integral case the solar neutrino background is strongly suppressed since less than 1 capture is expected within T$_{\rm e,\rm peak} \pm$~3~$\sigma_T$, assuming a resolution of 0.5~keV (FWHM). 

\begin{figure}[ht!]
\begin{center}
\includegraphics[scale=0.6]{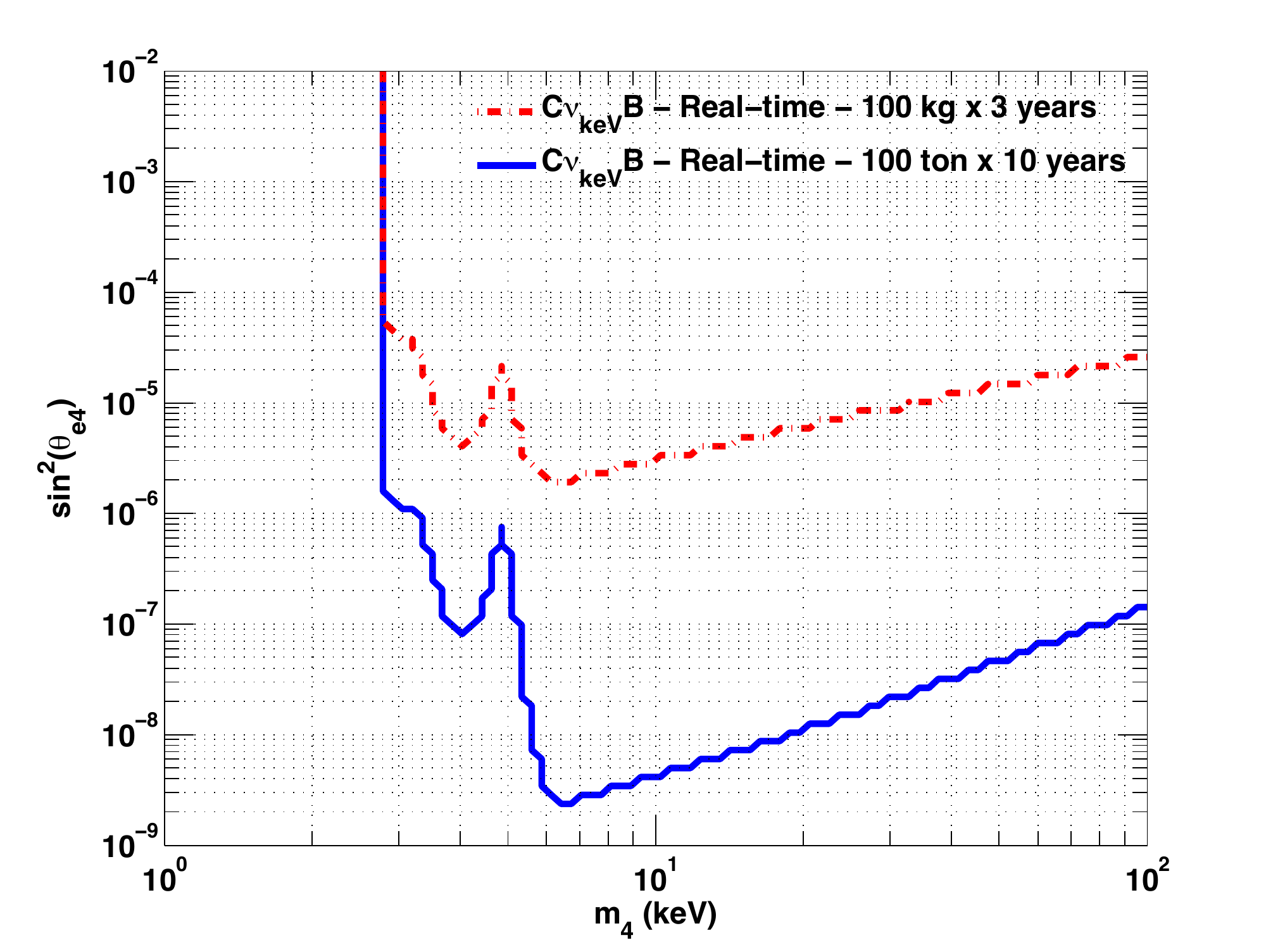}
\caption{\label{f:directsensitivity} 
90\% C.L. sensitivity for keV neutrino detection, in real time, for two exposures, 300~kg$\cdot$year and 1000~ton$\cdot$year (of $^{163}$Dy), assuming a resolution of 0.5~keV (FWHM).
Here $m_4$ is to be identified with the DM particle mass $M$ and $\theta_{e4}$ corresponds to $\theta_e$.
}
\end{center}
\end{figure}

In this configuration the real-time approach is entangled with the integral approach. Assuming the Dy-based detector was not purified from holmium atoms, the captures cosmic neutrino captures integrated over a geologic timescale would produce $^{163}$Ho atoms resulting in subsequent EC-decays affecting the low energy part of the spectrum as displayed in figure~\ref{f:directexp}. Figure~\ref{f:directsensitivity} shows the expected sensitivity for \rnb\, real-time detection for two different exposures, 300~kg$\cdot$year and 100~ton$\cdot$year, assuming a resolution of 0.5~keV (FWHM). Using a few 100~kg of $^{163}$Dy a 90\% sensitivity down to $\sin^2\theta_e \sim10^{-6}$ is attainable for $\M $ varying from 2.83~to 100~keV. Mixing angles as low as $\sin^2\theta_e \sim10^{-9}$ could in principle be explored with 100~tons of $^{163}$Dy, provided detector backgrounds are less than $10^{-6}$ counts/kg/day/keV, therefore about two order of magnitude below the expected reach of forthcoming DM experiments.

\subsubsection{Sterile Neutrino Scattering}
\label{exp-int-int}
Another option for the direct detection of the DM particle present in our galaxy is to search for sterile neutrino scattering (entangled to the mixing with an active neutrino flavour) in large-scale detectors. The density, energy distribution, and cross-sections of sterile neutrino DM is significantly different from WIMPs in typical CDM scenarios. Nevertheless, the ability of existing direct DM search experiments (such as Xenon~\cite{Aprile:2012nq} and LUX~\cite{Akerib:2013tjd}) to detect sterile neutrinos via elastic scattering is currently being investigated.

\begin{figure}
\centering
\includegraphics[width=.75\textwidth]{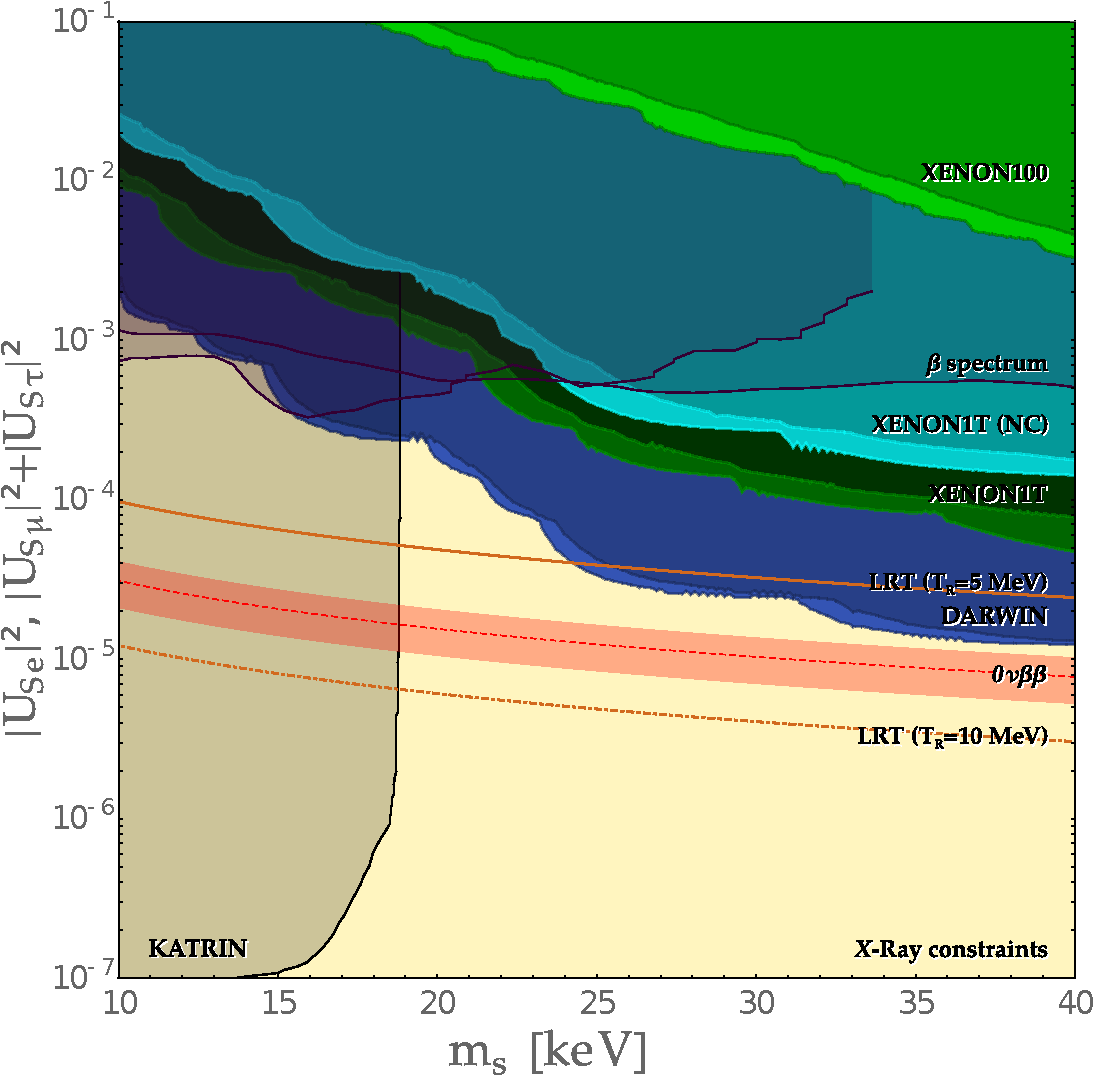}
\caption{\label{f:XenonExclusion} Light/Dark Green: Sensitivity on sterile neutrino WDM parameters for XENON100/XENON1T as a 
function of $\M $ and $|\theta_e|^2$, at $90\%$ and $99.9\%$ C.L.; Turquoise denoted ``XENON1T (NC)'': limit on $|U_{S\mu}|^2 + |U_{S\tau}|^2$ if the sterile neutrino does not couple to electron neutrinos and only has neutral currents; Blue: Equivalent for DARWIN; Purple: Current limits from analysis of $\beta$ spectrum of different radioisotopes 
 \cite{Holzschuh:1999vy,Holzschuh:2000nj}; Black: Expected statistical sensitivity of a modified KATRIN setup (Fig.\ 11 in \cite{Mertens:2014nha}); Red dashed: Limits coming from $0\nu\beta\beta$ experiments \cite{Rodejohann:2011mu}; Orange solid (and dot-dashed): Excluded area for production in case of a low reheating temperature (LRT) of $T_R=5$ MeV ($T_R=10$ MeV) \cite{Gelmini:2004ah}; Yellow: Constraint from $X$-ray searches \cite{Horiuchi:2013noa,Kusenko:2009up}. This figure has been taken from~\cite{Campos:2016gjh}.
 Here $m_s$ is to be identified with the DM particle mass $M$ and $U_{S\alpha}$ corresponds to $\theta_\alpha$.
 }
\end{figure}

Taking XENON100, XENON1T, and DARWIN as examples, the sensitivity for search keV-scale sterile neutrinos through  inelastic scattering with bound electrons is evaluated~\cite{Campos:2016gjh} and shown in Figure~\ref{f:XenonExclusion}. 

Other interesting ideas, such as a detection via an atomic excitation, are discussed in~\cite{Divari:2017coh}, cf. also \cite{Adhikari:2016bei}.

As a main result, the XENON1T experiment could explore a mass range between 10 and 40 keV while limiting
 the square of the mixing angle down to $\sim 5\times10^{-5}$.  An increased of the exposure by two orders of magnitude, as in the case of the DARWIN experiment, would allow to explore mixing angles squared down to $10^{-5}$.

\subsection{Detection through Sterile Neutrino Production}
\label{exp-prod}
The energy released in $\beta$-decay can be large enough for the production of a keV-scale sterile neutrino mass eigenstates. This fact can lead to two detectable signals of a sterile neutrino: 1) a  deformation of the beta decay spectrum, 2) a missing energy signal, via a full kinematic reconstruction of the decay.

\subsubsection{Beta Decay Spectroscopy }
\label{sec:beta-decay-spect}
\paragraph{General idea.}
The $\beta$-decay spectrum is given as a weighted superposition of the spectra corresponding to each neutrino mass eigenstate $m(\nu_i)$ the electron neutrino flavour is composed of. Since the mass splittings between the three light neutrino mass eigenstates are so small, no current $\beta$-decay experiment can resolve them. Instead, an effective light neutrino mass $m(\nu_{\mathrm{e}})^2 = \sum_{i=1}^{3} |(V_\nu)_{ei}|^2 m(\nu_i)^2$ is assumed. 

If the electron neutrino contains an admixture of a neutrino mass eigenstate with a mass $\M $ in the keV range, the different mass eigenstates will no longer form one effective neutrino mass term. In this case, due to the large mass splitting, the superposition of the $\beta$-decay spectra corresponding to the light effective mass term $m(\nu_{\mathrm{e}})$ and the heavy mass eigenstate $\M $, can be detectable. The differential spectrum can be written as
\begin{equation}
\label{eq:tritiumspec}
 \frac{d\Gamma}{dE} = \cos^2\theta \frac{d\Gamma}{dE}(m(\nu_{\mathrm{e}})) + \sin^2\theta \frac{d\Gamma}{dE}(\M ),
\end{equation} 
where $\theta$ describes the active-sterile neutrino mixing, and predominantly determines the size of the effect on the spectral shape~\cite{Shro80}. 

\begin{figure}
\centering
\includegraphics[width = 0.49\textwidth]{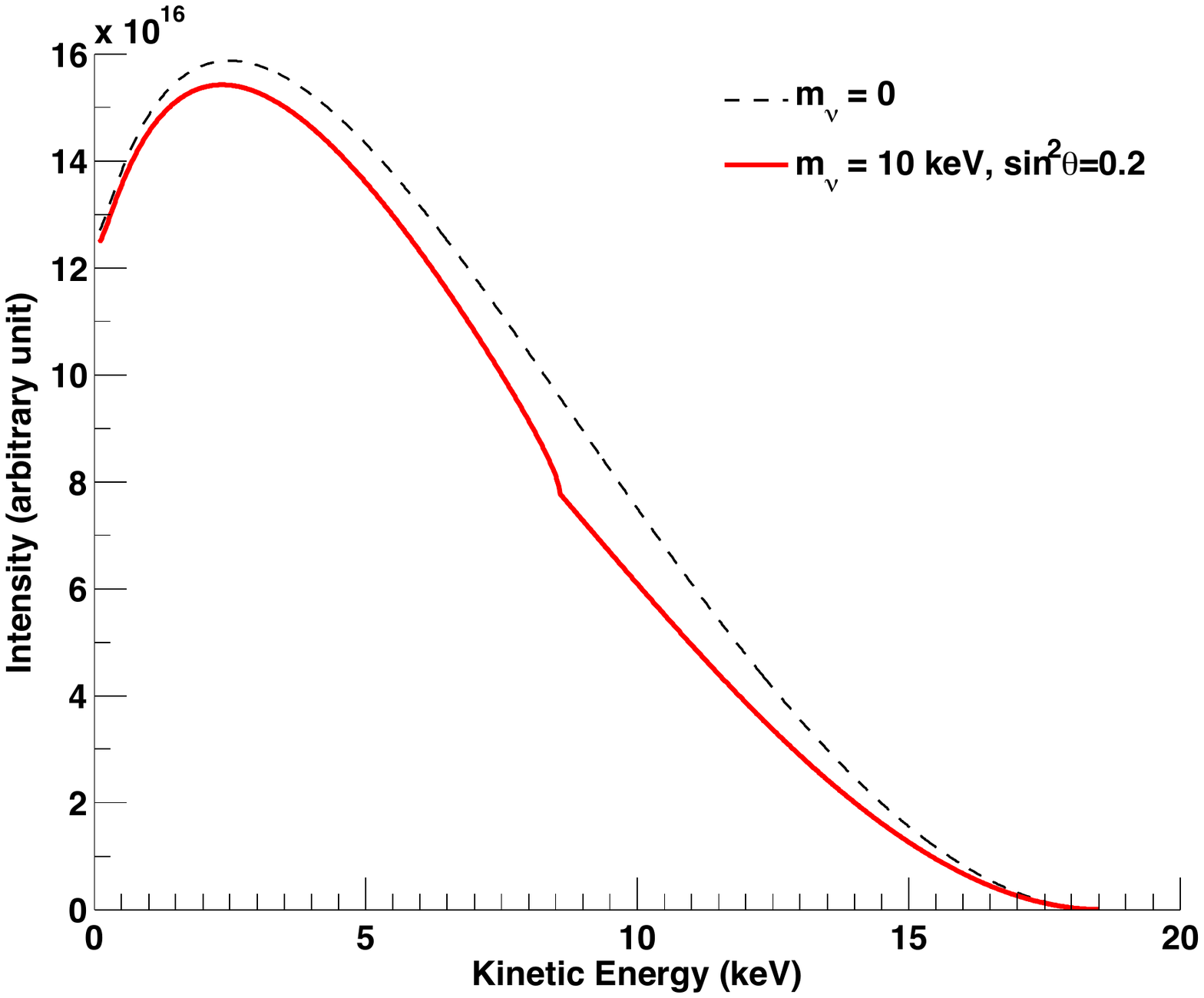}
\includegraphics[width = 0.49\textwidth]{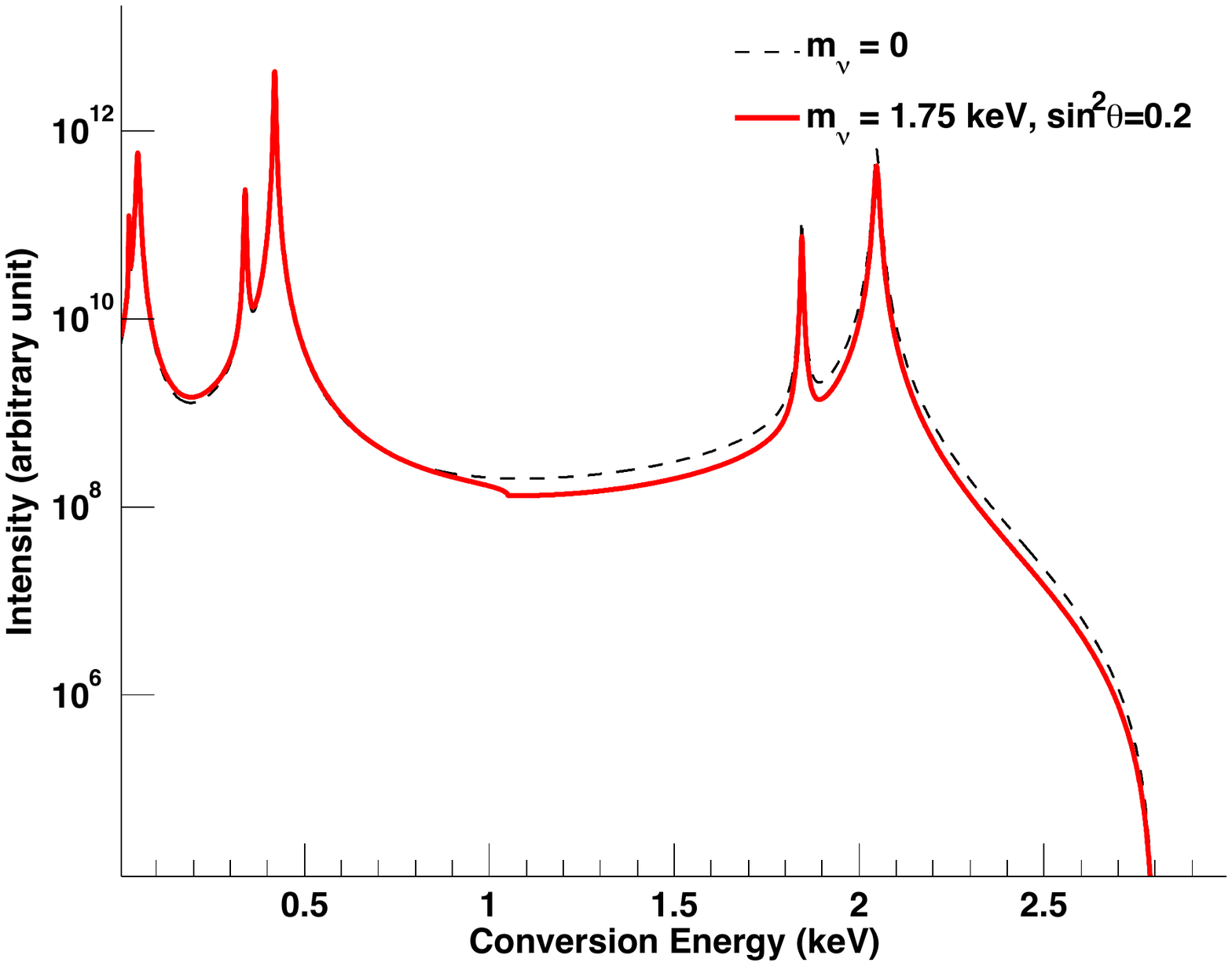}
\caption{\label{fig:spectrum}Left: tritium beta decay spectrum with the signature of a 10~keV sterile neutrino and an nonphysically large mixing angle of $\sin^2(\Theta) = 0.2$. Right: holmium-163 electron-capture spectrum with a signature of a 1.75~keV sterile neutrino and a mixing angle of $\sin^2\theta_e = 0.2$}
\end{figure}

The two isotopes under considerations for a direct determination of the absolute neutrino mass scale are tritium and holmium-163~\cite{DeRujula:1982qt}. Both isotopes allow in principle also for a search for sterile neutrinos. Figure~\ref{fig:spectrum} shows the signature of a sterile neutrino in tritium and holmium beta decay spectra with perfect energy resolution and no energy smearing from atomic, thermal or scattering effects. 

As tritium $\beta$-decay is a super-allowed decay the spectral shape is smooth and a precise theoretical description is possible. Furthermore, the life time of tritium of 12.4 years is rather short allowing for high signal rates with comparably low source densities, which minimises source related systematic effects. Finally, with an endpoint energy of 18.6~keV, tritium $\beta$ decay provides access to heavy neutrinos in a wide mass range~\cite{Mertens:2014osa, Mertens:2014nha}. 

Electron capture of holmium-163 has an endpoint energy of 2.8~keV, which correspondingly restricts the sterile neutrino search to a mass range of up to 2.8~keV. The energy spectrum of the double-forbidden decay of 163-Ho depicts several peaks corresponding to the atomic shells from which the electron is captured, see figure~\ref{fig:spectrum}. A precise theoretical description of the spectral shape is actively being pursued at the moment~\cite{PhysRevC.91.045505,ADR_ML_2015}. 

\begin{figure}[]
\begin{center}
\includegraphics[scale=0.6]{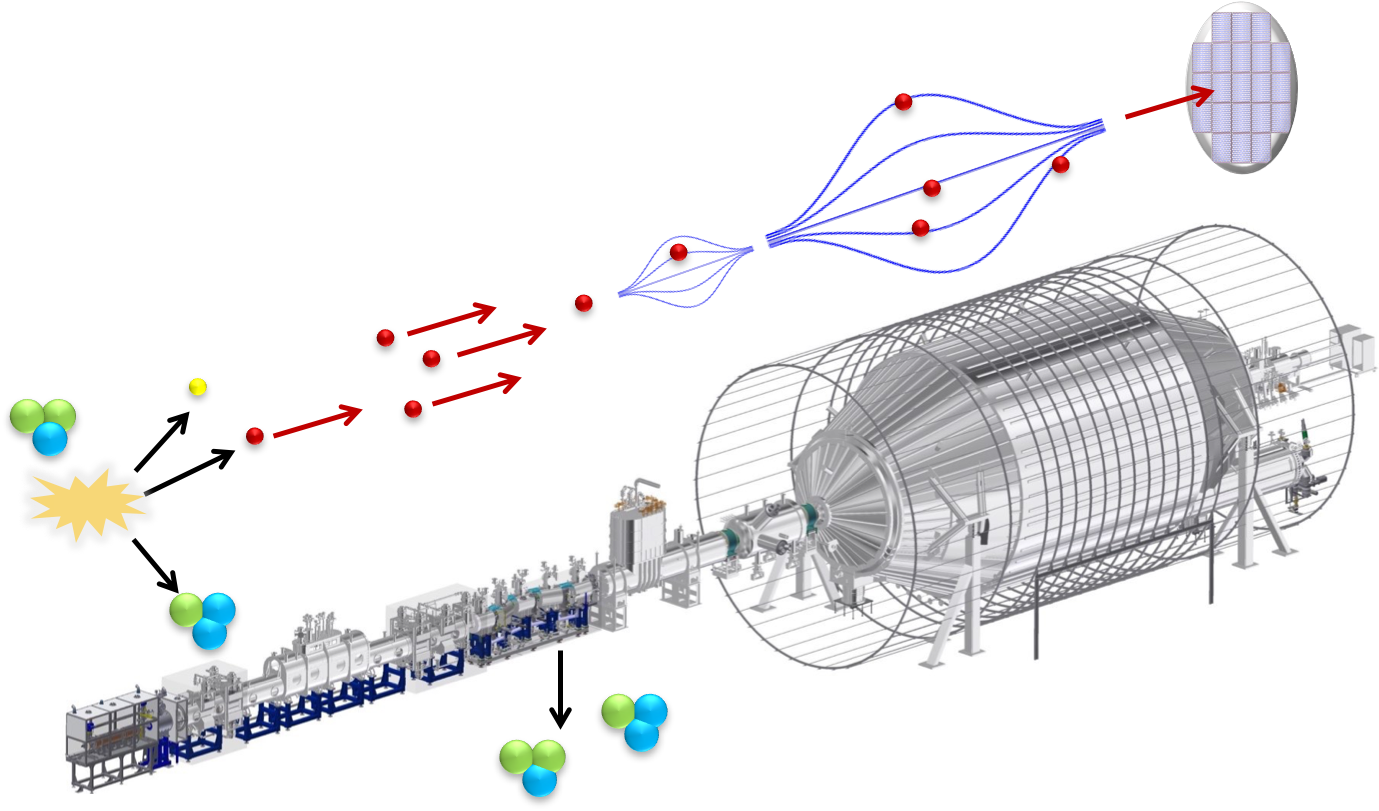}
\caption{\label{fig:KATRIN} 
Experimental setup of KATRIN: electrons created via $\beta$-decay in the 10-m long windowless gaseous tritium source are guided along magnetic field lines~\cite{Arenz:2018jpa} towards a large tandem spectrometer section, while a cryogenic and differential pumping system removes remaining tritium. The main spectrometer act as ultra-precise electrostatic filter~\cite{Arenz:2018ymp}, transmitting only those electrons with sufficient kinetic energy. By counting the number of transmitted electrons with a focal plane detector~\cite{Amsbaugh:2014uca} for different filter voltages the integral tritium $\beta$-decay spectrum is obtained. As the signature of a keV-scale sterile neutrino would appear further away from the endpoint $\mathrm{E}_0 = 18.6$~keV, the filter voltage has to be significantly reduced, which in turn increases the rate of transmitted electrons dramatically. Consequently, a novel multi-pixel detector system (displayed at the far right of the graph) will be required to search for sterile neutrinos with KATRIN. The design of such system is investigated in the framework of the TRISTAN project. The current baseline design is a silicon drift detector with 3500 pixels and a diameter of approximately 20 cm.
}
\end{center}
\end{figure}

\paragraph{Existing and proposed experiments}
Following ongoing and future experiments have the potential to search for sterile neutrinos via a characteristic distortion in the beta decay spectrum:
\begin{itemize}
\item Together with the Mainz experiment~\cite{Kra05}, the Troitsk nu-mass experiment currently holds the best limit on the effective electron anti-neutrino mass of $m_e < 2.05$ eV, from a direct measurement~\cite{PhysRevD.84.112003, PhysRevC.91.045505}. Currently, the measurements with the Troitsk nu-mass experiment are continued in a much wider energy range to set up limits on sterile neutrinos in a keV mass range~\cite{1748-0221-10-10-T10005}. In 2017, the best laboratory-based limit on keV-scale sterile neutrinos in a mass range of 0.1 - 2~keV was published~\cite{Abdurashitov2017} by the collaboration.

\item The KATRIN Experiment~\cite{Angrik:2005ep, Arenz:2018kma} is designed to measure the mass of the active neutrinos with a sensitivity of 200~meV (90\% CL). To do so it will measure the tritium $\beta$ decay spectrum close to its endpoint with unprecedented precision via the MAC-E-Filter technology. An extension of KATRIN's measurement interval to the entire tritium $\beta$ decay phase space, will require significant upgrades of the experimental setup shown in fig.~\ref{fig:KATRIN}. In the framework of the TRISTAN project this experimental extension of KATRIN is investigated~\cite{Mertens:2014osa,Mertens:2014nha,Dolde2017127}. In this context, also a possible time-of-flight measurement mode of the KATRIN experiment could be advantageous~\cite{Steinbrink:2017ung}.

\item A new idea to determine the neutrino mass via tritium beta decay is provided by the Project-8 experiment~\cite{Monreal:PhysRevD80051301:2009}. Here, the electron's energy is measured via its cyclotron radiation in a magnetic field. Recently, the Project-8 collaboration has successfully proven the feasibility of this new approach~\cite{PhysRevLett.114.162501}. In principle, as in the case of KATRIN, also here an extension of the region of interest, that would allow to search for keV-scale sterile neutrinos is possible. This approach is discussed in~\cite{Adhikari:2016bei}.

\item Ptolemy, a next-generation experiment, with the goal of detecting relic neutrinos, is currently being prototyped~\cite{betts2013development,HORVAT2017130}. Again, the apparatus could be utilised to also search for sterile neutrinos in the tritium beta decay spectrum, as discussed in~\cite{Adhikari:2016bei}.

\item The Electron Capture on Holium (ECHo) Experiment~\cite{Gastaldo:2013wha} is designed to measure the neutrino mass by a ultra-high precision of the endpoint region of holmium decay via micro-calorimeters. The main advantage in view of a sterile neutrino search is that ECHo (in contrast to e.g. KATRIN) automatically measures the entire energy spectrum, i.e. no major experimental upgrade is needed. Other experiments based on electron capture of holmium are the Holmes~\cite{HOLMES} and Numecs~\cite{NuMECS} experiment. 

\end{itemize}

\paragraph{The TRISTAN project}
As an example of a sterile neutrino search in $\beta$-decay, we would like to highlight some details on the technical realisation of the TRISTAN project. 

As mentioned above, an extension of the measurement interval of the KATRIN experiment requires a number of upgrades of the experimental setup, shown in fig.~\ref{fig:KATRIN}. 
Most notably a new detector and read-out system is needed. This system is required to handle extremely high counting rates (up to $10^{10}$ cps) and provide an excellent energy resolution (300~eV at 20~keV) at the same time. These requirements are met by the so-called Silicon-Drift-Detector technology~\cite{Gatti:1984uu}. 

Unlike, in the normal KATRIN measurement mode, such high-resolution SDD would allow to obtain a differential tritium $\beta$ spectrum. The combination of differential (energy-resolving detector) and integral (counting detector) spectra will be of key importance to mitigate a number of instrumental systematic uncertainties. The baseline design of the final TRISTAN detector is a 3500-pixel SDD system with a low-ADC-nonlinearity, high-sampling-rate waveform-digitiser. 

Currently, a 7-pixel prototype detector system produced by the semiconductor laboratory of the Max Planck Society (HLL)~\cite{HLL} and equipped with read-out electronics provided by the company XGLab~\cite{XGLab}, is being characterised. The measurements demonstrate excellent energy resolution of 130~eV at 6~keV and short shaping times (i.e. high rates), high linearity, and homogeneity among the pixels~\cite{Brunst:2018vka}. 

TRISTAN will prospectively be integrated in the KATRIN beamline in 2025, after the direct neutrino mass measurement program is completed. Prior to that, TRISTAN is being implemented at the Troitsk nu-mass experiment, where first differential and integral tritium-$\beta$-decay spectra were recorded and are currently being analysed.

\subsubsection{Full Kinematic Reconstruction}
Another idea, discussed in~\cite{PhysRevD.75.053005, Adhikari:2016bei, Smith:2016vku}, is a full kinematic reconstruction of a beta decay. In this case the sterile neutrino could be detected via the missing energy of a single $\beta$-decay event, by measuring the momenta of all other decay products.  

As in the above-mentioned spectral measurement, both the  $\beta$-decay~\cite{Adhikari:2016bei} and the electron capture~\cite{Smith:2016vku} process can be considered. The former requires a measurement of the momentum of the emitted electron and the recoiling daughter ion (assuming the mother nucleus was at rest), whereas the latter entails a precise determination of the momenta of emitted X-rays, Auger electrons, and the recoiling ion. In case of a  $\beta$-decay, the mass of the sterile neutrino would be given by
\begin{equation} 
  \M^2 = (Q-E_e-E_p)^2-(\textbf{p}+\textbf{k})^2
  \;,
\end{equation} 
where $E_e=\sqrt{m_e^2+\textbf{k}^2}-m_e$ and $E_p=\sqrt{M_p^2+\textbf{p}^2}-M_p$ are electron and recoil ion kinetic energies, respectively.  

For a neutrino mass measurement, the challenges of measuring energy and
momentum of daughter ion and electron with uncertainties of less than  1~eV, together
with the low rate near the endpoint, prevented a successful realisation of this idea. However, in case of a keV-scale sterile neutrino emission, the problems of reaching sufficiently
small uncertainties of the observables, as well as that of avoiding
high $\gamma$-factors, are very much relaxed. 

One of the main challenges is to achieve sufficient luminosity, while at the same time 1) keep the source small enough to allow for precise time of flight measurements of the recoiling nucleus, and 2) not too dense to prevent possible interactions within the source.  A possibility could be to create a supersonic gas jets with particle densities of about $10^{11}-10^{12}\mathrm{\,cm}^{-3}$ and temperatures of the order of 0.1~K or sources with magneto--optical traps with densities
of $10^{10}\mathrm{\,cm}^{-3}$ and temperature $\sim$0.1\,mK. These
methods would allow to generate $10^6-10^8$ beta decays per year for the source size of about 1\,mm$^3$. The COLTRIMS technique~\cite{0953-4075-30-13-006,DORNER200095} would allow to project
the total flux of the very slow recoil ions onto a multichannel plate
detector by help of a weak electric drift field. The ion's momentum would be
reconstructed from the coordinates of the hit channel and the time of
flight via the excellent angular and temporal resolution of this device. A detailed discussion of possible technical realisations can be found in~\cite{Adhikari:2016bei} and ~\cite{Smith:2016vku}.

\section{Conclusion}
\label{Conclusion}

\begin{figure}
\begin{center}
\includegraphics[width = 0.85\textwidth]{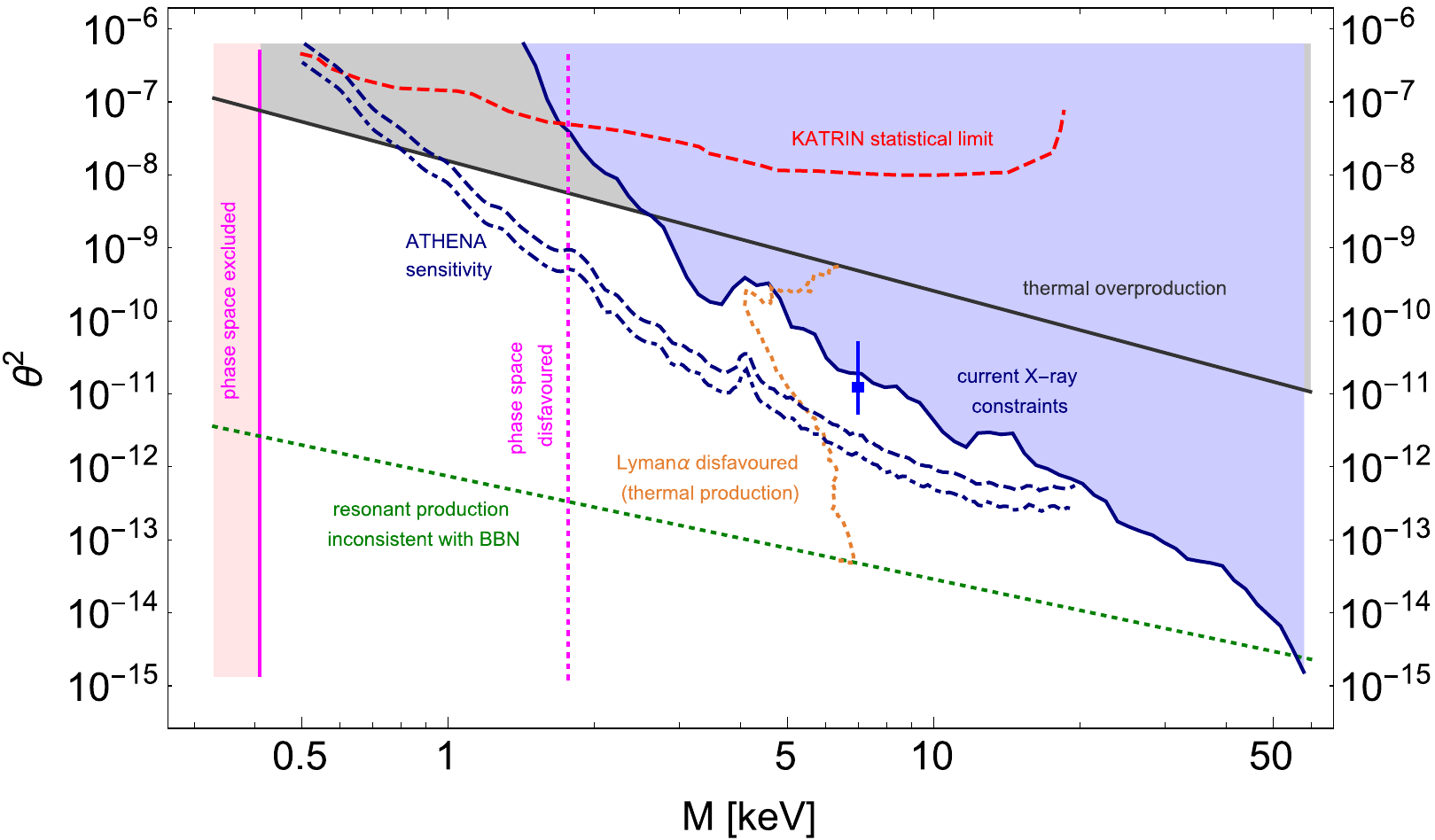}
\end{center}
\caption{\label{fig:SummaryPlot}  
 \textbf{Constraints on sterile neutrino DM.} 
 The solid lines represent the most important constraints that are largely \textbf{model independent}, 
 i.e., they can be derived for a generic SM-singlet fermion $N$ of mass $M$ and a mixing angle $\theta$ with SM neutrinos, without specification of the model that this DM candidate is embedded in.
  The  \emph{model independent phase space bound}  (solid purple line) is based on Pauli's exclusion principle (c.f. Section~\ref{sec:PhaseSpace}). 
  The bounds based on the \emph{non-observation of X-rays} from the decay $N\to \nu\gamma$ (violet area, see Section \ref{sec:decaying-dark-matter} for details) assume that the decay occurs solely through mixing with the active neutrinos with the decay rate given by eq.~(\ref{gamma}). 
  In the presence of additional interactions, these constraints could be stronger, see e.g.~\cite{Barry:2014ika}.
   All X-ray bounds have been smoothed and divided by a factor 2 to account for the uncertainty in the DM density in the observed objects.
They are compared to two estimates of the ATHENA sensitivity made in ref.~\cite{Neronov:2015kca}.
  The blue square marks the interpretation of the $3.5$ keV excess as decaying sterile neutrino DM~\cite{Bulbul:2014sua,Boyarsky:2014jta}. 
    All other constraints depend on the sterile neutrino production mechanism.
  As an example, we here show different bounds that apply to \textbf{thermally produced sterile neutrino DM}, cf. section~\ref{Sec:ThermalProduction}.
    The correct DM density is produced 
     for any point along black solid line 
     via the non-resonant mechanism 
    due to $\theta$-suppressed weak interactions (\ref{NWW}) alone (Section~\ref{Subsec:Nonres}). 
Above this line the abundance of sterile neutrinos would exceed the observed DM density.
We have indicated this \emph{overclosure bound} by a solid line because it applies to any sterile  neutrino, i.e., singlet fermion that mixes with the SM neutrinos.
It can only be avoided if one either assumes significant deviations from the standard thermal history of the universe or considers a mechanism that suppresses the neutrino production at temperatures of a few hundred MeV, well within the energy range that is testable in experiments, cf. e.g. \cite{Bezrukov:2017ike}. 
  For parameter values between the solid black line and the dotted green line, the observed DM density can be generated by resonantly enhanced thermal production (Section~\ref{Subsec:Res}).
    Below the dotted green line the lepton asymmetries  required for this mechanism to work are ruled out because they would alternate the abundances of light elements produced during BBN~\cite{Serpico:2005bc}.
 The dotted purple line represents the lower bound from phase space arguments that takes into account primordial distribution of sterile neutrinos, depending on the production mechanism~\cite{Boyarsky:2008ju}.
  As a \emph{structure formation bound} we choose to display the conservative lower bound on the mass of resonantly produced sterile neutrinos, based on the BOSS Lyman-$\alpha$ forest data~\cite{Baur:2017stq} (see 
 Section \ref{sec:structure-formation} for discussion).
  The structure formation constraints depend very strongly on the production mechanism (Section \ref{sec:production}). 
  The dashed red line shows the sensitivity estimate for the TRISTAN upgrade of the KATRIN experiment (90\% C.L., ignoring systematics, c.f.\ Section~\ref{exp-int}).
}
\end{figure}

Heavy sterile neutrinos are a very well motivated DM candidate. They overcome the shortcomings that exclude the SM neutrinos as viable DM candidates and e.g. appear in seesaw models that explain the light neutrino masses.
In practice there are, however, many constraints on the properties of sterile neutrinos as DM candidates. In the this article, we have presented a review of the physics behind those constraints and an update on the current status of the field.

The constraints can qualitatively be grouped into model dependent bounds and model independent bounds.
We refer to bounds as "model independent" if they only rely on the assumptions that the particles in question are massive fermions that carry no charges under any of the gauge interactions in the SM and that they mix with the known neutrino flavours $\nu_\alpha$. These properties can in fact be understood as the defining features of a heavy sterile neutrino. We display the most important model dependent bounds in figure \ref{fig:SummaryPlot}. 
Here we have assumed that a single species of sterile neutrinos $N$ with mass $M$ and total mixing $\theta^2$ comprises the entire observed DM density; all cosmological and astrophysical constraints would weaken if one relaxes these assumptions.

A model independent lower bound on the mass $M$ known as \emph{phase space constraint} comes from the fact that the particle number densities that would be required to explain the observed DM mass densities would be in conflict with Pauli's exclusion principle for very small $M$, cd. Sec.~\ref{sec:PhaseSpace}
For given $M$, there is an upper bound on $\theta$ from \emph{indirect DM searches} for X-ray emission from DM dense objects. 
We have reviewed the status of such searches in section \ref{sec:decaying-dark-matter}, including an update on the status of the unexplained 3.5 keV emission line that was reported in various observations (section \ref{sec:3_5kev}).
Both, the phase space constraint and the X-ray constraints, do not make any assumptions except for the validity of the known laws of gravity to establish the DM densities in astrophysical objects. 

If one in addition assumes a standard thermal history of the universe after big bang nucleosynthesis, which is suggested by the observation of light element abundances in the intergalactic medium and the Cosmic Microwave Background, 
and assumes that the particles and interactions during that epoch were the same as seen in laboratory experiments performed at the same energies, 
then one can impose another upper bound on the mixing angle $\theta$ for given mass $M$ from the requirement not to produce too much DM thermally through the weak interaction.
All other constraints (e.g. from the requirements that the lifetime of $N$ exceeds the age of the universe, that $N$ does not affect the reionisation of the universe and that the contribution to the effective number of relativistic degrees for freedom during big bang nucleosynthesis remains within observational limit), cf. section \ref{OtherConstraints}, are subdominant.

Within a given model one can impose even stronger constraints. These are usually connected to the mechanism of DM production in the early universe, which depends on the way how the sterile neutrinos are embedded into a more fundamental theory of nature. 
In order to be consistent with observation, it is not only necessary to produce the right amount of DM, but also to give the $N$ an initial momentum distribution that is consistent with the formation of the observed structures in the universe via gravitational collapse, cf. section \ref{sec:structure-formation}. This in particular imposes an upper bound on the free streaming horizon. We review different production mechanisms in section \ref{sec:production}. 
A minimal (unavoidable) amount of $N$ is produced thermally through oscillations and decoherent scatterings mediated by the weak interaction and the mixing $\theta$, cf. section \ref{Sec:ThermalProduction}. 
While the resulting spectrum tends to be colder than a thermal distribution, it is generally too warm to reproduce the observed small scale structures unless the production rate is resonantly enhanced due to a level crossing between the active and sterile neutrino dispersion relations in the primordial plasma known as Mikheyev-Smirnov-Wolfenstein (MSW) effect, cf. section \ref{Subsec:Res}.
The $\nu$MSM introduced in sec.~\ref{sec:numsm} is an example for a minimal model that can generate the matter potentials required for this crossing.
If the $N$ have new gauge interactions in addition to the $\theta$-suppressed weak interaction, these can significantly increase the thermal production rate without the a need for a MSW resonance, cf. section \ref{Sec:GaugeProduction}. Finally, the $N$ may also be produced non-thermally in the decays of heavy particles, cf. section \ref{Sec:DecayProduction}. All of these mechanisms tend to produce non-thermal momentum distributions. This and the fact that the deviations from Cold Dark Matter scenarios are expected to occur deeply in the non-linear regime of structure formation makes it highly challenging to translate astronomical observations into constraints on $M$ and $\theta$, cf. section \ref{sec:structure-formation}. In figure \ref{fig:SummaryPlot} we show a conservative estimate of the current constraints on the minimal thermal production mechanism via the weak interaction.

Sterile neutrinos can be further searched for directly and indirectly in the future.
Indirect searches on one hand involve future searches for X-ray emission. 
In particular, ATHENA or a re-built HITOMI satellite (also known as \emph{XARM}) would have a spectral resolution that could distinguish a narrow emission line from a broad feature in the X-ray spectrum. This will be crucial for the interpretation of the 3.5 keV excess or any other signal that may be discovered in the future.
On the other hand, one can expect a significant improvement of the structure formation bounds. Future optical telescopes will be able to resolve small scale structures at larger distances and provide better spectroscopical data on the Lyman$\alpha$ forest, while the Square Kilometer Array will open a new window of radio astronomy that allows to map the distribution of matter in the universe by 21cm observations. Moreover, a better understanding of baryonic effects in simulations of structure formation will reduce the systematic uncertainties in the translation of sterile neutrino properties into an observable matter power spectrum. 
In addition to these indirect searches, there are several proposals to search for DM sterile neutrinos directly in the laboratory, which we reviewed in detail in section \ref{exp-int}. 
These involve the possibility of sterile neutrino captures by nuclei (section \ref{exp-int-capt}) as well as scatterings off nuclei (section \ref{exp-int-int}), which somewhat resemble conventional direct searches for WIMPs.
Another possibility is to search for the impact of the sterile neutrino in $\beta$ decay spectra. This could e.g. be done with an upgrade of the KATRIN experiment (section \ref{sec:beta-decay-spect}). We display the statistical sensitivity of such a search in figure \ref{fig:SummaryPlot}. For low masses, it could indeed outperform the indirect searches for X-ray emission.

In summary, we conclude that sterile neutrinos are a theoretically very well-motivated DM candidate that can be searched for indirectly and directly.
If the sterile neutrinos are solely produced by the weak interaction, the combination of different constrains confines the allowed parameter region in all directions in the $M$-$\theta$ plane, cf. figure \ref{fig:SummaryPlot}.
Similar constraints can be obtained for the other production mechanisms (with the exception of the lower bound on $\theta$, which can be avoided if the DM production does not rely on mixing). This makes sterile neutrinos a very testable DM candidate, and at least the minimal models can be ruled out or confirmed within foreseeable time.

\section*{Acknowledgements}
MaD would like to thank George Fuller, Mikko Laine, Fedor Bezrukov,  Alexander Merle and Aurel Schneider for very valuable input during previous collaborations on which part of the present discussion is based.
This project has received funding from the European Research Council (ERC) under the European Union's Horizon 2020 research and innovation programme (GA 694896).
This research was supported by the DFG cluster of excellence
\emph{Origin and Structure of the Universe} (www.universe-cluster.de).

\bibliographystyle{elsarticle-num}
\bibliography{kevnudm}

\end{document}